\documentclass[fleqn,usenatbib]{mnras}

\usepackage{newtxtext,newtxmath}
\usepackage{textcomp}

\usepackage[T1]{fontenc}
\usepackage{ae,aecompl}

\DeclareRobustCommand{\VAN}[3]{#2}
\let\VANthebibliography\thebibliography
\def\thebibliography{\DeclareRobustCommand{\VAN}[3]{##3}\VANthebibliography}

\usepackage{graphicx}	
\usepackage{graphbox} 

\title[The outskirts of halo globular clusters]{A search for stellar structures around nine outer halo globular clusters in the Milky Way}

\author[S. Zhang et al.]{
Shumeng Zhang$^{1}$\thanks{E-mail: u5692595@alumni.anu.edu.au}
Dougal Mackey$^{1}$,
and Gary S. Da Costa$^{1}$
\\
$^{1}$Research School of Astronomy and Astrophysics, Australian National University, Canberra, ACT 2611, Australia}

\date{Accepted XXX. Received YYY; in original form ZZZ}

\pubyear{2022}

\begin{document}
\label{firstpage}
\pagerange{\pageref{firstpage}--\pageref{lastpage}}
\maketitle

\begin{abstract}
We use deep imaging from the Dark Energy Camera to explore the peripheral regions of nine globular clusters in the outer halo of the Milky Way. Apart from Whiting 1 and NGC 7492, which are projected against the Sagittarius stream, we see no evidence for adjacent stellar populations to indicate any of these clusters is associated with coherent tidal debris from a destroyed host dwarf. We also find no evidence for tidal tails around any of the clusters in our sample; however, both NGC 1904 and 6981 appear to possess outer envelopes. Motivated by a slew of recent {\it Gaia}-based discoveries, we compile a sample of clusters with robust detections of extra-tidal structure, and search for correlations with orbital properties. While we observe that clusters with tidal tails are typically on moderately or very eccentric orbits that are highly inclined to the Galactic plane and often retrograde, these are neither necessary nor sufficient conditions for the formation of extra-tidal structure. That many objects with tidal tails appear to be accreted leads us to speculate that this lack of consistency may stem from the inhomogeneous dynamical history of the Milky Way globular cluster system. Finally, we note that clusters with prominent stellar envelopes detected in ground-based imaging (such as NGC 1851 and 7089) are now all known from {\it Gaia} to possess long tidal tails -- experimental confirmation that the presence of an extended envelope is indicative of tidal erosion.
\end{abstract}




\section{Introduction}
Globular clusters are important tracers of the Galactic halo.  As compact, massive stellar systems, those that formed in now-accreted dwarf satellites were able to withstand the tidal destruction of their hosts and now inhabit the outskirts of the Milky Way.  The kinematics and chemical compositions of these clusters provide clues to their origin, and information about both the properties of their now-defunct parent galaxies and the history of mass assembly in the Galaxy \citep[e.g.,][]{searle:78,mackey:04,forbes:10,kruijssen:19,massari:19}. It is well established that such clusters can act as "signposts" indicating the possible location of coherent stellar debris from their host -- this is clearly seen, for example, in the case of Sagittarius, where stripped clusters such as Pal 12 and Whiting 1 remain deeply embedded in the tidal stream from the disrupting dwarf \citep[e.g.,][]{md:02,bellazzini:03,carraro:07}. Moreover, a substantial fraction of the remote globular clusters in Andromeda (M31) appear to trace the plethora of stellar streams and overdensities seen in the outer halo of this galaxy \citep[e.g.,][]{mackey:10,mackey:19a,huxor:14}. 

To date, searches for such debris around Galactic globular clusters using ground-based wide-field imaging have proven largely fruitless \citep[e.g.,][]{carballo:14,sollima:18}. Recent results from {\it Gaia}-based studies of the Milky Way field halo, at least in the Solar vicinity, suggest a possible explanation for this -- while multiple distinct accretion events can be clearly identified in kinematic and metallicity phase space \citep[e.g.,][]{helmi:18,belokurov:18,myeong:19,koppelman:19,naidu:20}, it appears that, with the exception of the ongoing merger of Sagittarius, these all occurred very long ago in Galactic history.  As a consequence, debris from the accreted systems (including field stars and globular clusters) is now spatially well mixed in the inner halo of the Milky Way.  Nonetheless, {\it Gaia} information on stellar populations further out in the Galactic halo is much sparser, and more distant clusters remain worthwhile targets for this type of search. Further motivation is provided by the availability of wide-field imagers on large telescopes -- such as the Dark Energy Camera (DECam) on the 4m Blanco Telescope, MegaCam on the 4m Canada-France-Hawaii Telescope, and Megacam on the 6.5m Magellan Clay Telescope -- that allow the outskirts of clusters to be contiguously observed to well beyond their limiting radii\footnote{In this paper we will typically refer to the {\it Jacobi radius} -- i.e., the radius of the approximately spherical equipotential surface where the gravitational forces due to the cluster, and to the Milky Way, are equal.} and to depths several magnitudes below the main-sequence turn-off.

In addition to the above, any stellar system evolving in an external tidal field experiences the gradual erosion of member stars as internal relaxation processes lead them to drift across the Jacobi (equipotential) surface and escape. The expected observational consequence is the formation of tidal tails, since in general the escaped stars remain on similar orbits to the parent system for some period of time. Tidal tails have been observed for many Galactic globular clusters -- the most famous examples being the long tails of Pal 5 \citep[e.g.,][]{odenkirchen:03} and NGC 5466 \citep[e.g.,][]{grillmair:06a,belokurov:06}. Such tails, if they can be detected, offer important leverage for both the study of stellar dynamics in a tidal field \citep[e.g.,][]{kupper:10,piatti:20b} and for measuring the properties of the Galactic dark matter halo \citep[e.g.,][]{koposov:10,koposov:19,bonaca:21}. 

As with many problems in near-field astrophysics, the availability of {\it Gaia} astrometry has led to a slew of new discoveries in this area of research, with many globular clusters now known to possess long (but generally extremely faint) tidal tails \citep[e.g.,][]{carballo:19,carballo:20,kaderali:19,sollima:20,ibata:19a,ibata:19b,ibata:21,palau:21,bonaca:21,kuzma:21}. However, {\it Gaia} studies are typically limited to relatively nearby clusters such that traditional ground-based wide-field imaging surveys are still useful for exploring more distant systems.

In this paper, we present deep panoramic DECam imaging of nine globular clusters in the outer halo of the Milky Way: NGC 1904, NGC 2298, NGC 6864, NGC 6981, NGC 7492, Whiting 1, Pyxis, AM 1, and IC 1257. Our aim is to conduct a homogeneous search for (i) coherent tidal debris from a putative destroyed host galaxy, and (ii) extra-tidal structure such as tidal tails. This continues our successful programme surveying the low surface-brightness outskirts of stellar systems in the Galactic halo, which has revealed extended stellar envelopes around several globular clusters including NGC 1261, NGC 1851, NGC 5824, and NGC 7089 \citep{kuzma:16,kuzma:18}, tidal tails emanating from the very remote globular clusters Pal 15 and Eridanus \citep{myeong:17}, and the tidal distortion of the low-luminosity dwarf galaxies Hercules, Sextans, and Bo\"{o}tes I \citep{roderick:15,roderick:16a,roderick:16b}. 

In Section \ref{s:data} we describe our observations and data reduction, in Section \ref{s:analysis} we present results for each of the nine clusters, and in Section \ref{s:discussion} we compile a set of new information on cluster tidal tails from {\it Gaia} with the aim of exploring different orbital properties as predictors of tail formation. Finally, we present a summary of our conclusions in Section \ref{s:conclusions}.

\section{Observations and Data reduction}
\label{s:data}
\subsection{Observations}
Basic information about our target clusters is shown in Table \ref{t:targets}. All nine sit at Galactocentric radii larger than $\approx 13$\ kpc, placing the entire sample in the outer halo of the Milky Way. Each target was observed with DECam on the 4m Blanco telescope at Cerro Tololo Inter-American Observatory in Chile. This instrument comprises a $62$ CCD mosaic, which spans a $3$ deg$^2$ field of view and has a pixel scale of $0.27\arcsec$\ pixel$^{-1}$ \citep{flaugher:15}.

\begin{table*}
\centering
\caption{Positional and distance information for our nine target globular clusters. Shown are the equitorial and Galactic coordinates of the cluster centre, $($RA$_0,\,$Dec$_0)$ and $(\ell_0,\,b_0)$ respectively; the distance from the Sun and from the Galactic centre, $D$ and $R_{\rm gc}$ respectively; and the position angle (east of north) of the Galactic centre with respect to the cluster, $\phi_{\rm gc}$. The coordinates come from \citet{harris:10}, and the line-of-sight distances from \citet{baumgardt:21}. We calculate $R_{\rm gc}$ assuming the Sun lies $8.178$\ kpc from the Galactic centre \citep{gravity:19}, and $\phi_{\rm gc}$ assuming the Galactic centre has coordinates $\alpha_{\rm gc} = 17^{\rm h}\,45^{\rm m}\,40.04^{\rm s}$, $\delta_{\rm gc} = -29\degr\,00\arcmin\,28.1\arcsec$.}
\label{t:targets}
\begin{tabular}{llccccccc}
\hline
Cluster & Other & RA$_0$ & Dec$_0$ & $\ell_0$ & $b_0$ & $D$ & $R_{\rm gc}$ & $\phi_{\rm gc}$ \\
Name & Name & \multicolumn{2}{c}{(J2000.0)} & \multicolumn{2}{c}{(degrees)} & (kpc) & (kpc) & (deg) \\ 
\hline
Whiting 1 &  & $02^{\rm h}\,02^{\rm m}\,57.0^{\rm s}$ & $-03\degr\,15\arcmin\,10\arcsec$ & $161.22$ & $-60.76$ & $30.59\pm1.17$ & $35.15$ & $234.7$ \\
AM 1 &  & $03^{\rm h}\,55^{\rm m}\,02.3^{\rm s}$ & $-49\degr\,36\arcmin\,55\arcsec$ & $258.34$ & $-48.47$ & $118.91\pm3.40$ & $120.28$ & $204.2$ \\
NGC 1904 & M79 & $05^{\rm h}\,24^{\rm m}\,11.1^{\rm s}$ & $-24\degr\,31\arcmin\,29\arcsec$ & $227.23$ & $-29.35$ & $13.08\pm0.18$ & $19.09$ & $185.8$ \\
NGC 2298 &  & $06^{\rm h}\,48^{\rm m}\,59.4^{\rm s}$ & $-36\degr\,00\arcmin\,19\arcsec$ & $245.63$ & $-16.00$ & $9.83\pm0.17$ & $15.08$ & $164.9$ \\
Pyxis &  & $09^{\rm h}\,07^{\rm m}\,57.8^{\rm s}$ & $-37\degr\,13\arcmin\,17\arcsec$ & $261.32$ & $+7.00$ & $36.53\pm0.66$ & $38.61$ & $136.9$ \\
IC 1257 &  & $17^{\rm h}\,27^{\rm m}\,08.5^{\rm s}$ & $-07\degr\,05\arcmin\,35\arcsec$ & $16.54$ & $+15.15$ & $26.59\pm1.43$ & $19.27$ & $169.3$ \\
NGC 6864 & M75 & $20^{\rm h}\,06^{\rm m}\,04.7^{\rm s}$ & $-21\degr\,55\arcmin\,16\arcsec$ & $20.30$ & $-25.75$ & $20.52\pm0.45$ & $14.30$ & $250.0$ \\
NGC 6981 & M72 & $20^{\rm h}\,53^{\rm m}\,27.7^{\rm s}$ & $-12\degr\,32\arcmin\,14\arcsec$ & $35.16$ & $-32.68$ & $16.66\pm0.18$ & $12.53$ & $241.7$ \\  
NGC 7492 &  & $23^{\rm h}\,08^{\rm m}\,26.6^{\rm s}$ & $-15\degr\,36\arcmin\,41\arcsec$ & $53.39$ & $-63.48$ & $24.39\pm0.57$ & $23.57$ & $243.6$ \\
\hline
\end{tabular}
\end{table*}

\begin{table}
\centering
\caption{Observing log for the nine target clusters.}
\label{t:log}        
\begin{tabular}{lclll}
\hline
Cluster & Filter & Exposures & Date & Seeing \\
\hline

Whiting 1 & $g$ & $5\times360$s & 2013 Sep 26 & $1.2-1.3\arcsec$ \\
          & $i$ & $5\times360$s & 2013 Sep 25-26 & $1.1-1.3\arcsec$ \\

AM 1 & $g$ & $4\times900$s & 2014 Feb 17 & $1.0\arcsec$ \\
     & $i$ & $11\times600$s & 2014 Feb 18-19 & $0.8-1.2\arcsec$ \\

NGC 1904 & $g$ & $3\times250$s & 2014 Feb 26 & $1.0-1.3\arcsec$ \\
         & $i$ & $3\times360$s & 2014 Feb 26 & $0.9-1.0\arcsec$ \\

NGC 2298 & $g$ & $3\times250$s & 2014 Feb 26 & $1.0\arcsec$ \\
         & $i$ & $3\times360$s & 2014 Feb 26 & $0.9\arcsec$ \\

Pyxis & $g$ & $5\times360$s & 2014 Feb 25 & $1.4-2.0\arcsec$ \\
      & $i$ & $5\times400$s & 2014 Feb 25 & $1.2-1.3\arcsec$ \\

IC 1257 & $g$ & $3\times250$s & 2014 Feb 26 & $1.1\arcsec$ \\
        & $i$ & $3\times360$s & 2014 Feb 26 & $1.1-1.2\arcsec$ \\

NGC 6864 & $g$ & $3\times300$s & 2013 Jul 13 & $1.3\arcsec$ \\
         & $i$ & $3\times300$s & 2013 Jul 13 & $1.1-1.3\arcsec$ \\

NGC 6981 & $g$ & $3\times300$s & 2013 Sep 25 & $1.1\arcsec$ \\
         & $i$ & $3\times300$s & 2013 Sep 25 & $1.0-1.1\arcsec$ \\

NGC 7492 & $g$ & $3\times300$s & 2013 Jul 12 & $1.3-1.4\arcsec$ \\
         & $i$ & $3\times300$s & 2013 Jul 12 & $1.2-1.4\arcsec$ \\
\hline
\end{tabular}
\end{table}

Table \ref{t:log} shows the observing log. Observations were conducted across four different runs between July 2013 and February 2014 (programs 2013A-0617, 2013B-0617, and 2014A-0621, all PI: D. Mackey; and 2014A-0620, PI: A. Casey). Conditions were generally very good, with seeing around $\sim 1\arcsec$ (the single exception being Pyxis, for which the $g$-band images have FWHM $1.4-2.0\arcsec$). For each cluster at least three $g$-band and three $i$-band exposures were taken, with dithering between the exposures to help remove bad pixels and cosmic rays, and fill the inter-chip gaps. Exposure times were set such that the signal-to-noise ratio S/N $=10$ for stars at least $\sim 2$ magnitudes below the main-sequence turn-off (MSTO). However, for some difficult targets like AM 1 (which sits at a very large distance) or IC 1257 (which has high foreground reddening), the exposures were only sufficient to reach the MSTO. 

\subsection{Photometry}
The raw images were reduced by the DECam community pipeline \citep{valdes:14}, and we obtained the processed data products from the NOAO science archive. The DECam community pipeline performs standard tasks including debiasing and dark correction, flat-fielding, illumination correction, cosmic ray and bad pixel masking, astrometric calibration, distortion correction, and remapping. While it also creates final stacked images for each pointing, we preferred to photometer the individual corrected and resampled frames on a CCD-by-CCD basis as described below.

Our photometry procedure followed that outlined in \citet{zhang:21} \citep[see also][]{koposov:15,koposov:18}, utilising the {\sc swarp}, {\sc sextractor}, and {\sc psfex} software packages \citep{bertin:96,bertin:10,bertin:11}. For a given target and filter we treated each of the $62$ DECam CCDs individually. We first aligned the single CCD frames and then created a coadded image using {\sc swarp}. Next, we ran {\sc sextractor} to identify the location of each source in this deep image. We then returned to the individual aligned CCD frames and used the brightest non-saturated sources in the detection list in combination with {\sc psfex} to compute a point spread function (PSF) model for each frame. Finally, we fed the full list of detected sources to {\sc sextractor} running in double image mode, and conducted forced photometry for all sources on every individual aligned CCD frame, using the appropriate PSF model for each.

For a given cluster and filter this resulted in $62$ lists of source detections, with $1-N$ photometric measurements per source (where $N$ is the total number of exposures for that particular target and filter). Before combining these measurements, we calibrated them to the SkyMapper third data release (DR3). We elected to use SkyMapper because its Southern Sky Survey \citep{wolf:18} provides uniform coverage in $g$ and $i$ for all clusters in our sample. At the time of writing DR3 is available only to the Australian astronomical community; however, its photometric system is the same as that constructed for the publicly-available second data release \citep[see][DR3 simply contains additional imaging]{onken:19}. 

To calibrate the photometry, we concatenated the $62$ lists and matched all stars to those in SkyMapper DR3 with magnitudes in the range $g = 15-19$, or $i =  16.5-19.0$ as appropriate.  Here, the bright end is set by the saturation level of our photometry, and the faint end by the precision of the SkyMapper photometry. We also chose only stars with SkyMapper colours $0.4 < (g-i) < 1.0$ to minimize the effect of non-zero colour terms between the DECam filters and SkyMapper. Across the relevant colour interval (matching the main-sequence turn-off and upper main sequence for the ancient, metal-poor populations found in our targets), these non-zero terms induce systematics at a level comparable to the size of our individual uncertainties (a few $\times 0.01$ mag) and we therefore decided not to correct them. In any case, (i) for some pointings large portions of the DECam frame were only sparsely populated, meaning that it would not have been possible to determine robust colour terms for all targets; and (ii) our analysis for each pointing (presented in Section \ref{s:analysis}) is almost entirely differential, rendering these small corrections unnecessary.

We used the matched stars to determine the mean offset across a single pointing for transforming our instrumental DECam measurements in a given filter to the SkyMapper photometric system. We also computed additional per-CCD corrections from the residuals after this mean offset had been applied, in order to correct for e.g., small differences in gain across the mosaic. After calibrating all measurements for a given star, we calculated the final magnitude and uncertainty via an error-weighted average.  Other photometric parameters of interest, such as the shape parameters {\tt fwhm}, {\tt ellipticity}, {\tt spread\_model}, and {\tt spreaderr\_model} provided by {\sc sextractor}, were also averaged.

This process produced two lists per cluster, containing the final $g$- and $i$-band photometry respectively. To obtain a final catalogue of stellar sources for each cluster we first performed an internal cross-match on each of these two lists to identify objects detected on multiple CCDs, which can happen due to the dither pattern used to cover the inter-chip gaps on the DECam mosaic. Measurements for any duplicate detections were combined in a weighted average. We then cross-matched the $g$- and $i$-band lists.  To try and eliminate non-stellar sources we utilised the criterion described by \citet{koposov:15} \citep[see also][]{desai:12,myeong:17}, keeping only objects for which $|${\tt spread\_model}$| <$ {\tt spreaderr\_model} $+\,\,0.003$ in both filters. Finally, we corrected our photometry for the effects of interstellar absorption on a star-by-star basis by using the extinction coefficents provided by \citet{wolf:18} (i.e., $2.986$ for the $g$-band and $1.588$ for the $i$-band) in combination with the \citet{schlegel:98} dust maps. Note that the \citet{wolf:18} coefficients implicitly account for the necessary rescaling of these maps discussed by \citet{schlafly:11}. 

\subsection{Determining the cluster membership}
\label{ss:membership}
\begin{figure*}
\centering
\includegraphics[trim=0mm 0mm 24mm 0mm, clip, height=60mm]{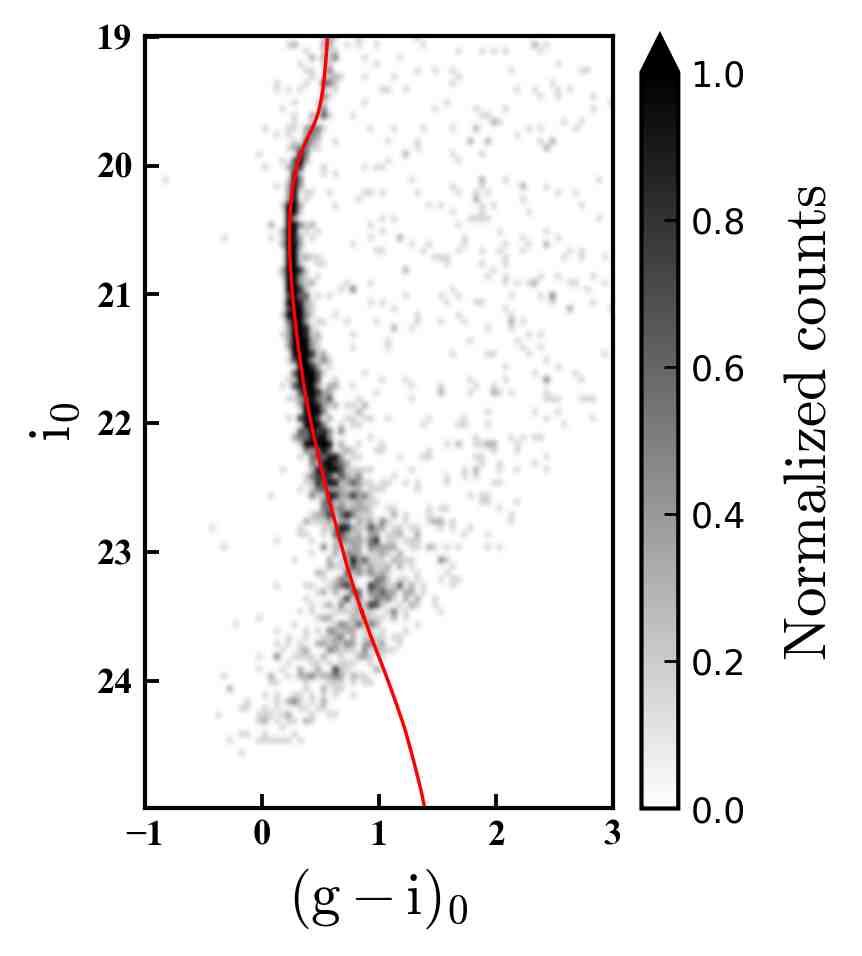}
\includegraphics[trim=0mm 0mm 24mm 0mm, clip, height=60mm]{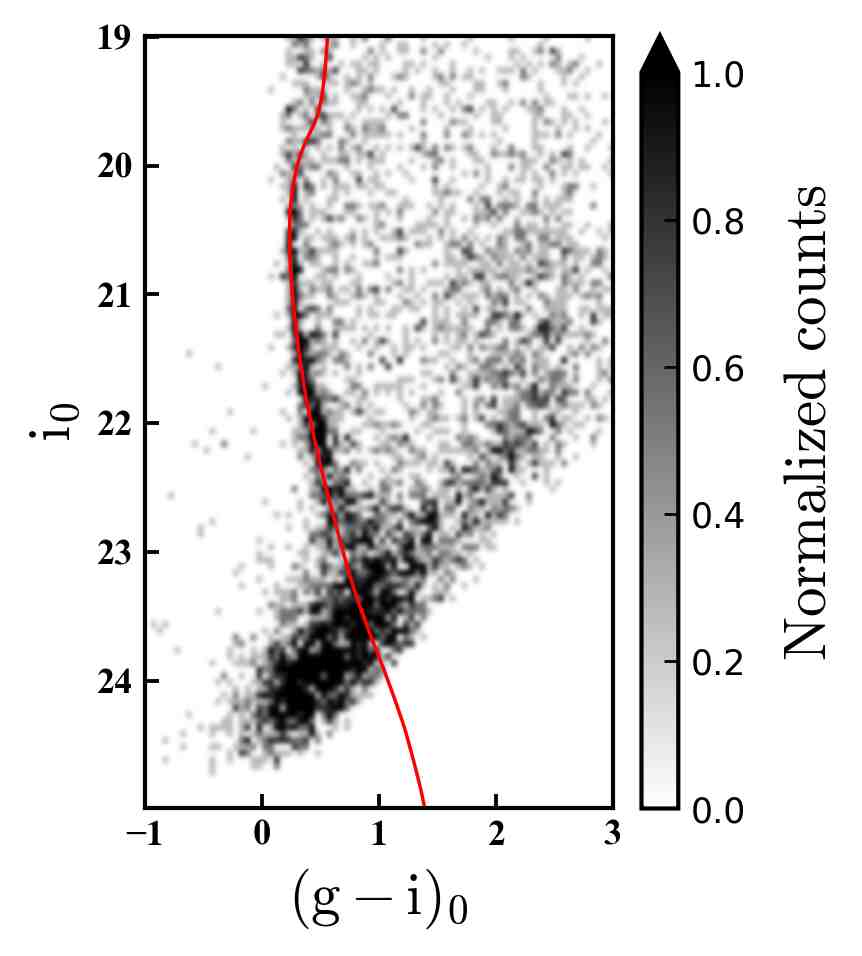}
\includegraphics[trim=0mm 0mm 24mm 0mm, clip, height=60mm]{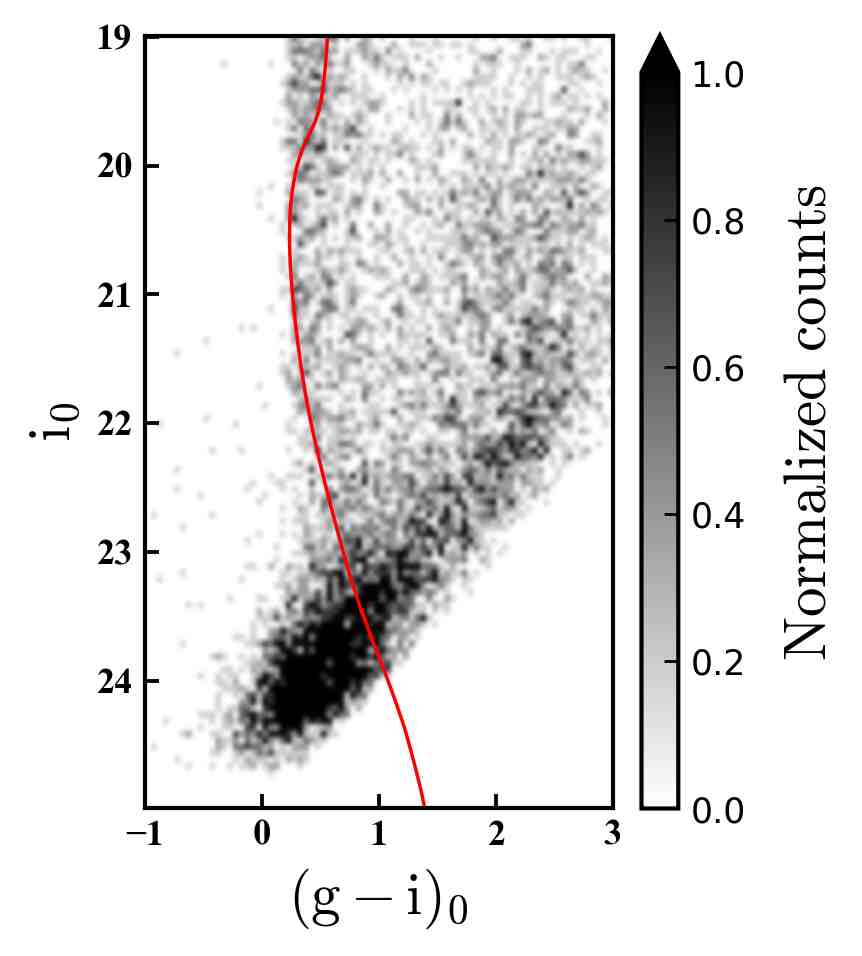}
\includegraphics[height=60mm]{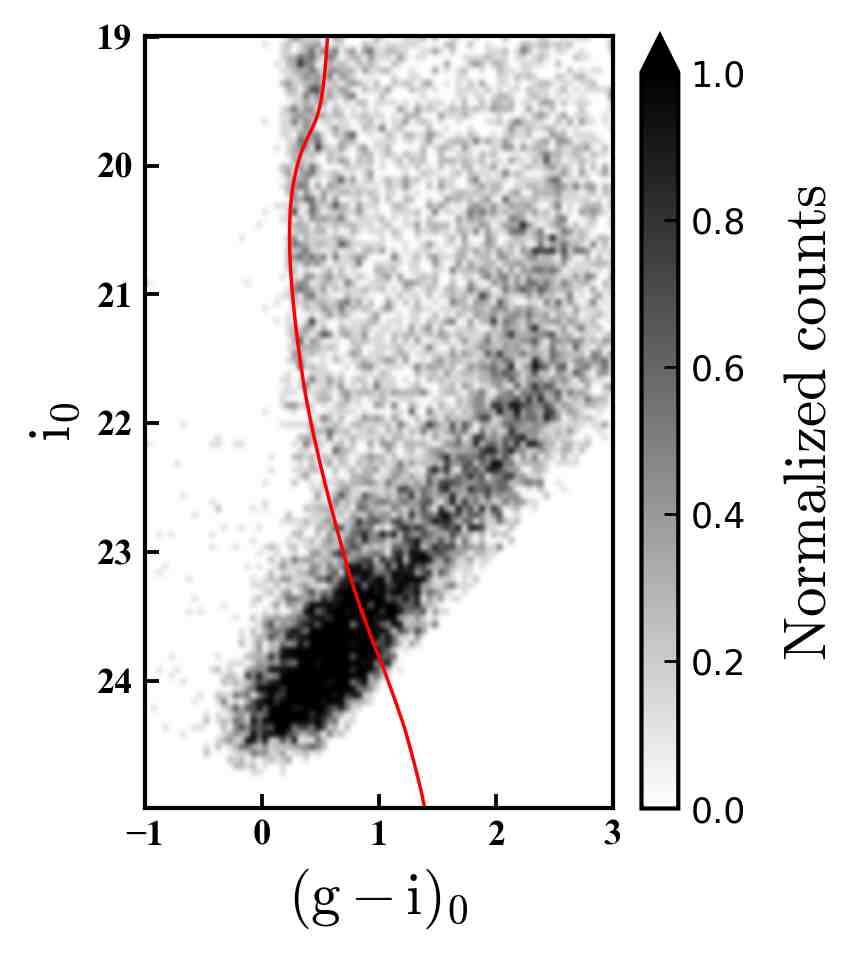}
\caption{Smoothed Hess diagrams for stars across a range of angular distances from the centre of NGC 6864. From left to right, stars in the radial range $0-5\arcmin$, $5-10\arcmin$, $10-15\arcmin$, and $15-20\arcmin$. The cluster's main sequence and main-sequence turn-off are clearly visible in the first two panels, but not in the second two panels (i.e., not at radii larger than $\approx 10\arcmin$). The red line marks the fiducial sequence determined for this cluster as described in Section \ref{ss:membership}.}
\label{f:cmd6864}
\end{figure*}

Figure \ref{f:cmd6864} shows example colour-magnitude diagrams (CMDs) at increasing angular distances from the cluster NGC 6864. The CMD for stars at small radii ($0-5\arcmin$) is, unsurprisingly, dominated by cluster members. The main-sequence turn-off (MSTO) can be seen near $i_0 \approx 20.5$, and our photometry additionally covers at least the top two magnitudes of the main sequence. Also present are contaminating populations due to Milky Way stars lying along the same line-of-sight, which predominantly sit to the red of the cluster main sequence. The main sequence is still clearly evident in the next annulus outwards ($5-10\arcmin$); however at radii larger than $\sim 10\arcmin$ it is absent, or at least sufficiently weak that it is not easily visible relative to the contaminants. The two outer annuli show that in addition to red colours, contaminating populations are also present at locations on the CMD coincident with the primary cluster sequences.

To explore the spatial distribution of stars in the outskirts of a given cluster, we require a method that can be used to identify the likely cluster members and separate them from the contaminants across the field of view. Following \citet{kuzma:16,kuzma:18} we introduce a simple membership criterion based on the proximity of a star to the prominent cluster sequences seen in the CMD \citep[see also the dwarf galaxy studies of][]{roderick:15,roderick:16a,roderick:16b,zhang:21}. For a given cluster we first find an isochrone from the MIST library \citep{choi:16} that closely matches the ridgeline of the main sequence and MSTO observed for stars in the inner regions of the cluster (e.g., the left-hand panel of Fig. \ref{f:cmd6864}). For most targets this means using stars within $\sim5\arcmin$ of the centre, but for smaller objects such as Whiting 1, AM 1, and IC 1257, it means using stars only in the inner $\sim1-2\arcmin$. As an initial estimate we start with the listed metallicity in \citet{harris:10} and assume an ancient age $\sim 12$\ Gyr \citep[except for Whiting 1, which is known to be somewhat younger, see][]{carraro:07}. We then vary the age and metallicity until the isochrone provides a close fit to the ridgeline of the main sequence and MSTO, as judged by eye. We emphasise that the aim here is not to obtain a precise age and metallicity measurement for each cluster, but simply to generate an accurate empirical representation of the cluster sequences.

For a given cluster we used this fiducial track to calculate a weight value, $w$, for each star across the field. The weight value is defined using the Gaussian distribution $N(x; \mu, \sigma)$, where $\mu(i_0)$ represents the colour of the track as a function of magnitude; $\sigma(i_0)$ is the mean photometric uncertainty in colour as a function of magnitude, derived from the distribution of individual stellar photometric errors produced by {\sc sextractor}; and $x=\Delta(g-i)_0$ is the horizontal distance between a star and the fiducial sequence. As such, $w$ quantifies the likelihood that the star is a member of the cluster, based on its position on the CMD relative to the primary cluster sequences, scaled by the photometric uncertainty. The calculation is normalized to have $w=1.0$ on the cluster ridgeline. 

We tune the member selection for each cluster by varying the threshold weight above which stars are considered likely to belong to the system. In general we adopt $w>0.3$ but this limit can be increased (made stricter) for clusters with heavy field contamination. We also need to impose bright and faint limits for selecting cluster members. This is done individually for each cluster, and depends on the depth of the photometry and our desire to only include regions of the CMD where the signal from cluster members is maximised relative to contamination. In general it is straightforward to select the bright limit: we always aim to retain stars around the MSTO since these are quite blue in colour, and their separation from contaminating populations on the CMD is hence comparatively large. However, the sub-giant branch (SGB) is redder and has many fewer stars, and so is typically not strongly enhanced relative to the field. As a consequence, our bright limit in most cases sits $\approx 0.5-1.0$\ mag above the cluster MSTO.

Selecting the faint limit is more complicated. Its value is constrained largely by the detection efficiency of the photometry software, which can vary (mildly) across the field-of-view due to small systematic changes in the image quality. Differences in the line-of-sight extinction across the field-of-view can also produce noticeable changes in the detection completeness at faint magnitudes. Selecting a faint limit without due care therefore has the potential to introduce spatial variations in the density of cluster members that are artificial in nature, but mimic the appearance of low surface-brightness structures near our target systems. 

In lieu of running extensive (but very computationally demanding) completeness tests across all our DECam pointings, we developed the following method to determine an appropriate faint selection cut-off for each cluster. We first select all stars in the field-of-view with colours similar to those of stars on the main sequence and MSTO of the cluster in question, usually $0.25 \leq (g-i)_0 \leq 0.60$ (although the latter value may be as large as $\sim 1.0$ for clusters where the photometry traces several magnitudes of main sequence). Next, we split this catalogue into nine bins spanning a $3\times 3$ grid across the field-of-view. For each bin we construct a histogram of $g_0$ and a histogram of $i_0$, and compute the magnitude for which each histogram has fallen to $75\%$ of the peak value on the faint side\footnote{For the bin containing the cluster centre we also exclude all stars within approximately two half-light radii \citep[as listed by][]{harris:10} when constructing these histograms, to avoid the results being biased by crowding. Our present study is focused on the distribution of stars in the outskirts of the clusters rather than their centres.}. Together, these measurements provide strong guidance in selecting an appropriate field-wide faint limit -- in general we select a value that is at least several tenths of a magnitude brighter than the brightest cut-off in the set. For several of our targets, the CMDs show large populations of unresolved galaxies at very faint magnitudes. In such cases we make further adjustments (brightwards) to the faint selection limit to avoid these areas.

Targeted small-scale completeness tests across several clusters indicate that this method leads to a faint cut-off that is always brighter than the $\approx 80\%$ completeness level. More critically, the potential effects of spatially varying detection incompleteness are almost entirely eliminated. In general, the final selection region for the nearby, well-populated clusters in our sample covers several magnitudes of the upper main sequence (spanning a range in stellar masses $\sim 0.5-0.8 M_\odot$). However, this range is substantially smaller ($\la 1$\ mag) for the more distant and/or heavily reddened targets.

While the procedure described above does a good job of maximising the selection of cluster members relative to non-members, some level of contamination inevitably remains due to sources lying coincident with cluster populations on the CMD (as in Fig. \ref{f:cmd6864}). To explore this, we cross-matched selected member stars in NGC 1904, 2298, and 6981 with {\it Gaia} EDR3. As the closest three clusters in our sample, these are the only targets with a sufficient overlap between the {\it Gaia} catalogue and our selection region to allow a useful number of cross-matches. We found that for each cluster, a large majority of matched stars are grouped within $\sim 1\sigma$ of the systemic proper motion value inferred by \citet{vasiliev:21} (where $\sigma$ refers to the per-star proper motion measurement uncertainty). However, each cluster also has a small fraction of proper motion outliers, varying from $\approx 3\%$ for NGC 1904, to $\approx 10\%$ for NGC 2298 (approximately correlating with the observed density of field stars). This experiment confirms the presence of residual contamination among our selected members; we take care to make additional corrections for this intrinsic background during our subsequent analysis.

\subsection{Radial density profiles}
\label{ss:rdp}
One simple method of examining the spatial distribution of cluster-like populations is via a one-dimensional radial density profile. This can help provide the first indication as to whether a cluster exhibits signs of unusual spatial extension or extra-tidal structure. To construct the radial density profile for a given cluster in our sample, we first split the catalogue of likely members (i.e., stars with weight $w$ above the selected threshold) into circular annuli about the cluster centre. We adopt annulus widths that increase as a function of projected radius to reflect the usual decline in stellar density -- typically we use widths of $0.5\arcmin$ for $(0\arcmin \leq r < 5\arcmin)$; $1\arcmin$ for $(5\arcmin \leq r < 10\arcmin)$; $5\arcmin$ for $(10\arcmin \leq r < 20\arcmin)$; and $10\arcmin$ for $(20\arcmin \leq r < 60\arcmin)$. Because of the large, contiguous DECam field-of-view, no correction for incomplete annuli is required. We use the outermost few measurements to determine the residual contamination level in the catalogue, and subtract this mean background from all points. Note that this method implicitly assumes that these regions are largely free from cluster members, and could result in subtracting some part of the desired signal if not; however, this assumption appears generally valid for the present sample (i.e., the mean background density is approximately constant when measured in the outskirts of each field).

For many targets we are unable to reliably determine star counts at small radii due to severe central crowding, leading to an apparent turn-over in the density profile. We are also generally unable to correct this by transitioning to integrated-light measurements at small radii because the cluster centre is often saturated in our images. To obtain radially-complete profiles we therefore elected, where possible, to supplement our measurements with data from \citet{trager:95}, who presented surface-brightness profiles for a large sample of Galactic globular clusters. To join the two data sets together for a given system, we shift the \citet{trager:95} measurements vertically until they agree with the points in our profile across the region of radial overlap (excluding any points in our data that are obviously affected by crowding).  Although \citet{trager:95} measured surface-brightness, as opposed to our star counts, joining the two data sets in this way is allowable provided the effective mass-to-light ratio does not change significantly as a function of radius.  This is an acceptable approximation here because (i) the mass ranges probed by the two data sets are very similar, with ours focused near the top of the main sequence, and the surface-brightness profiles dominated by the most luminous cluster stars (which have only recently evolved off the main sequence); and (ii) the overlap region between the two data sets generally occurs at large radii.

Note that we were unable to use the above method for Pyxis, IC 1257, and Whiting 1, which are not included in the \citet{trager:95} sample.  However, these three clusters are also relatively sparsely populated, such that we were able to reliably measure inwards of $1\arcmin$ for each.

Milky Way globular clusters exhibit a wide variety of structures, especially in their outer regions. Historically, the most widely used models are those of \citet{king:66}. These are characterised in their outskirts by a tidal (or truncation) radius $r_t$, which arises because stars have a finite escape velocity due to the external tidal field. An analytic approximation to the shapes of the models in this family is given by \citet{king:62}:
\begin{equation}
n= k\, \left[\frac{1}{\sqrt{1 + (r/r_c)^2 }} - \frac{1}{\sqrt{1 + (r_t/r_c)^2 }} \right]^{2}\,, 
\label{e:king}
\end{equation}
where $n(r)$ is the surface density as a function of projected distance from the cluster centre, $r_c$ is the core radius, and $k$ is a coefficient that is proportional to the density at $r=0$. More recently, however, it has been recognised that King models often provide a poor fit to the very low surface-brightness outskirts of clusters. In particular, many clusters appear more radially extended than can be explained by a King model \citep[e.g.,][]{mclaughlin:05,carballo:12}; in such cases models with a more gentle truncation, such as those of \citet{wilson:75} or the more general lowered isothermal family described by \citet{gieles:15}\footnote{Which includes both the \citet{king:66} and \citet{wilson:75} models.}, provide a much better description \citep[e.g.,][]{deboer:19}. 

Nonetheless, there are some cases where even these more extended models cannot adequately fit the data.  Clusters with large tidal tails, such as Palomar 5 or NGC 5466, extend beyond their Jacobi radius and tend to show a power-law density decline in their outskirts \citep[e.g.,][]{odenkirchen:03,fellhauer:07}. Similarly, a few clusters such as NGC 1851 and NGC 7089 (M2) \citep[e.g.,][]{olszewski:09,kuzma:16,kuzma:18}, apparently possess extended stellar envelopes that cannot be completely described by any lowered isothermal model; again, these examples tend to exhibit power-law profiles at large radii.

As a simple indicative measure, we fit our radial profiles using Eq. \ref{e:king}. Combined with our more informative two-dimensional density maps (described below), these help provide a straightforward test as to whether a given target might exhibit an extended envelope or tail-like structure in its outskirts. We emphasise that the main aim of the present work is not to conduct a detailed investigation of globular cluster structures -- far more sensitive and complete efforts already exist in that regard \citep[e.g.,][]{miocchi:13,deboer:19} -- but rather to search for spatially coherent groupings of members at or beyond the nominal "edge" of each system. 
      
\subsection{Two-dimensional density distributions}
\label{ss:2dmap}
Our primary tool in searching for extra-tidal structure is the two-dimensional distribution of cluster-like stars across the field-of-view. We create a raw map by dividing the catalogue of members for a given cluster into a grid of square bins across the field-of-view. In general we adopt a bin size of $0.004\degr = 14.4\arcsec$, which, after some experimentation, we found to strike a good balance between bins that are too small (which increases noise) or too large (which loses detail). In sub-dividing and examining the area around a cluster, we utilise a coordinate system with its origin at the cluster centre $({\rm RA}_0, {\rm Dec}_0)$, and with a correction for spherical distortion in the direction of Right Ascension. That is, the $x$-direction of the map has coordinates $({\rm RA}-{\rm RA}_0)\cos({\rm Dec})$ and the $y$-direction has coordinates $({\rm Dec}-{\rm Dec}_0)$.

\begin{figure}
    \centering
    \includegraphics[width=0.95\columnwidth]{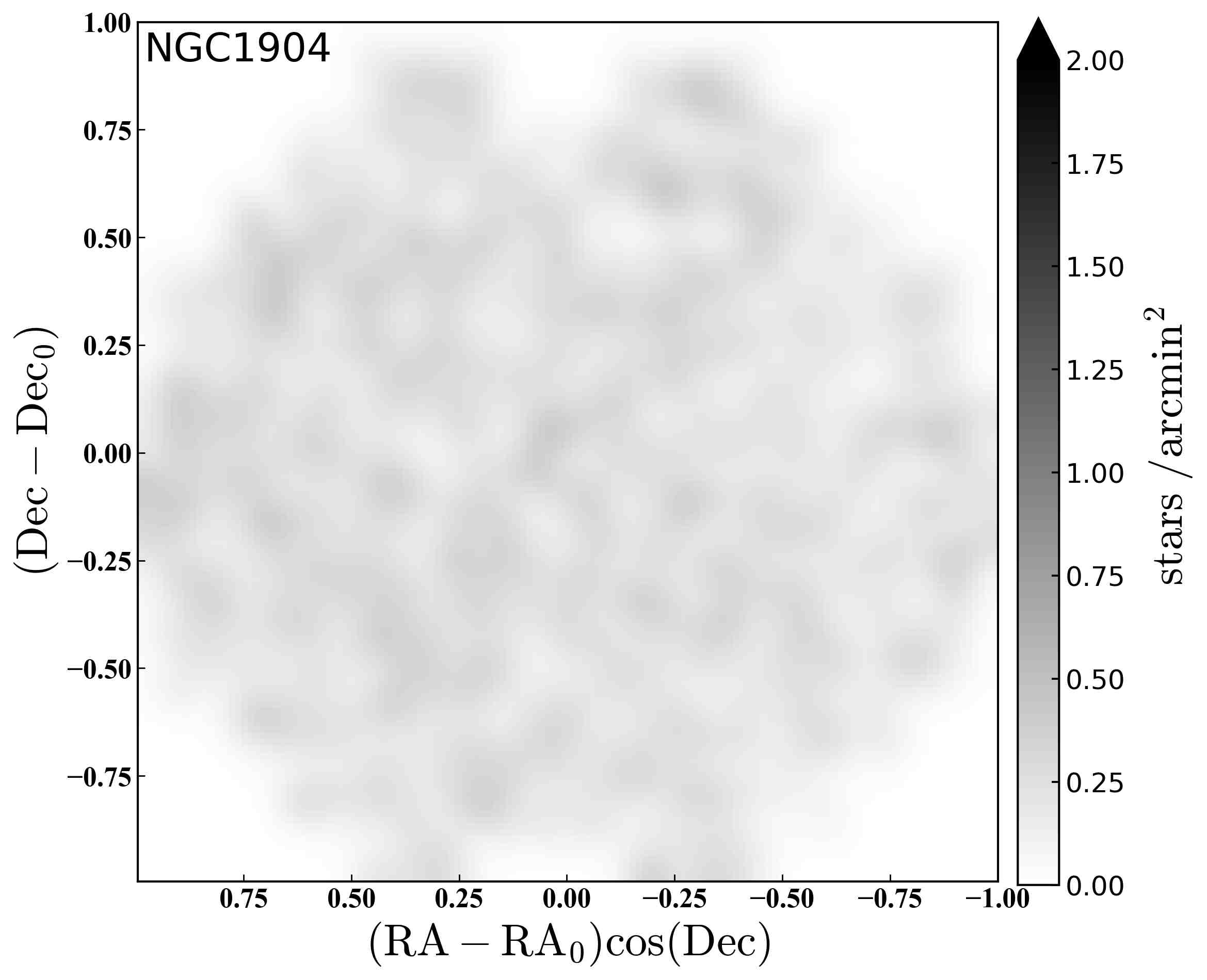}
    \caption{Example contamination map for NGC 1904, created as described in Section \ref{ss:2dmap}, and corresponding to step (iii) of our correction procedure. The distribution of non-members for this target is flat across the field-of-view, with evenly spread peaks and troughs and no large-scale structure. This is characteristic of almost all clusters in our sample (with the exception of Pyxis and IC 1257, as highlighted in the text). Note that in creating this map for display, we have adopted a larger bin size (of side $0.01\degr$) than that typically utilised for our analysis.}
    \label{f:contaminants}
\end{figure}

As noted above, residual contamination is still an issue; we attempt to correct for this by using a method similar to that introduced in, e.g., \citet{roderick:15,roderick:16a,roderick:16b,kuzma:16,kuzma:18,zhang:21}. First, we compile a list of contaminating sources with weight $w$ below the selected threshold. To ensure that this sample follows, as far as possible, a similar colour-magnitude distribution to the member sample, we further limit the selection to a rectangular region on the CMD with edges defined by the red, blue, bright, and faint limits of the member catalogue. We then divide this list into the same set of bins as for the member sample, to create a contamination map. We use this map to correct the member map according to the following procedure:
\begin{enumerate}
\item{The first step is to create a version of the contamination map that is scaled to a mean bin density of $1.0$.}
\item{We next divide the member map by this normalized contamination map, to account for any large-scale spatial variations in sensitivity (this is akin to a flat-fielding procedure).}
\item{Then we create a second version of the contamination map, rescaled such that the mean density in the outer parts of the field-of-view (typically at radii beyond $\sim 45\arcmin$) is equal to the mean density across the same region of the flat-fielded member map. Note that this radial limit is well outside the approximate expected tidal radii \citep[as listed by][]{harris:10}, which are $\la 15\arcmin$ for all clusters.}
\item{The final step is to subtract the scaled contamination map from step three from the flat-fielded member map, to produce a new map where the mean density at large radii is approximately zero, and the residual contamination has been removed.} 
\end{enumerate}
The above procedure works best when the contaminating populations are distributed approximately uniformly across the field of view, without steep density gradients. This is the case for a majority of clusters in our sample. In Figure \ref{f:contaminants} we show an example contamination map for NGC 1904, corresponding to step (iii) in our correction procedure. The distribution of non-members is flat, with no outlying density peaks or troughs, and little or no large-scale structure. This is typical for all targets except Pyxis and IC 1257, which sit at low Galactic latitude and have relatively heavy foregrounds. The contamination maps for these two clusters exhibit large-scale density gradients and mild patchiness, while their corrected density maps display residual variations towards the field edges. This limits the extent to which we can reliably detect cluster members at large angular radii.

\begin{figure*}
    \begin{minipage}{0.99\textwidth}
        \centering
        \includegraphics[align=c,height=6.6cm]{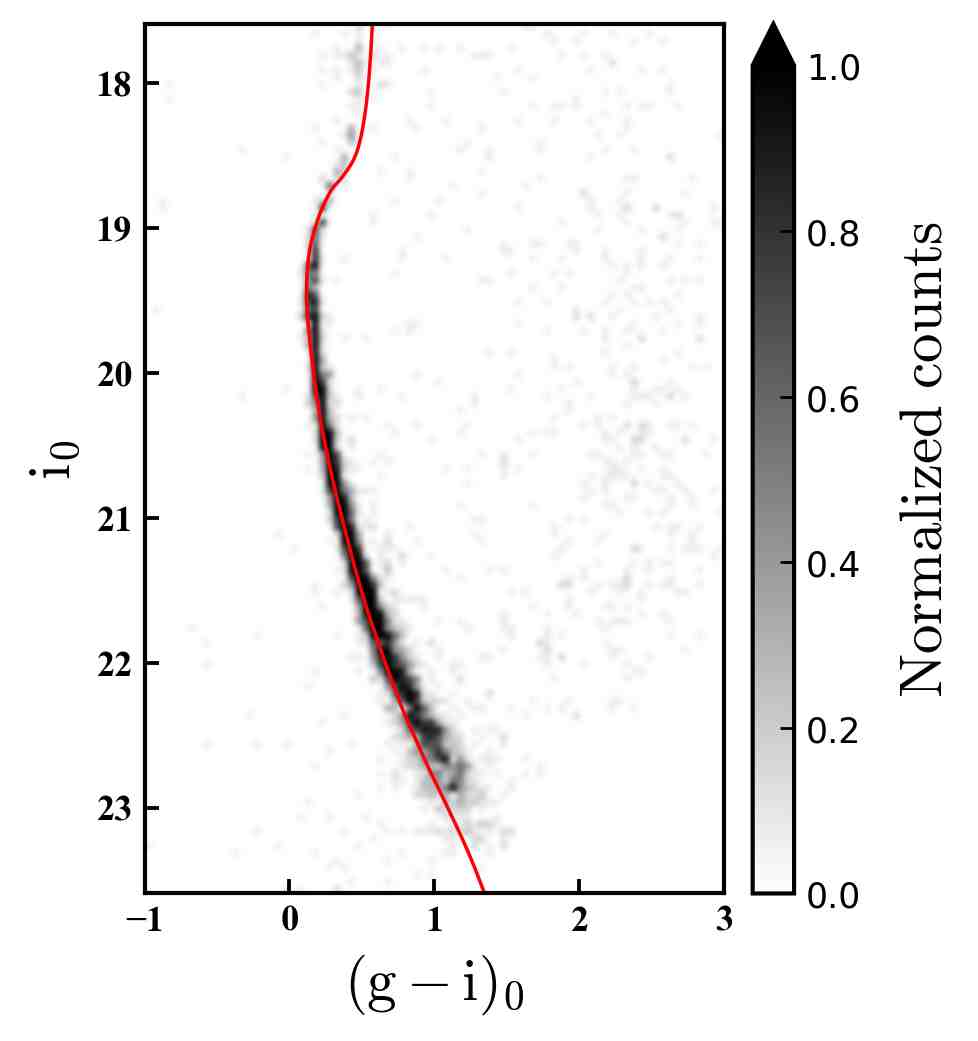}
        \hspace{-1mm}
        \includegraphics[align=c,height=6.6cm]{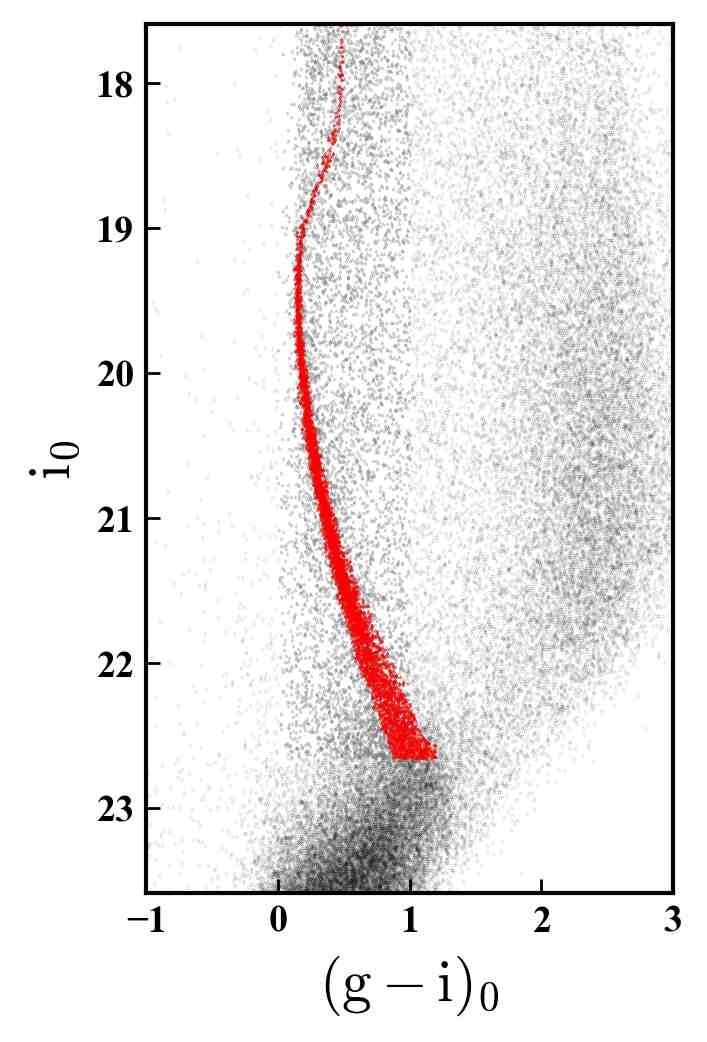}
        \hspace{1mm}
        \includegraphics[align=c,width=6.5cm]{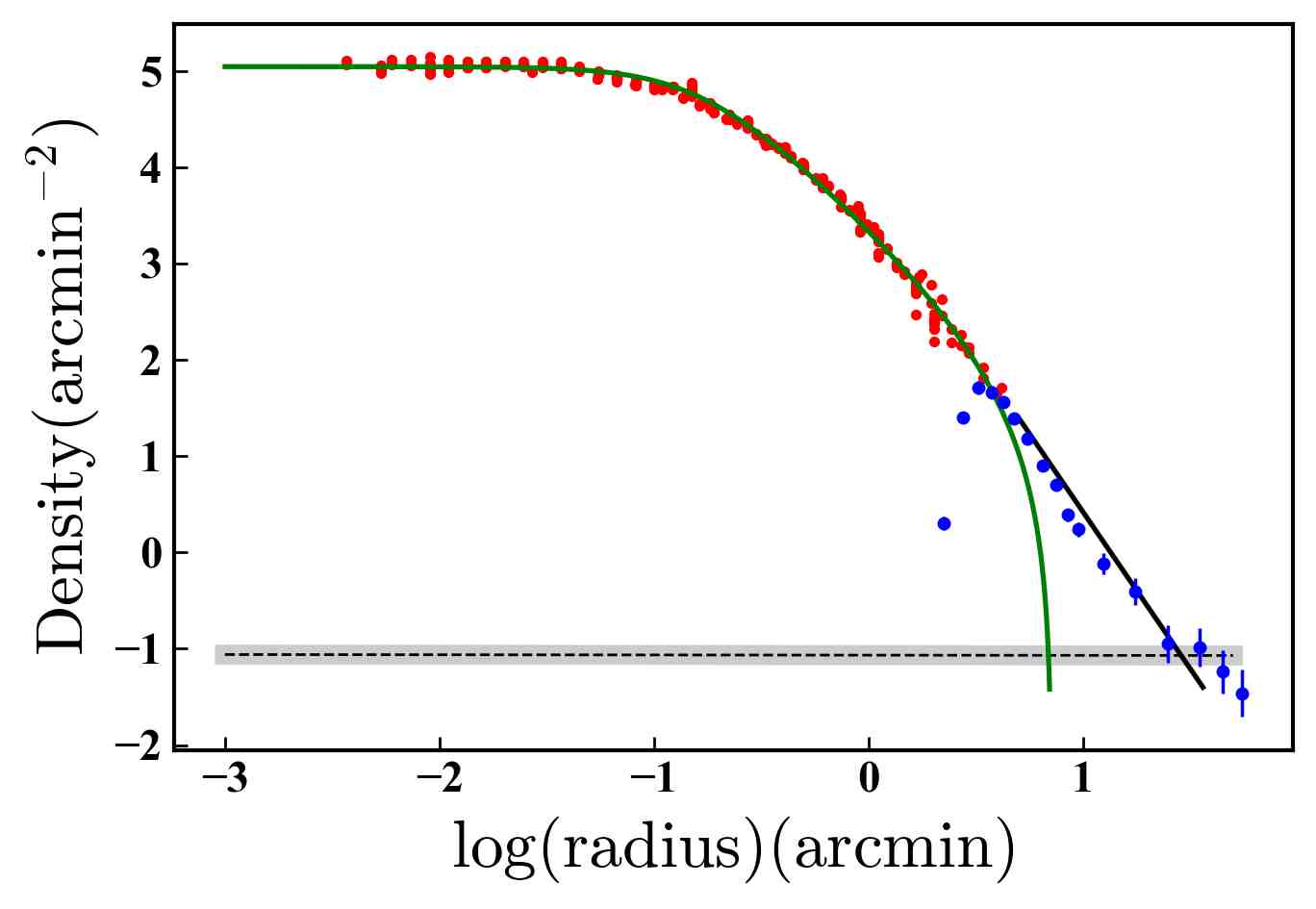}
    \end{minipage}\\
    \vspace{2mm}
    \begin{minipage}{0.99\textwidth}
        \centering
        \includegraphics[align=c,height=7.1cm]{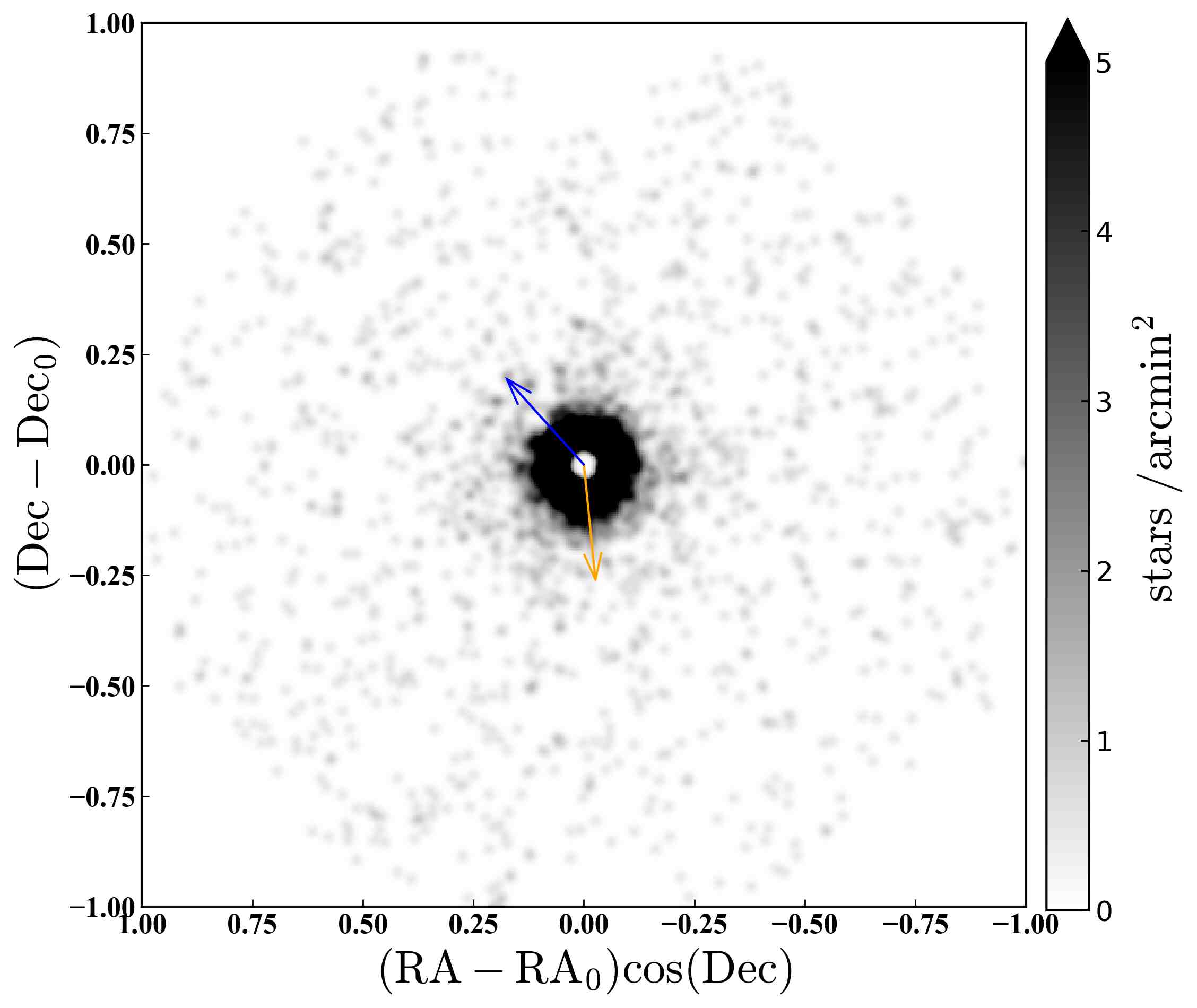}
        \hspace{2mm}
        \includegraphics[align=c,height=7.1cm]{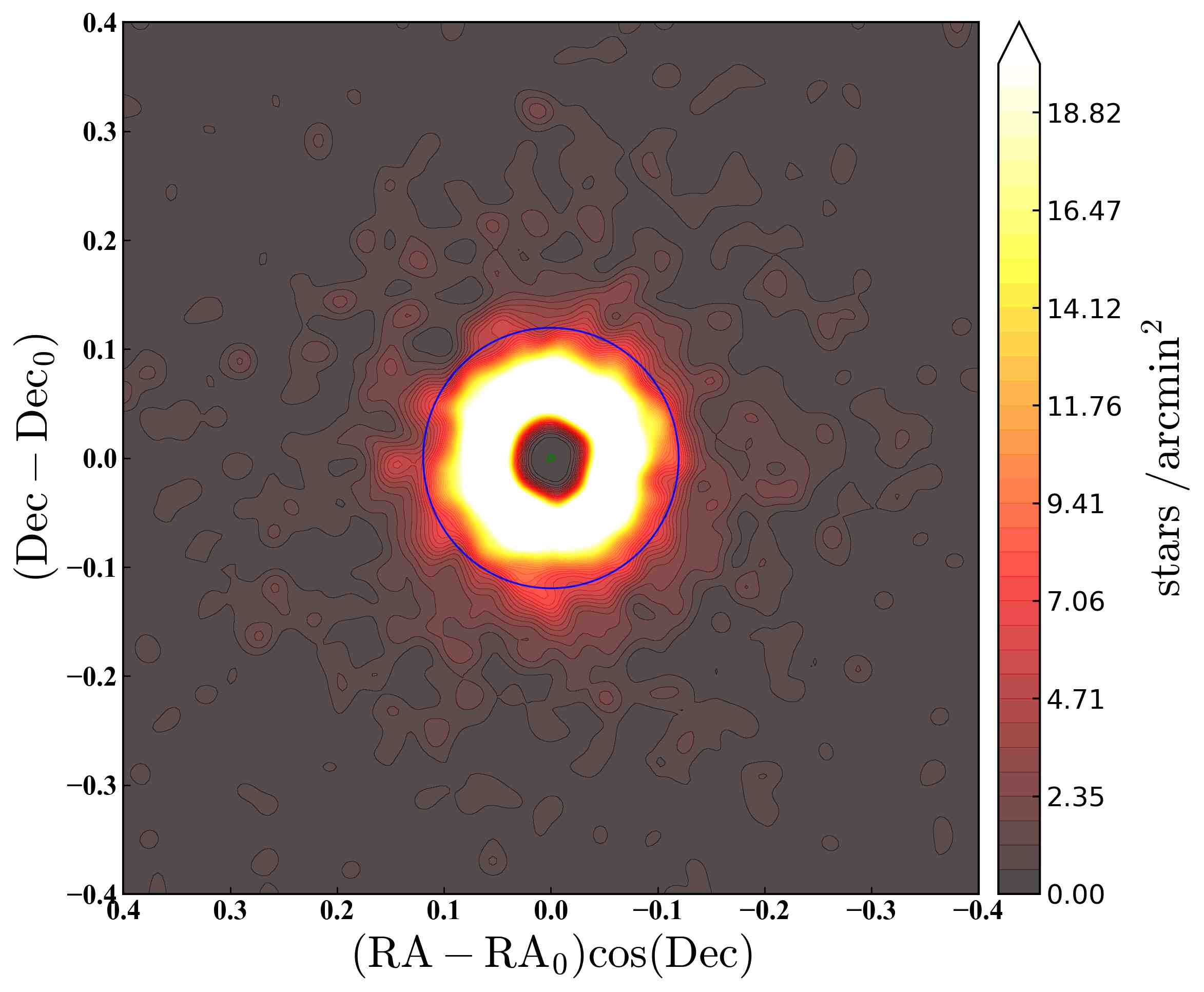}
    \end{minipage}
\caption{Structural data for NGC 1904. The top-left panel shows the CMD for all stars within $0-7\arcmin$ of the cluster centre, along with our fiducial track for determining membership (red line). The top-centre panel shows the CMD for all stars in the DECam field-of-view, with those satisfying the membership criterion $w>0.3$ marked in red. Stars used for defining the contamination map are marked in grey; these occupy a rectangular region surrounding the member selection. The top-right panel shows the radial density profile, with our data marked in blue and literature data from \citet{trager:95} in red. The green line indicates our best-fit \citet{king:62} model, derived as described in the text (Section \ref{ss:NGC1904}), while the black line shows our power-law fit to the outer points (slope $-3.25$). The black dashed line indicates the measured background level, while the grey shaded region covers $\pm 1\sigma$\ about this level, where $\sigma$ is the standard deviation of measurements conducted in four quadrants. The lower-left panel shows the contamination-corrected density map for the entire DECam field-of-view surrounding NGC 1904. North is up and east is to the left. The orange and blue arrows indicate, respectively, the direction of the Galactic centre, and the on-sky projection of the cluster's orbital motion (as calculated in Section \ref{ss:orbits}). This map uses a bin size of $14.4\arcsec \times 14.4\arcsec$, smoothed with a Gaussian kernel of width $2.5$ bins. The central region of incompleteness, where stars cannot be resolved due to severe crowding, is evident. The lower-right panel shows the region closest to the cluster as a contour map. Here, the blue circle marks the tidal radius derived from our King-model fit ($r_t = 7.2\arcmin$), while the (tiny) green circle at the centre indicates the core radius. Both density maps clearly show the stellar envelope surrounding NGC 1904, extending $\ga 20\arcmin$ from its centre.}
\label{f:NGC1904}
\end{figure*}

\section{Results and Analysis}
\label{s:analysis}
In this Section we present results for each of the clusters in our sample. We first consider large, well-populated clusters (including a couple with many previous measurements in the literature), followed by our more difficult (and less well studied) targets.

\begin{figure*}
    \begin{minipage}{0.99\textwidth}
        \centering
        \includegraphics[align=c,height=6.6cm]{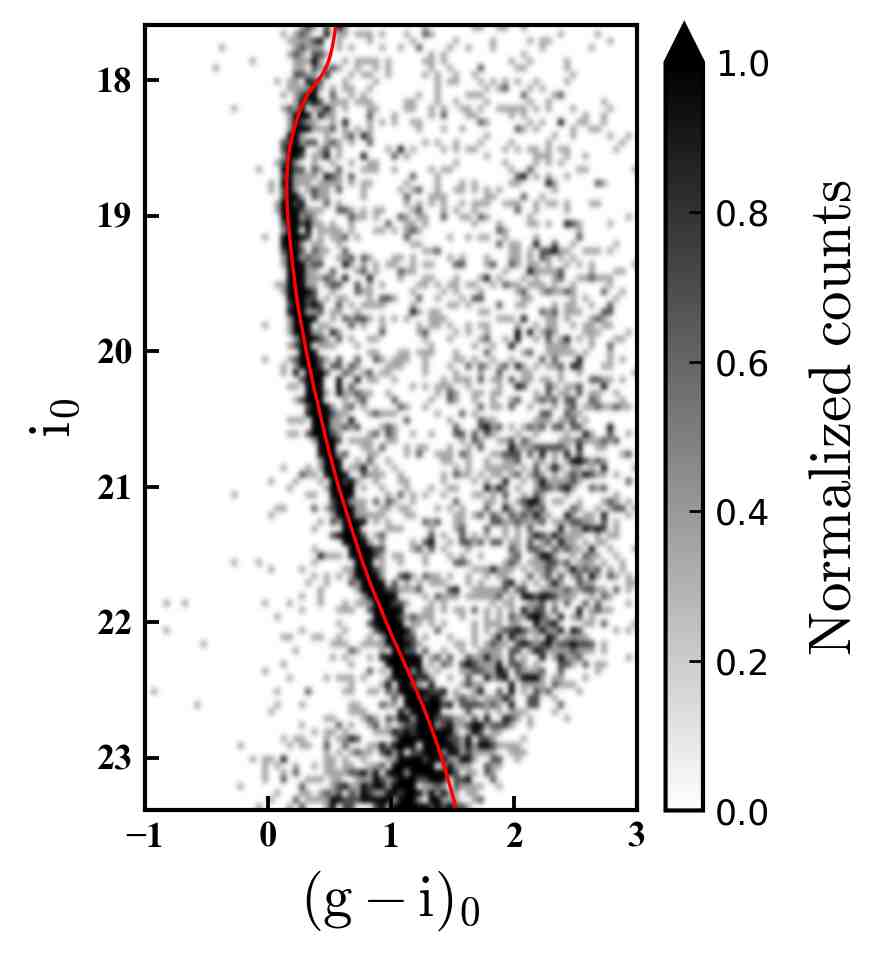}
        \hspace{-1mm}
        \includegraphics[align=c,height=6.6cm]{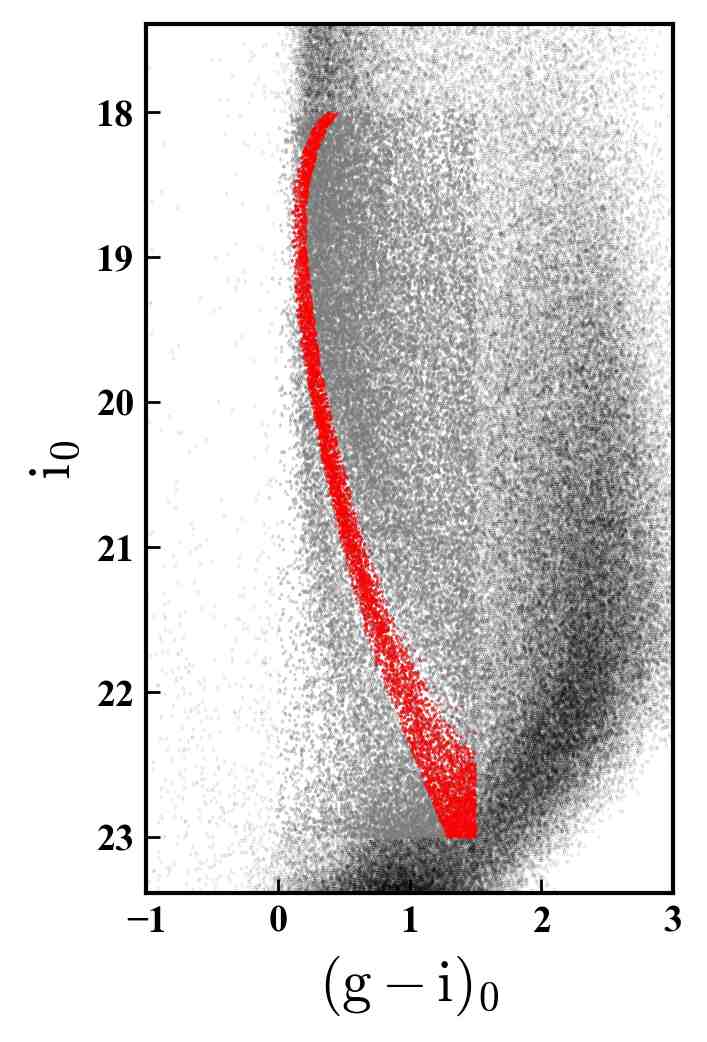}
        \hspace{1mm}
        \includegraphics[align=c,width=6.5cm]{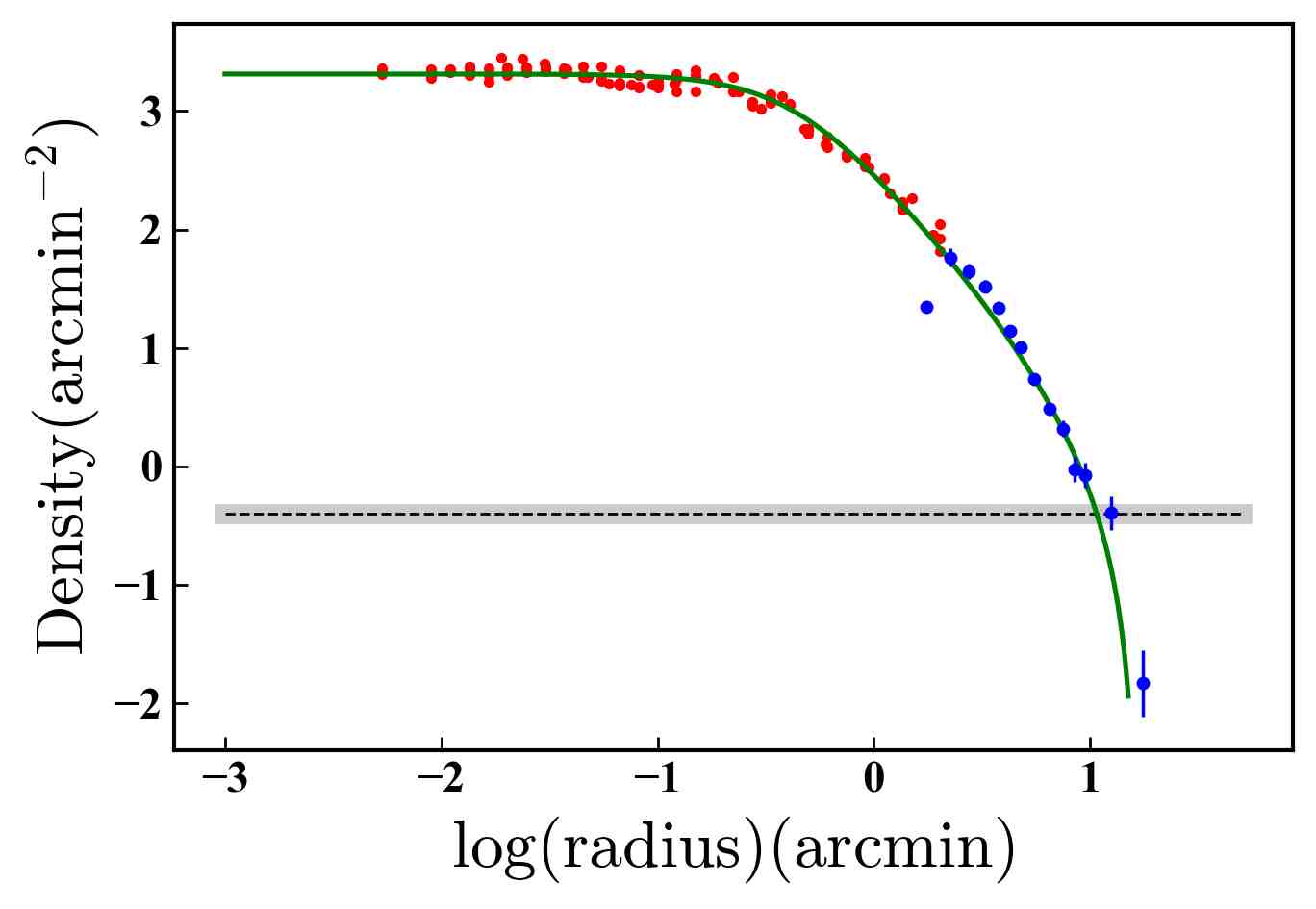}
    \end{minipage}\\
    \vspace{2mm}
    \begin{minipage}{0.99\textwidth}
        \centering
        \includegraphics[align=c,height=7.1cm]{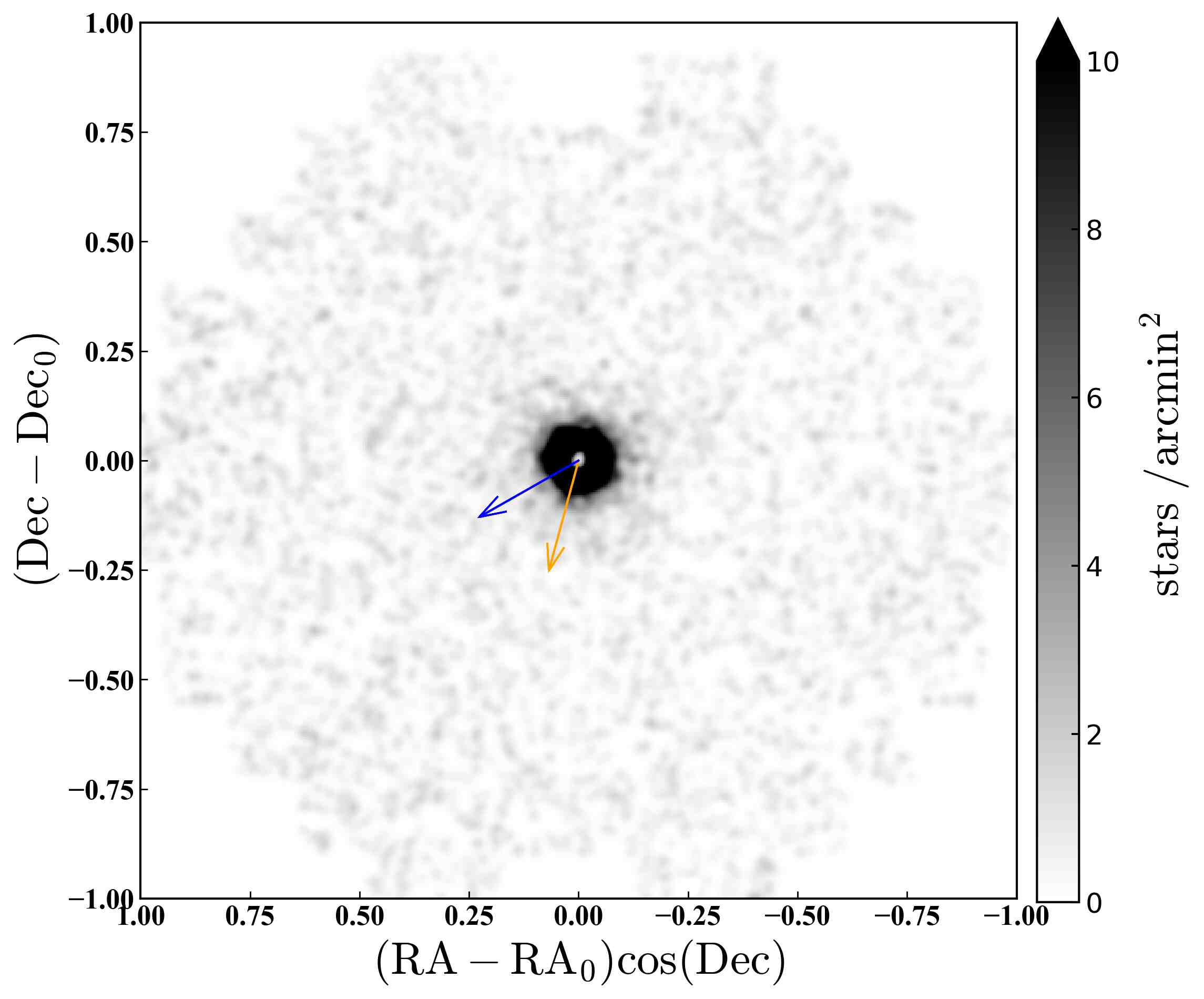}
        \hspace{2mm}
        \includegraphics[align=c,height=7.1cm]{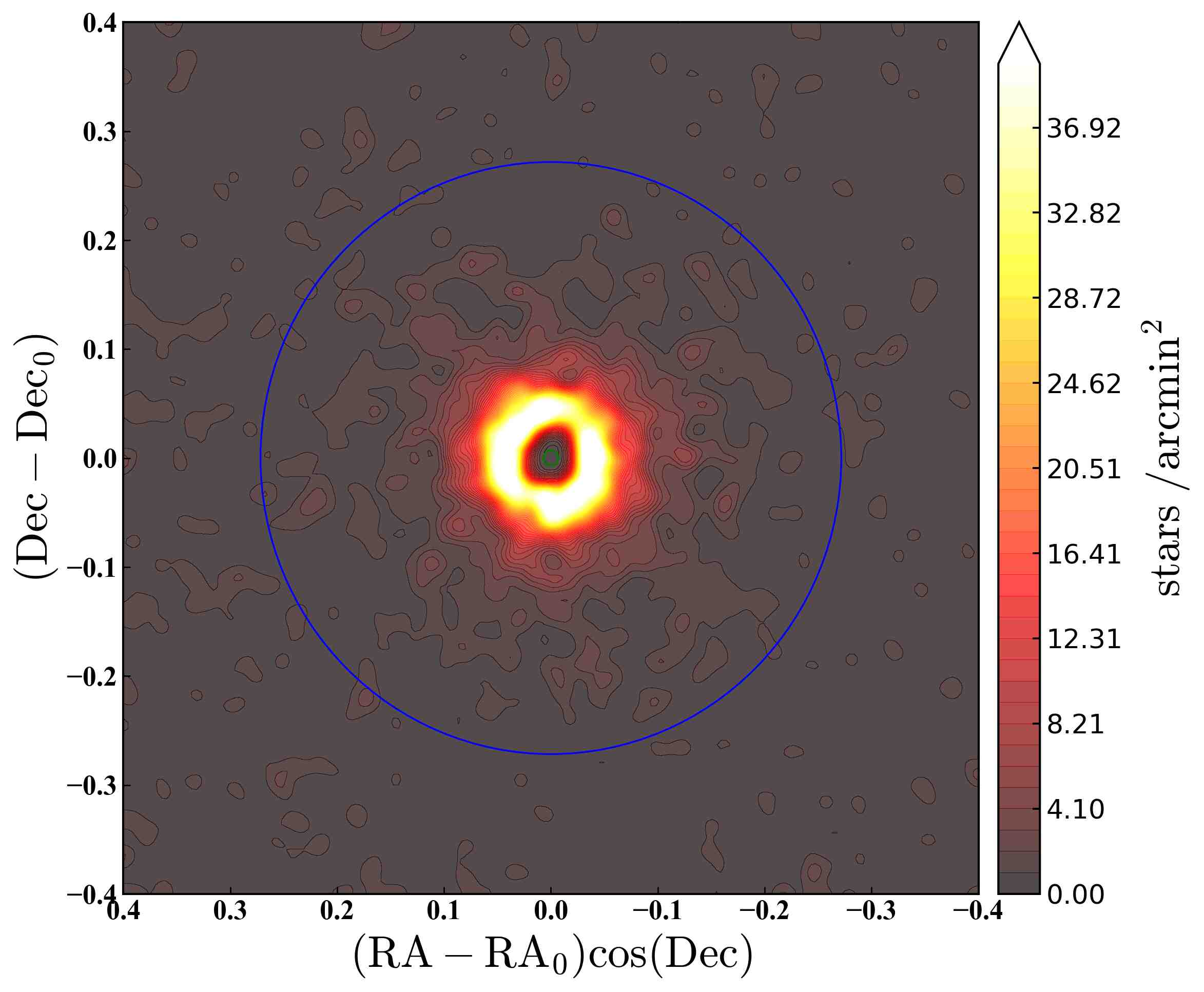}
    \end{minipage}
\caption{Same as Figure \ref{f:NGC1904}, but for NGC 2298.}
\label{f:NGC2298}
\end{figure*}    

\subsection{NGC 1904 (M79)} 
\label{ss:NGC1904}
NGC 1904 is located $13.08\pm0.18$\ kpc from the Sun \citep{baumgardt:21} and $\approx 19.1$\ kpc from the Galactic centre, placing it firmly in the outer halo of the Milky Way. Our measurements for this cluster are presented in Figure \ref{f:NGC1904}. Our photometry extends $\approx 3$\ mag below the MSTO, and, due to the relatively low level of field contamination, we are also able to use stars on the sub-giant branch to trace the cluster population. Both the radial profile and the two-dimensional density map show evidence for cluster stars to a radius of at least $\sim 20\arcmin$, which (as outlined below) is consistent with numerous previous studies. The outer structure of the cluster does not appear elongated in any particular direction -- i.e., the density distribution looks similar at all position angles. There is no evidence for secondary populations beyond $\sim 20\arcmin$ that could indicate the presence of tidal debris from a now-destroyed host dwarf.

Our radial density profile for NGC 1904 strongly suggests a power-law decline at large distances from the cluster centre. To explore this in more detail we first fit a model of the form of Eq. \ref{e:king} to the data.  Perhaps unsurprisingly, using all data points produces a poor fit -- both in the cluster's outskirts and towards its centre. To remedy this we systematically removed the outermost points, working inwards, until the best-fit model adequately matched the remaining data in the profile. This produces a core radius $r_c = 0.16\pm 0.01\arcmin$ and a nominal tidal radius\footnote{Note that this parameter does not hold any physical significance here; as noted in the text, the Jacobi radius for NGC 1904 is $\approx 42\arcmin$.} of $r_t = 7.2\pm0.2\arcmin$; the data begin to deviate from the expected shape at approximately this location. We fit a power-law to the data points sitting outside this radius and find a best-fit slope of $-3.2$.

The outskirts of NGC 1904 have been thoroughly explored by a number of previous studies, with most finding evidence for a structure that significantly departs from a King model. Both \citet{grillmair:95} and \citet{leon:00} suggested that the cluster might be surrounded by an extended stellar envelope (or "halo"), with \citet{grillmair:95} measuring $r_t \approx 11\arcmin$ and a power-law decline outside this radius. More recent studies using deeper data \citep[e.g.,][]{carballo:12,carballo:14,carballo:18a,miocchi:13} reach similar conclusions, with \citet{carballo:18a} reporting a power-law index of $-2.7$ (see their Table 2). 

\citet{deboer:19} use {\it Gaia} DR2 astrometry to construct a very clean sample of NGC 1904 members, and were able to trace the cluster out to nearly the Jacobi radius of $42.0\arcmin$ predicted by \citet{balbinot:18}. They too observe a significant departure from a King-like shape, finding that extended models such as the "{\sc limepy}" family from \citet{gieles:15} produce a much better fit to the profile at least out to $\sim 20\arcmin$. Beyond this, their profile shows a mild excess compared to all models, which might indicate the presence of an envelope-like feature $\sim 150$\ pc in radius, similar to the larger structures seen around NGC 1851 and NGC 7089 (both $\ga210$\ pc in radius). Recent very wide-field studies are consistent with this idea -- using data from the Dark Energy Survey (DES), \citet{shipp:18} suggest the presence of extra-tidal features around NGC 1904, extending up to $\sim1.5\degr$ ($\approx 340$\ pc) from its centre and following an approximately N-S orientation. The density map from the {\it Gaia} DR2 study of \citet{sollima:20} also appears to show connected structure extending $\sim50-100\arcmin$ both north and south of the cluster but no other convincing features out to $5\degr$, leading the author to conclude that no extended tidal tails are present. 

Our observations \citep[as well as those of][]{deboer:19}, are mildly inconsistent with the results from \citep{shipp:18} and \citet{sollima:20}; however, it is plausible that our method of normalizing the contamination map using the outer parts of our DECam field-of-view (i.e., at radii $\la 1\degr$) could suppress very low surface-brightness structures, especially in this case where the contamination is relatively sparse. Data from both DES and {\it Gaia} DR2 span much wider areas, likely mitigating this issue.

\begin{figure*}
    \begin{minipage}{0.99\textwidth}
        \centering
        \includegraphics[align=c,height=6.6cm]{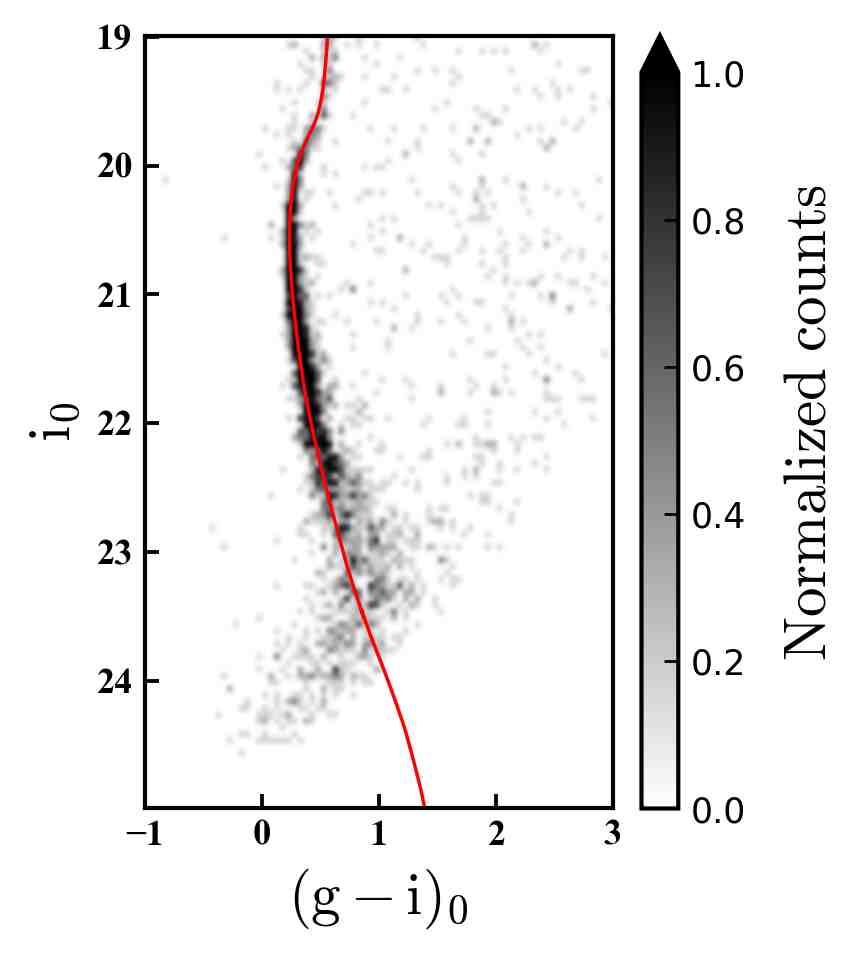}
        \hspace{-1mm}
        \includegraphics[align=c,height=6.6cm]{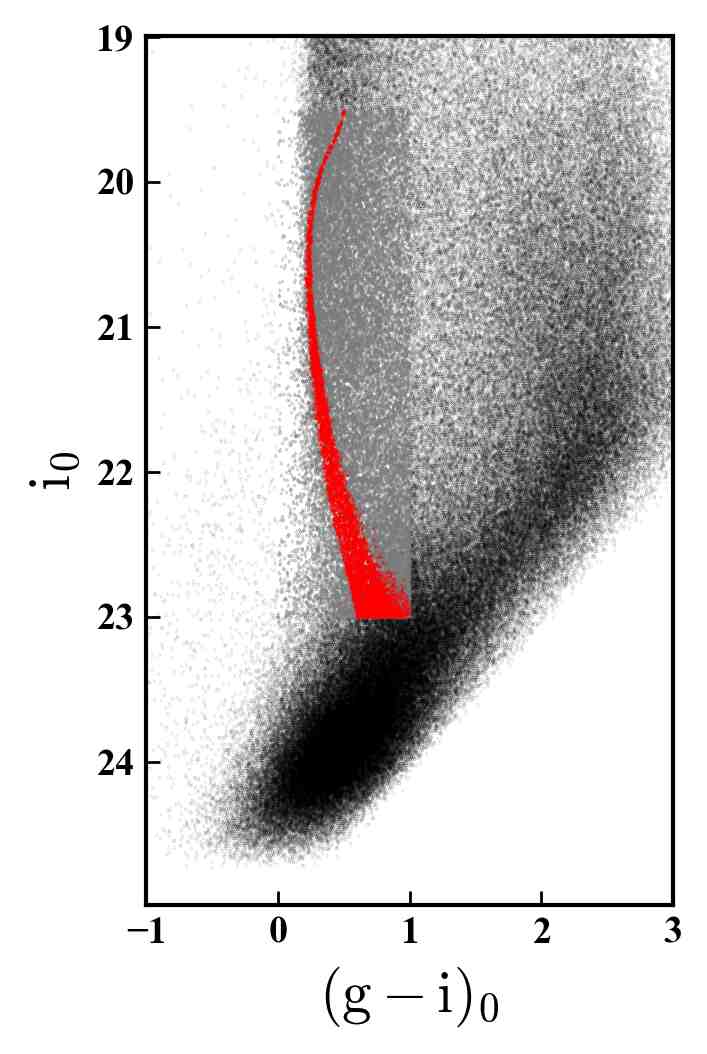}
        \hspace{1mm}
        \includegraphics[align=c,width=6.5cm]{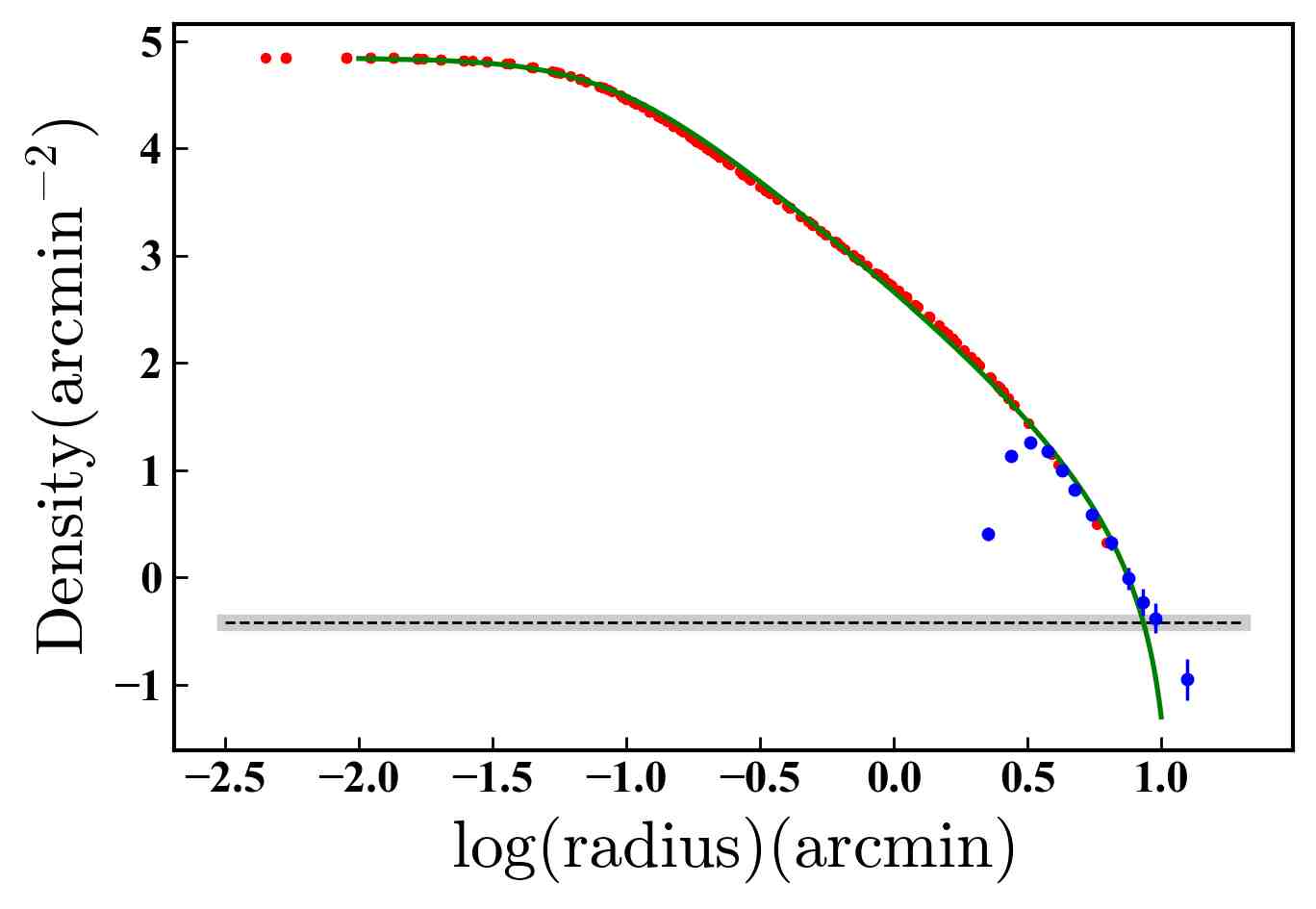}
    \end{minipage}\\
    \vspace{2mm}
    \begin{minipage}{0.99\textwidth}
        \centering
        \includegraphics[align=c,height=7.1cm]{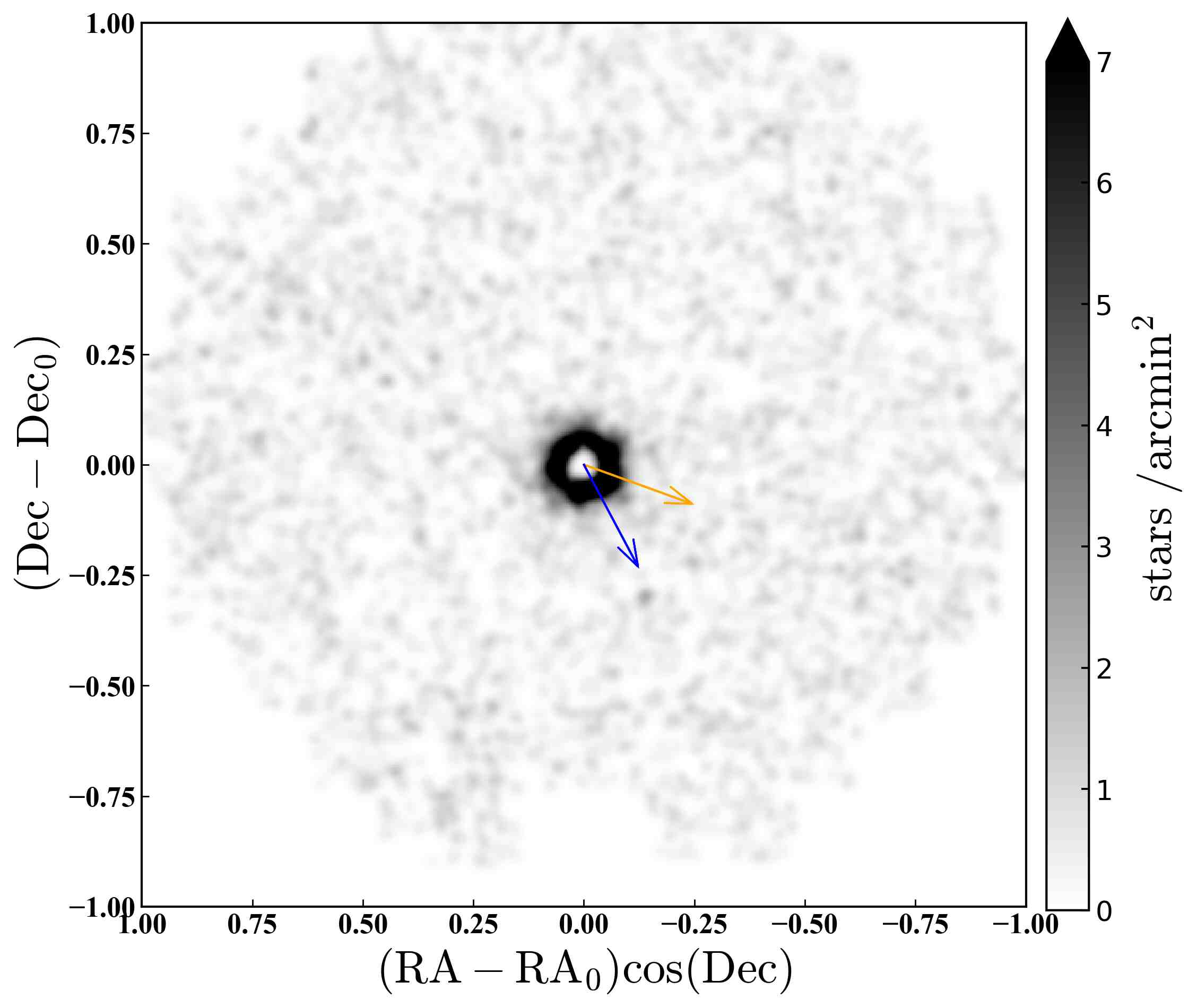}
        \hspace{2mm}
        \includegraphics[align=c,height=7.1cm]{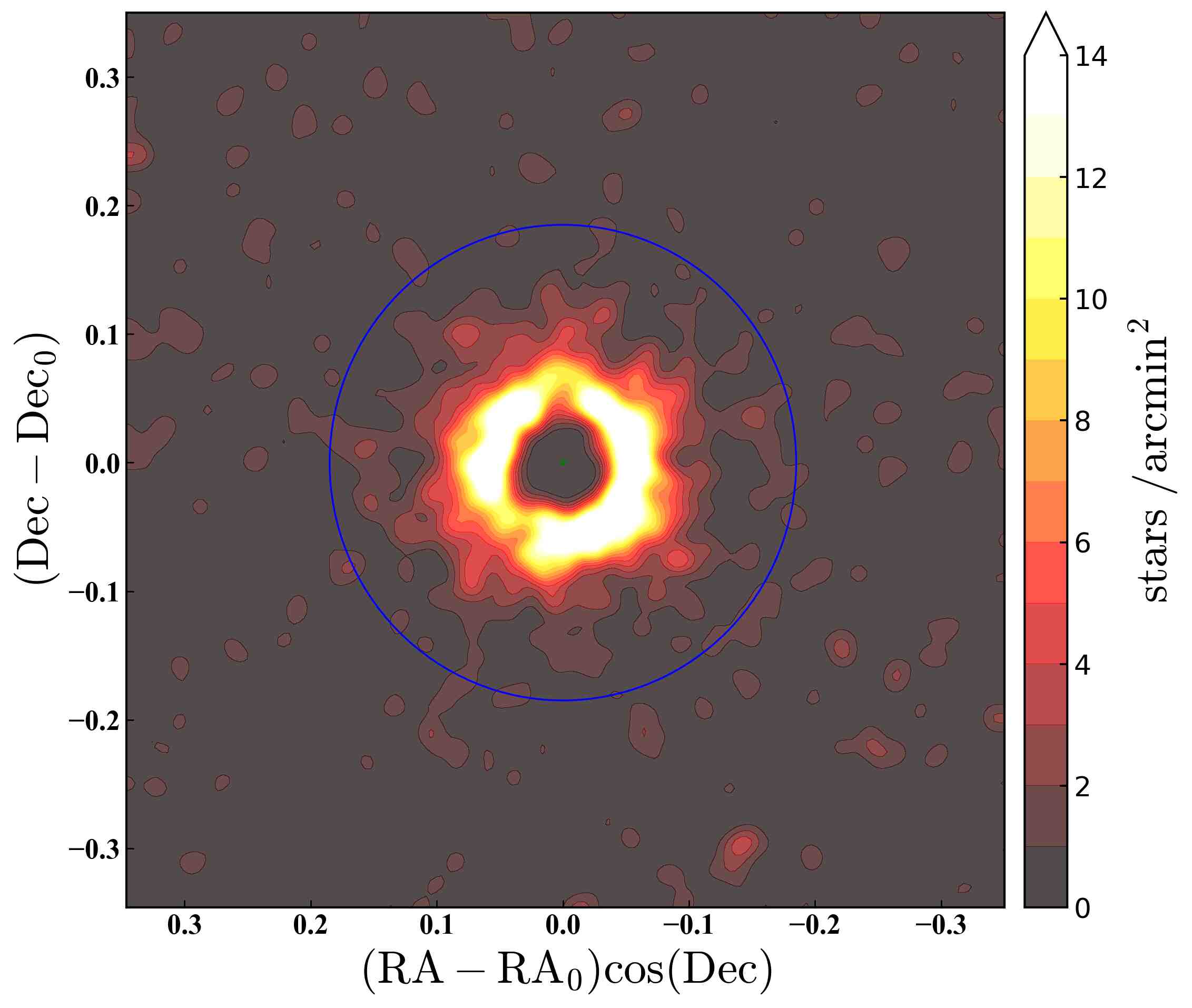}
    \end{minipage}
\caption{Same as Figure \ref{f:NGC1904}, but for NGC 6864 (M75).}
\label{f:NGC6864}
\end{figure*}

\subsection{NGC 2298}
\label{ss:NGC2298}
NGC 2298 is another well-studied globular cluster in the outer halo of the Milky Way (with a Galactocentric distance of $\approx 15.1$\ kpc, see Table \ref{t:targets}). Our photometry for this system covers the top five magnitudes of the main sequence and MSTO, as shown in Figure \ref{f:NGC2298}. Contamination is higher than for NGC 1904, but can be cleanly subtracted; no unexpected populations are present. Our measurements for NGC 2298 show no evidence for extended structure, either in its density map or its radial density profile.  We find that a King model provides a good description of the profile at all radii; the best-fit model has a core radius $r_c = 0.42 \pm 0.02\arcmin$, and tidal radius $r_t = 16.3\pm 0.2\arcmin$. This is somewhat smaller than the Jacobi radius of $27.8\arcmin$ calculated by \citet{balbinot:18}. The density map shows that, within the tidal radius, the stars are evenly distributed with azimuth -- that is, the isodensity contours do not appear to have an overall elongated or elliptical shape.

A few previous studies have suggested the presence of extra-tidal features around NGC 2298 \citep[e.g.,][]{leon:00}. In particular, \citet{balbinot:11} highlighted possible tails at a distance of $\sim 1$\ degree to the north-west, north-east, and south of the cluster. However, more recent investigations using deep ground-based wide-field data have not recovered similar structures. \citet{carballo:12,carballo:14} found that a King model adequately fit their radial profile, while the DECam-based study by \citet{carballo:18a} also came to a similar conclusion (albeit noting the possibility of low-significance asymmetries within the tidal radius). \citet{deboer:19}, using {\it Gaia} DR2 astrometry to efficiently select cluster members, were able to trace the radial profile of NGC 2298 to approximately the expected Jacobi radius, finding a weak departure from a King-like structure for data outside $\sim 10\arcmin$. The more extended \citet{wilson:75} and \citet{gieles:15} models were able to provide a good match to their measurements.

Perhaps surprisingly, given the apparent lack of distortion or obvious tidal features in the outskirts of NGC 2298, both the {\it Gaia} DR2 study of \citet{sollima:20} and the all-sky EDR3-based search from \citet{ibata:21} recently revealed extended tidal tails belonging to this cluster. The density map from \citet{sollima:20} clearly shows tail-like structures along a SE-NW axis, reaching at least $\sim 100\arcmin$ ($\ga 280$\ pc) from the centre of the cluster in both directions (i.e., to a distance of more than $\sim 4$ times the predicted Jacobi radius). The {\sc streamfinder} algorithm of \citet{ibata:21} appears even more sensitive, unveiling a $\sim 12\degr$ long tail, again with SE-NW orientation. We discuss the interesting discrepancy between these {\it Gaia}-based studies, and deep photometric work such as that presented here, in more detail in Section \ref{s:discussion}. 
	
\subsection{NGC 6864 (M75)}
\label{ss:NGC6864}
The good agreement between our results for NGC 1904 and 2298, and those of previous studies using comparable data sets \citep[e.g.,][]{carballo:18a,deboer:19} establishes the validity of our reduction and analysis techniques. We now proceed to consider increasingly less well studied clusters.

NGC 6864 is a luminous system sitting $14.3$\ kpc from the Galactic centre. Our selection region spans just over $3$\ magnitudes of the upper main sequence and MSTO (see Figure \ref{f:NGC6864}). As with NGC 2298, we see no evidence for an extended structure or extra-tidal features. A King model with $r_c = 0.09 \pm 0.01\arcmin$ and $r_t = 11.1 \pm 0.1\arcmin$ provides a good description of the radial profile except, possibly, for the outermost two points. However, the two-dimensional map does not reveal any significant signal outside the nominal $r_t$. 

The outer structure of NGC 6864 has previously been studied by \citet{grillmair:95}, who found that a King model described their observed profile well. However, with deeper data \citet{carballo:12,carballo:14} observed that a power-law model provided a better fit to their profile at large radii than a King model, while \citet{deboer:19}, using {\it Gaia} DR2 data, found that the extended \citet{wilson:75} and \citet{gieles:15} models were required to adequately describe their profile at radii outside $\sim 5\arcmin$ \citep[see also][]{miocchi:13}. Overall, this appears consistent with our outer two data points; however, none of the models or profiles extend even close to the Jacobi radius of $27.4\arcmin$ calculated by \citet{balbinot:18}. Most recently, \citet{piatti:22} claimed the detection of a weak envelope structure around NGC 6864 extending to $\sim 25\arcmin$; however, we cannot reproduce this result. 

\subsection{NGC 6981 (M72)}
\label{ss:NGC6981}

\begin{figure*}
    \begin{minipage}{0.99\textwidth}
        \centering
        \includegraphics[align=c,height=6.6cm]{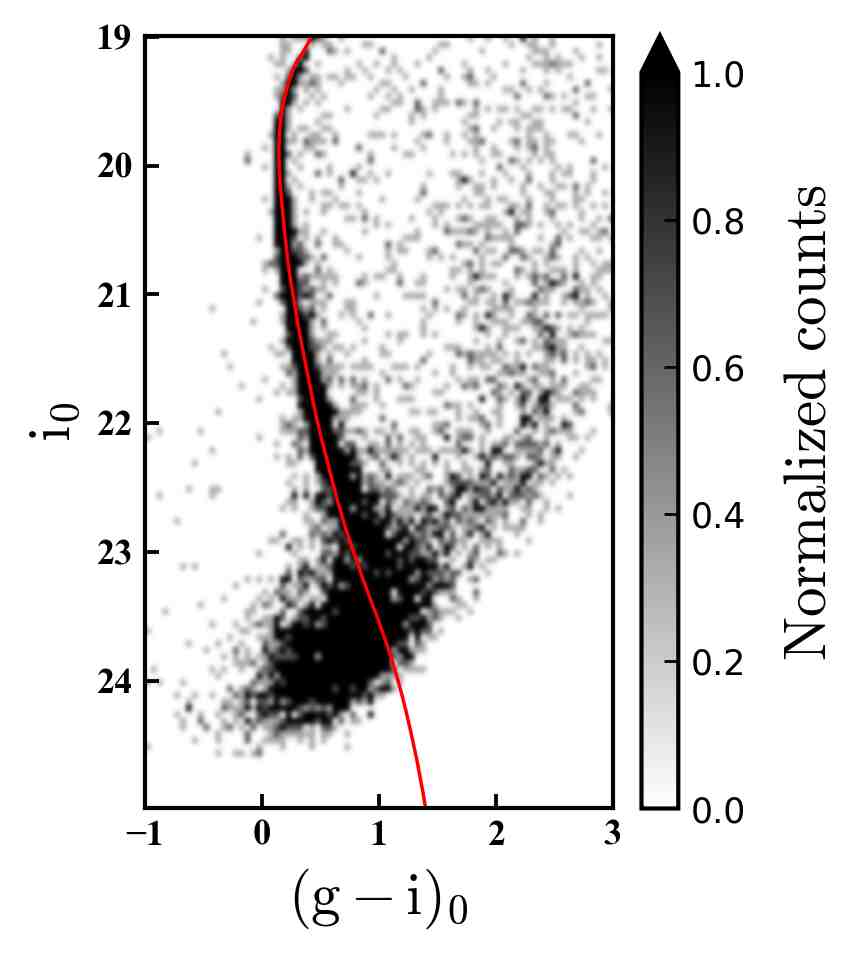}
        \hspace{-1mm}
        \includegraphics[align=c,height=6.6cm]{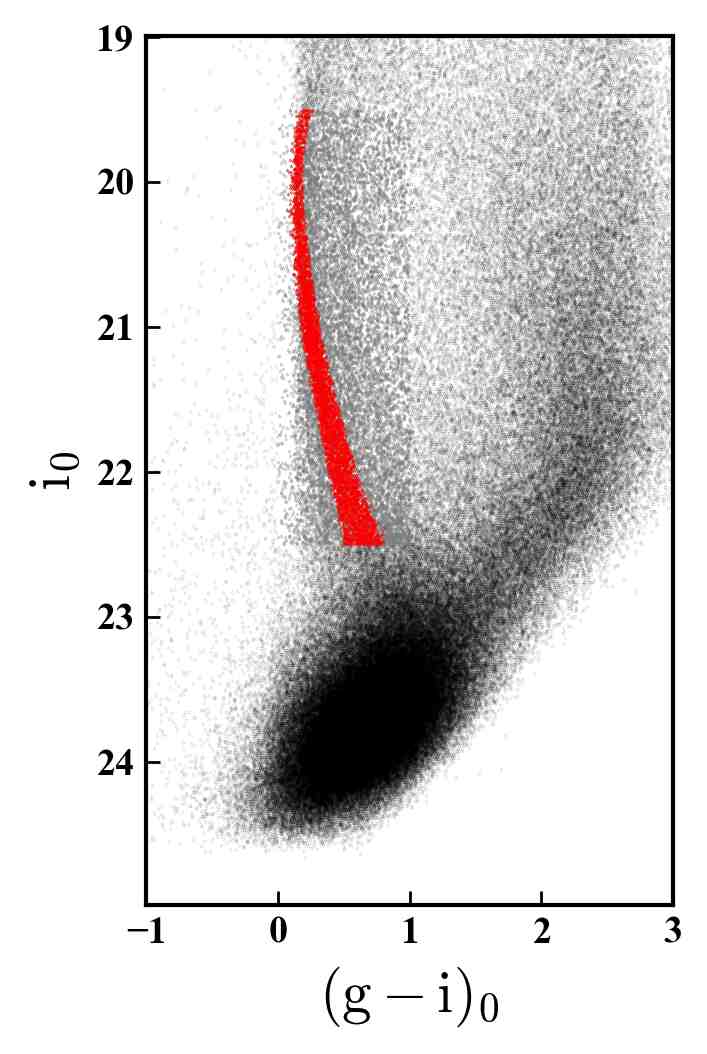}
        \hspace{1mm}
        \includegraphics[align=c,width=6.5cm]{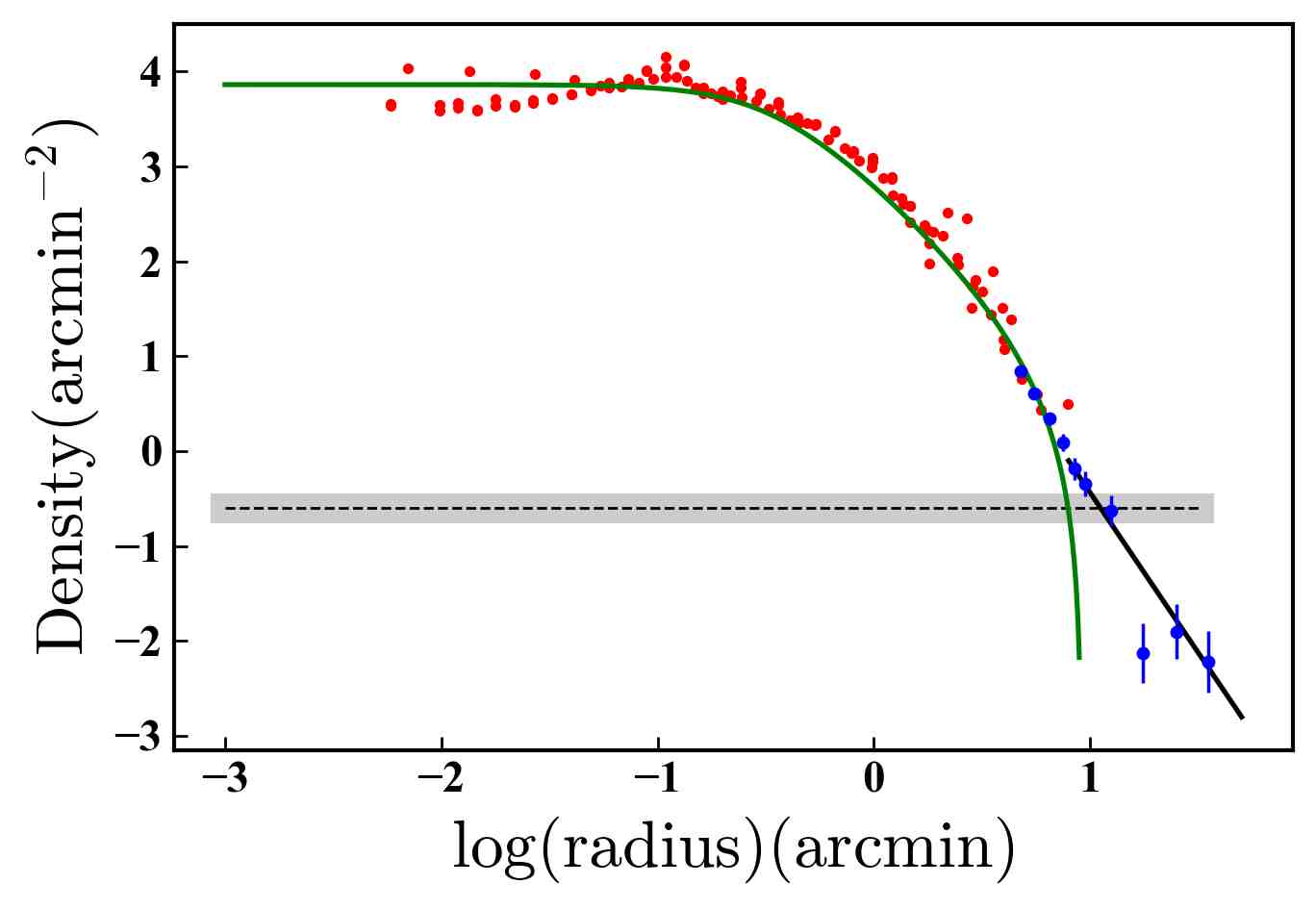}
    \end{minipage}\\
    \vspace{2mm}
    \begin{minipage}{0.99\textwidth}
        \centering
        \includegraphics[align=c,height=7.1cm]{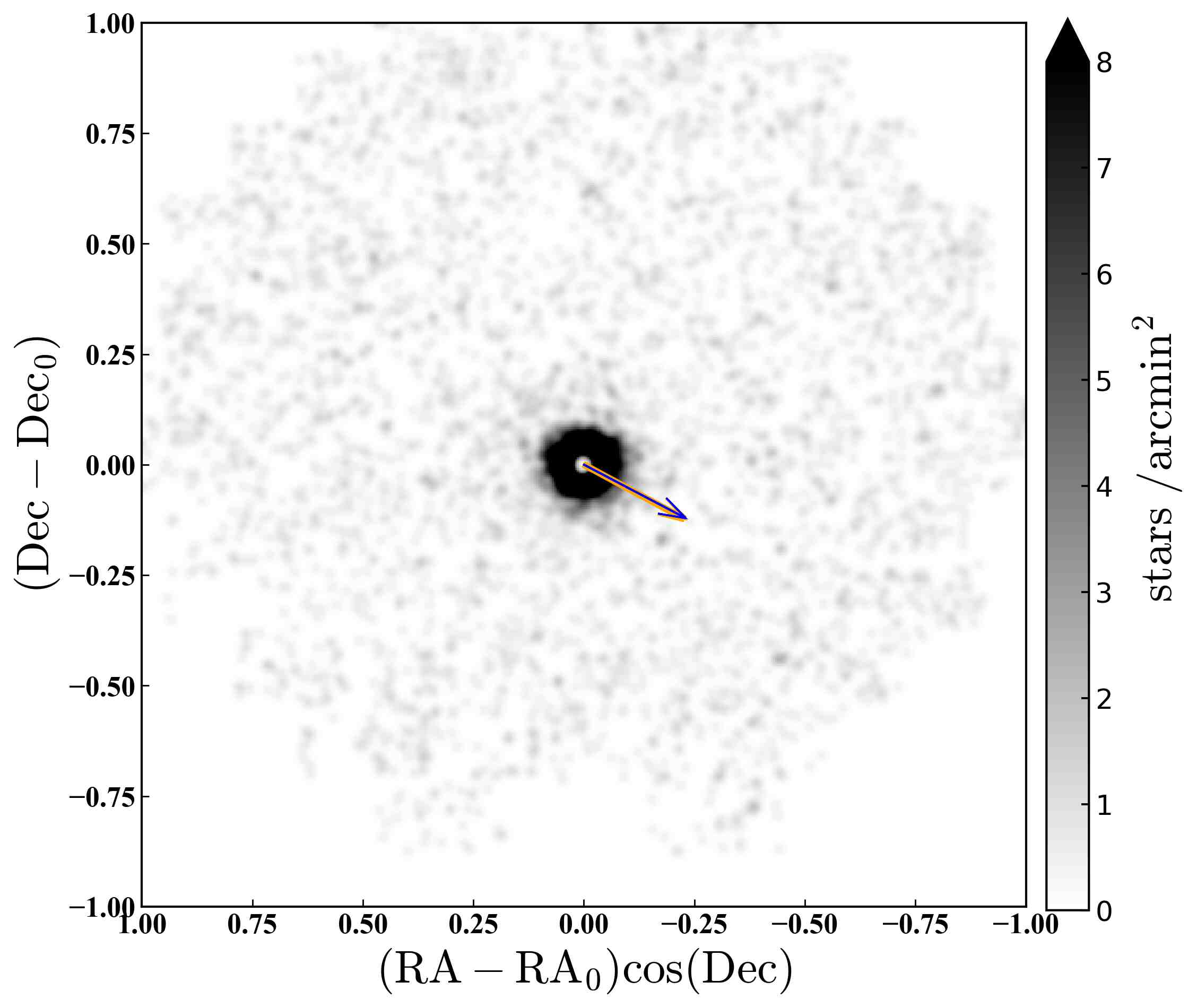}
        \hspace{2mm}
        \includegraphics[align=c,height=7.1cm]{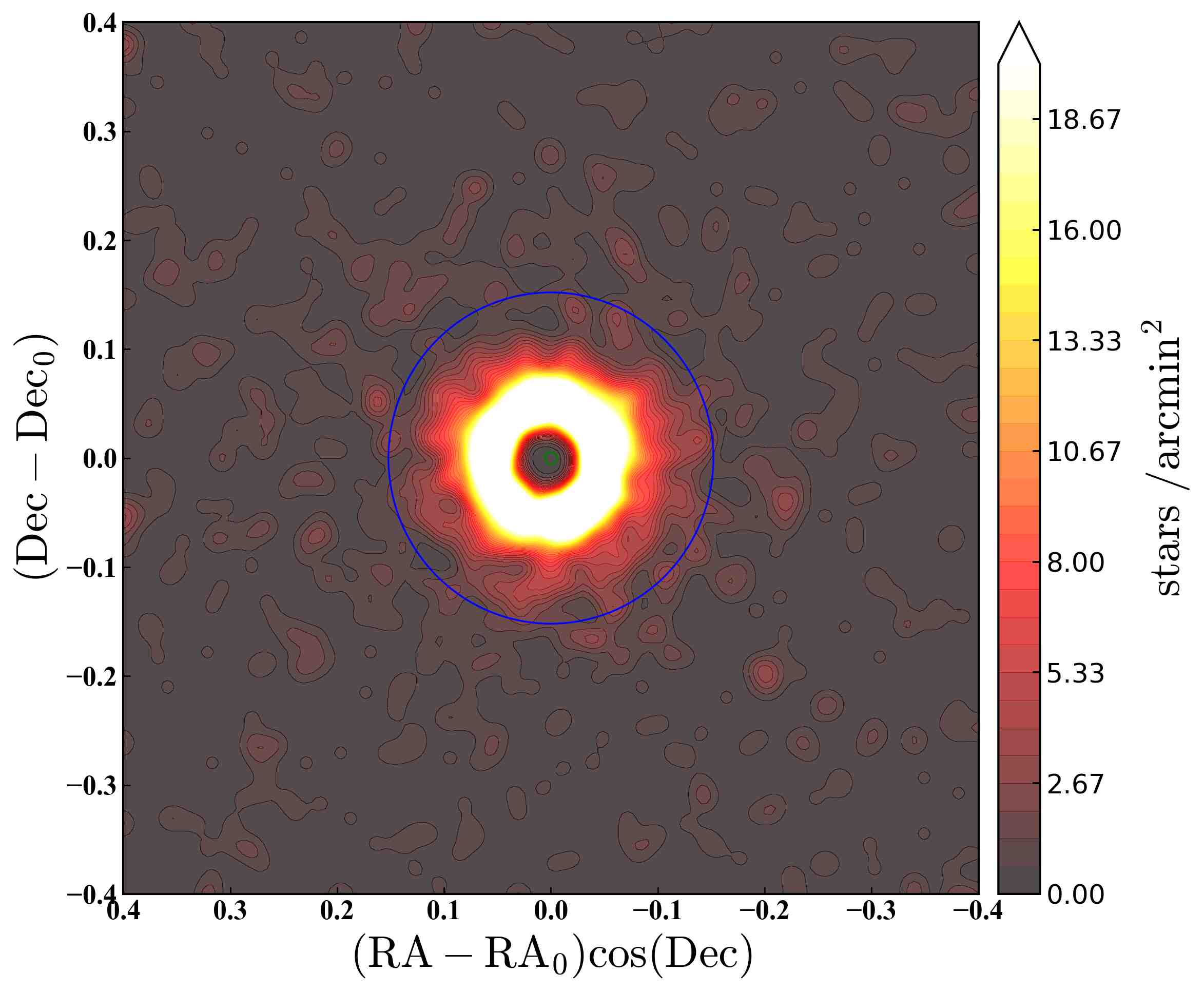}
    \end{minipage}
\caption{Same as Figure \ref{f:NGC1904}, but for NGC 6981 (M72).}
\label{f:NGC6981}
\end{figure*}    

NGC 6981 is located at a Galactocentric radius of $15.8$\ kpc. Our measurements for this cluster are presented in Figure \ref{f:NGC6981}; the selection region covers approximately $3$\ mag around the MSTO and upper main sequence. Our radial profile is suggestive of a power-law decline at large radii, similar to (but at lower significance than) the outskirts of NGC 1904. As before, we fit a King model by systematically removing the outermost data points until the model adequately represents the observations. The nominal tidal radius is $r_t = 9.1 \pm 0.2\arcmin$, and represents the approximate position where the profile begins to depart from a King-like shape. Outside this radius a power-law with index $-3.4$ fits the data well, out to nearly $\sim20\arcmin$ ($\approx 100$\ pc). This is commensurate with the Jacobi radius of $19.3\arcmin$ computed by \citet{balbinot:18}.

Our density map supports the idea that NGC 6981 possesses a low-luminosity envelope-type structure. Furthermore, by eye this appears to be preferentially extended to the north-east where it can be traced to at least $\sim15\arcmin$, compared with $\la10\arcmin$ at other position angles. This is approximately along the Great Circle defined by NGC 6981 and the Galactic centre ($\phi_{\rm gc} = 241.7\degr$ for this cluster, see Table \ref{t:targets}). Interestingly, \citet{piatti:21a} also observed evidence for weak extension of NGC 6981 in a similar direction, using an independent DECam data set and analysis methodology. Apart from this, the outskirts of NGC 6981 have been studied by \citet{deboer:19}, who were able to trace the profile to approximately $20\arcmin$ (i.e., the expected Jacobi radius). They found that outside $\approx 5\arcmin$ the data depart from a King-type shape and require the more extended Wilson/{\sc limepy} models to achieve an acceptable fit.

\subsection{NGC 7492}
\label{ss:NGC7492}

\begin{figure*}
    \begin{minipage}{0.99\textwidth}
        \centering
        \includegraphics[align=c,height=6.6cm]{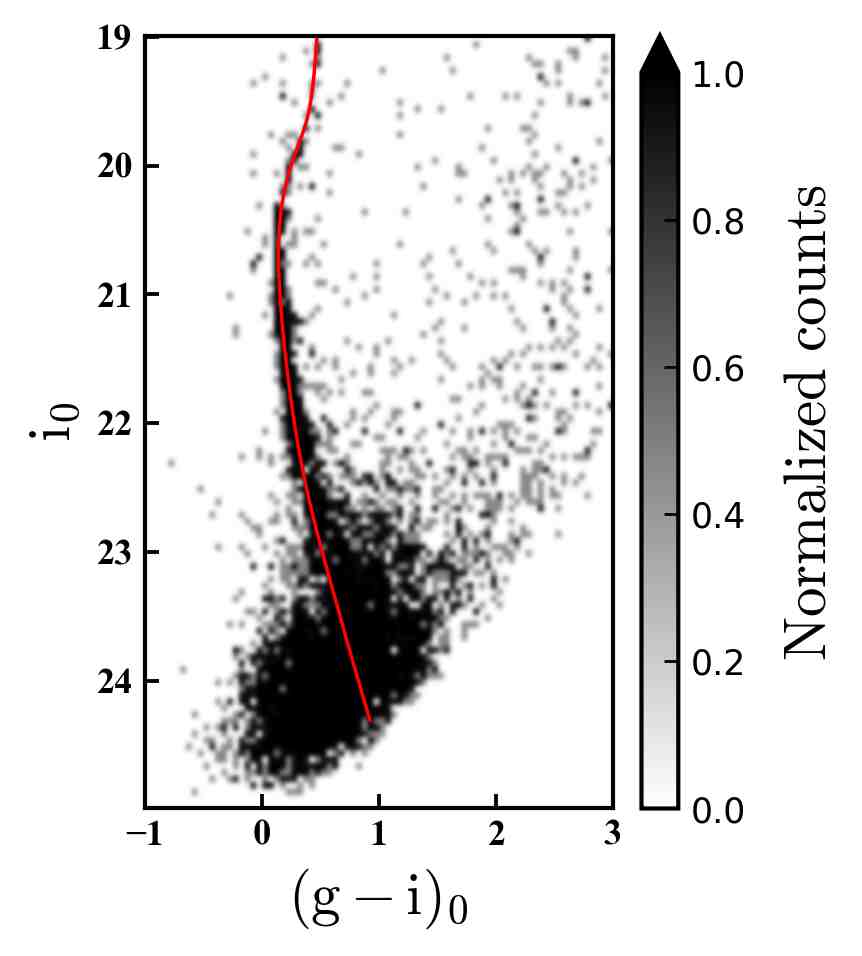}
        \hspace{-1mm}
        \includegraphics[align=c,height=6.6cm]{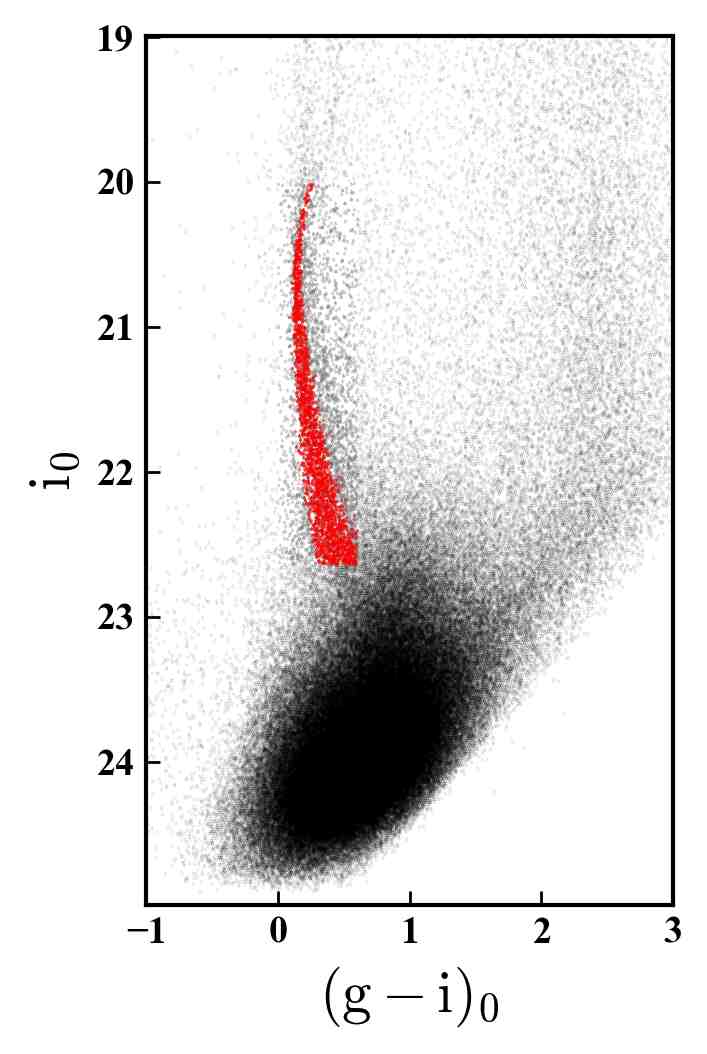}
        \hspace{1mm}
        \includegraphics[align=c,width=6.5cm]{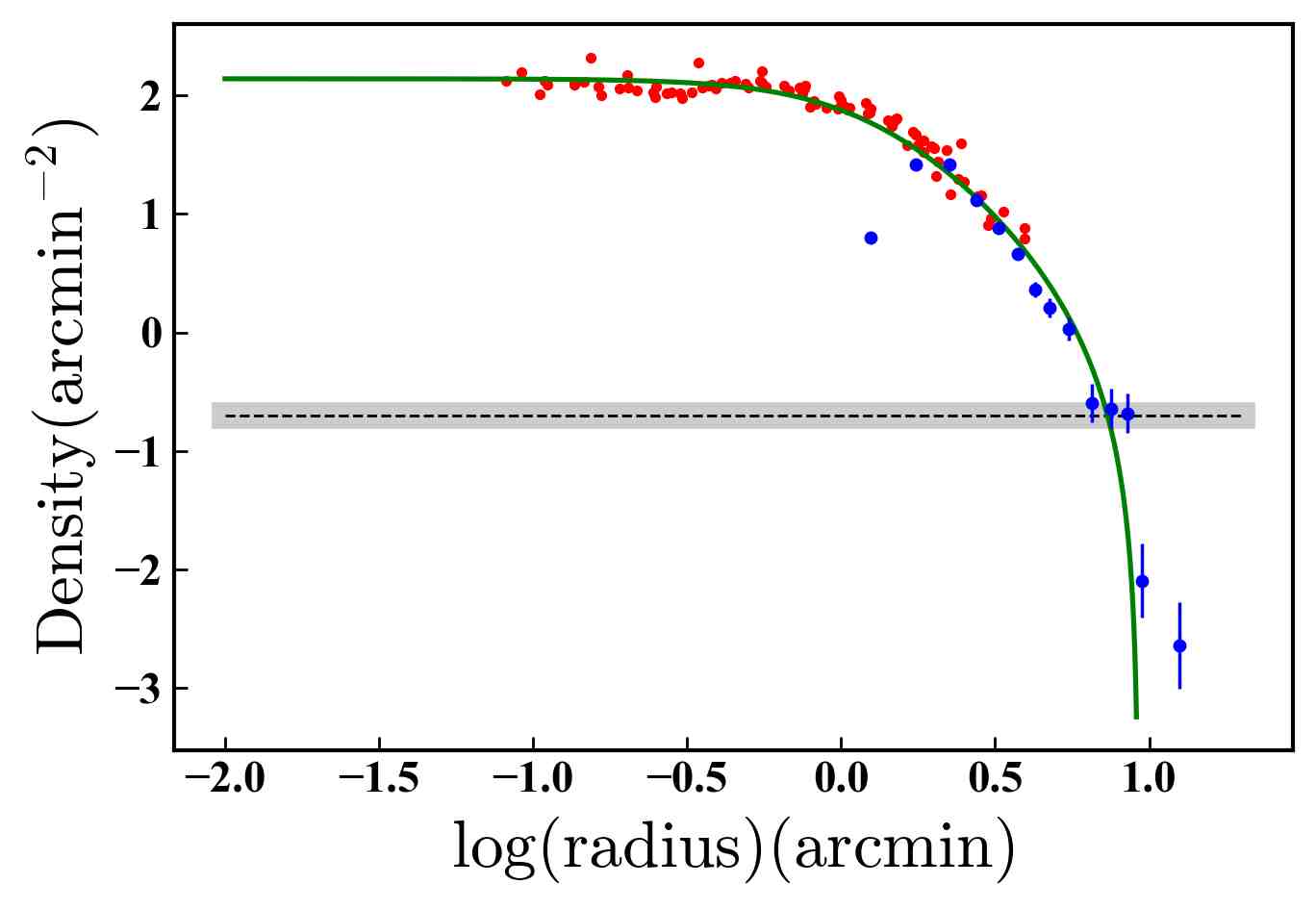}
    \end{minipage}\\
    \vspace{2mm}
    \begin{minipage}{0.99\textwidth}
        \centering
        \includegraphics[align=c,height=7.05cm]{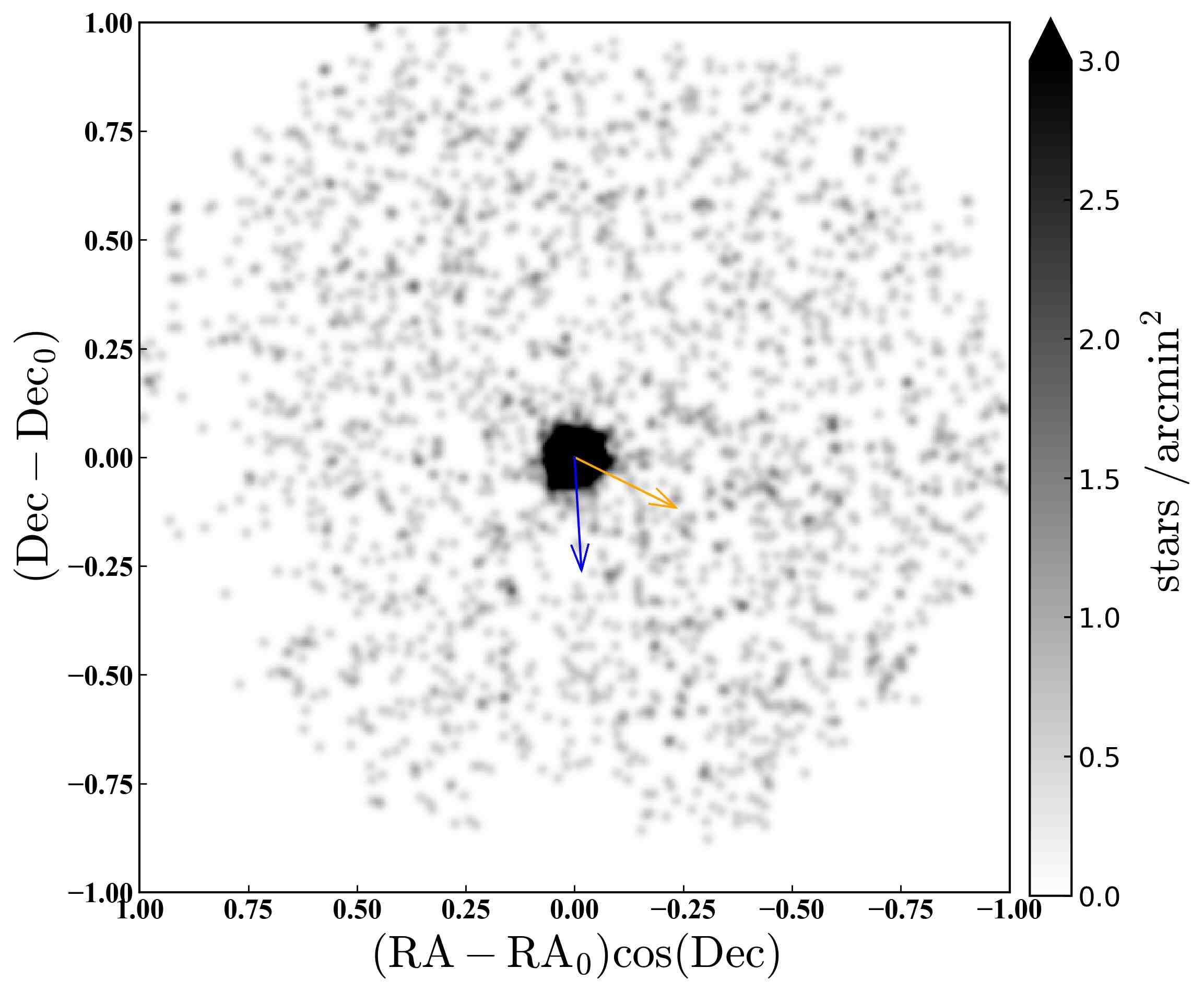}
        \hspace{2mm}
        \includegraphics[align=c,height=7.05cm]{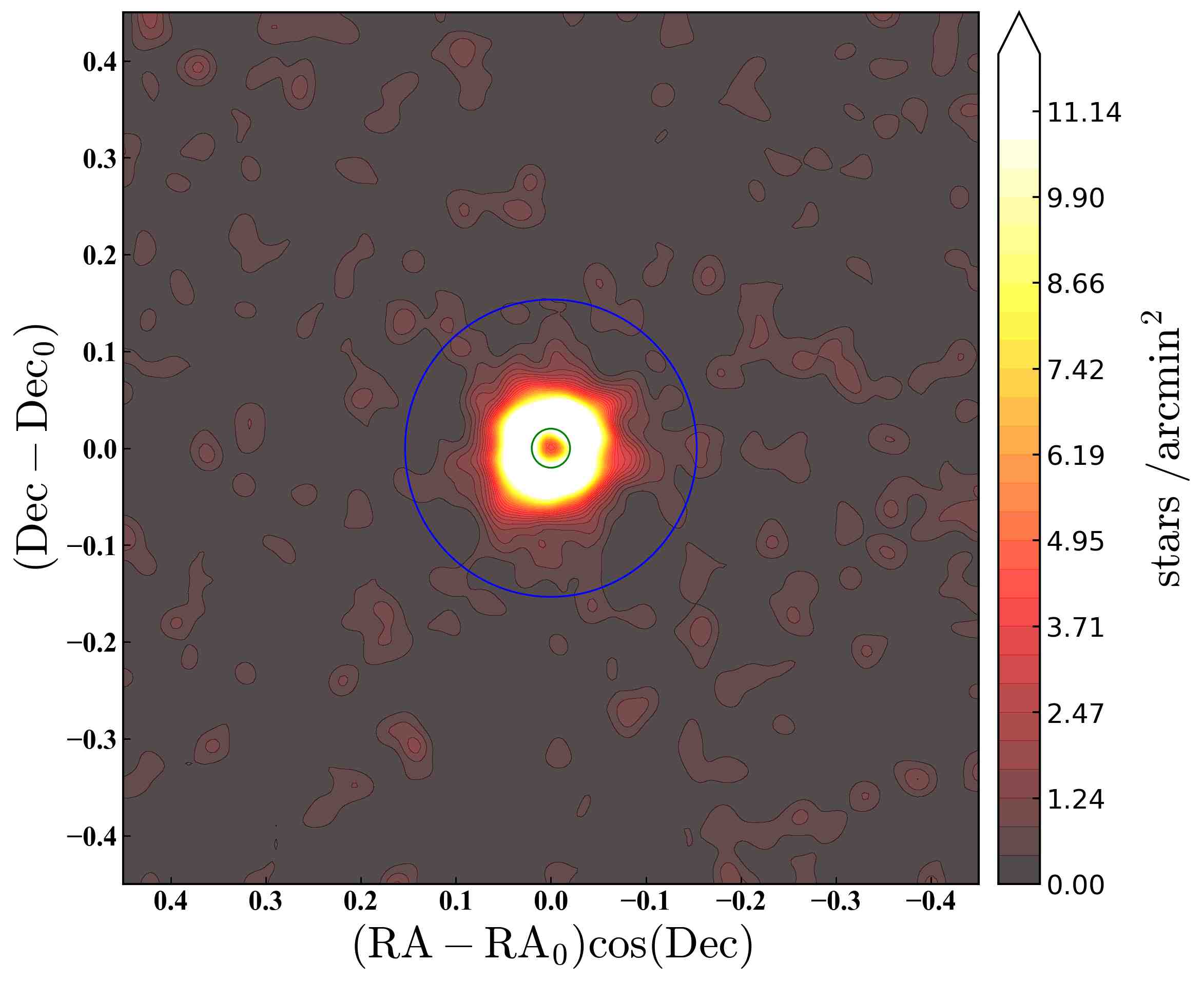}
    \end{minipage}
\caption{Same as Figure \ref{f:NGC1904}, but for NGC 7492.}
\label{f:NGC7492}
\end{figure*}    

NGC 7492 is one of the more remote globular clusters in our sample, sitting at $\approx 23.6$\ kpc from the Galactic centre. It is known to be projected against the stellar stream arising from the disruption of the Sagittarius dwarf galaxy \citep[e.g.,][]{carballo:14,sollima:18}; however spectroscopic measurements \citep{carballo:18b} show that it is kinematically quite distinct from the Sagittarius stream (and is therefore not an ex-Sgr member). 

Figure \ref{f:NGC7492} shows our results for NGC 7492.  The selection region spans $\sim 2.5$\ mag around the MSTO and upper main sequence. We clearly observe the surrounding Sagittarius populations; because these occupy a similar colour-magnitude locus to the cluster, we were especially careful in defining the membership and contamination limits for this target. The radial density profile can be traced to just past $\sim 10\arcmin$, approximately commensurate with the Jacobi radius $r_J = 13.4\arcmin$ predicted by \citet{balbinot:18}. The shape of the profile is well described by a King model with $r_c = 1.20 \pm 0.09\arcmin$ and $r_t = 9.2\pm 0.1\arcmin$, although the outermost two data points may suggest a mildly more extended structure. However, nothing obviously in excess of the expected background fluctuations is evident in the two-dimensional density map.

Previous studies of the outskirts of NGC 7492 present contradictory findings. Several early works \citep[e.g.,][]{leon:00,lee:04} suggested the presence of tidal tails around the cluster; more recently this was affirmed by the Pan-STARRS PS1 investigation of \citet{navarrete:17} who found evidence for tails extending to $\pm 1.5\degr$ along a NE-SW axis. On the other hand, using much deeper data from both MegaCam on the 4m Canada-France-Hawaii Telescope and Megacam on the 6.5m Magellan Clay Telescope, \citet{munoz:18a,munoz:18b} were unable to corroborate these results. They report an almost perfectly circular morphology with no evidence for any asymmetrical extended structure surrounding the cluster, entirely consistent with the results outlined here.

Several additional works have studied the radial density profile of NGC 7492. \citet{carballo:12} found that a power-law model with a steep fall-off ($\gamma\approx -4$) provided a better fit to their data than a King model, while \citet{deboer:19} found a \citet{king:66} model to be slightly more suitable than the more extended \citet{wilson:75} or \citet{gieles:15} models. However, neither work presented evidence for cluster structure beyond the Jacobi radius of $r_J = 13.4\arcmin$, again consistent with the results described here.

\citet{munoz:18b} suggests that the apparently stark differences between the outcomes of studies that do, or do not, find evidence for tidal tails around NGC 7492 might be attributable to the surrounding Sagittarius populations, which lie close to the cluster locus on the CMD and potentially complicate member selection. As such, this cluster is clearly a target worth revisiting with wide-field {\it Gaia} data \citep[similar to that presented by][for other targets]{sollima:20} and/or spectroscopic observations, both of which could help more robustly select cluster members. Interestingly, the spectroscopic investigation by \citet{carballo:18b} revealed a few possible kinematic members up to $\sim 30\arcmin$ from the centre of NGC 7492 both to the east/north-east \citep[in the direction of one of the tails described by][]{navarrete:17}, and to the north-west \citep[a region largely devoid of cluster structure in the maps of][]{navarrete:17}. Further study of this perplexing cluster is warranted.

\subsection{Whiting 1}
\label{ss:whiting1}

\begin{figure*}
    \begin{minipage}{0.99\textwidth}
        \centering
        \includegraphics[align=c,height=6.6cm]{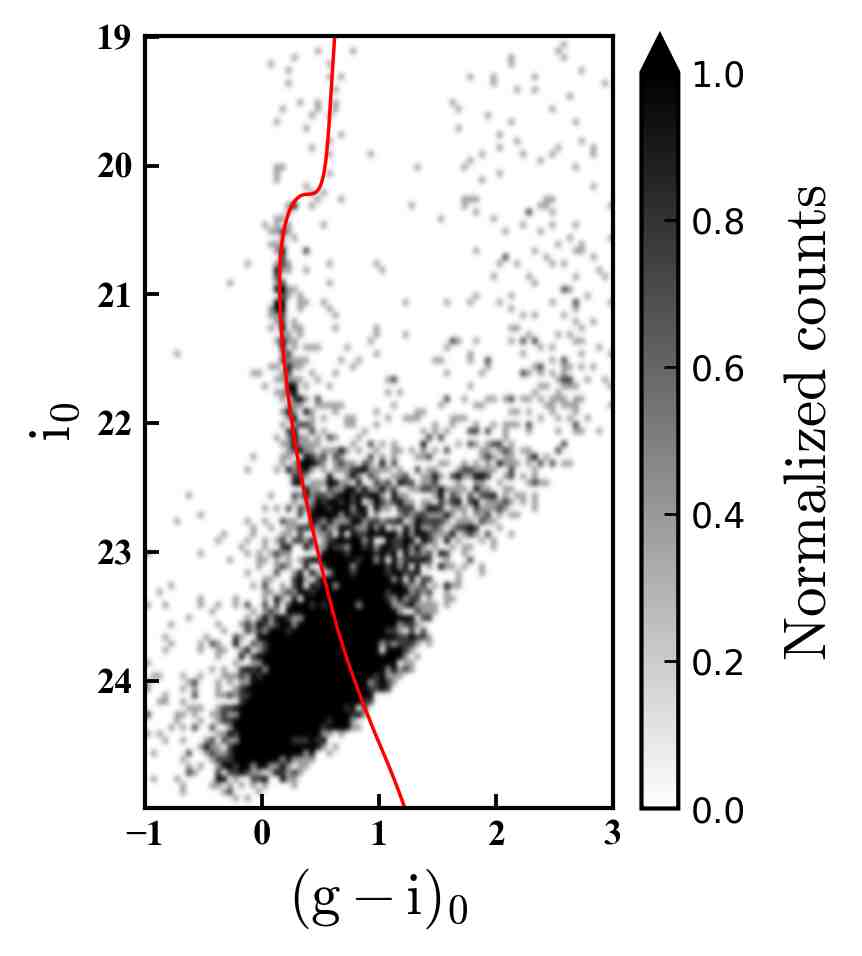}
        \hspace{-1mm}
        \includegraphics[align=c,height=6.6cm]{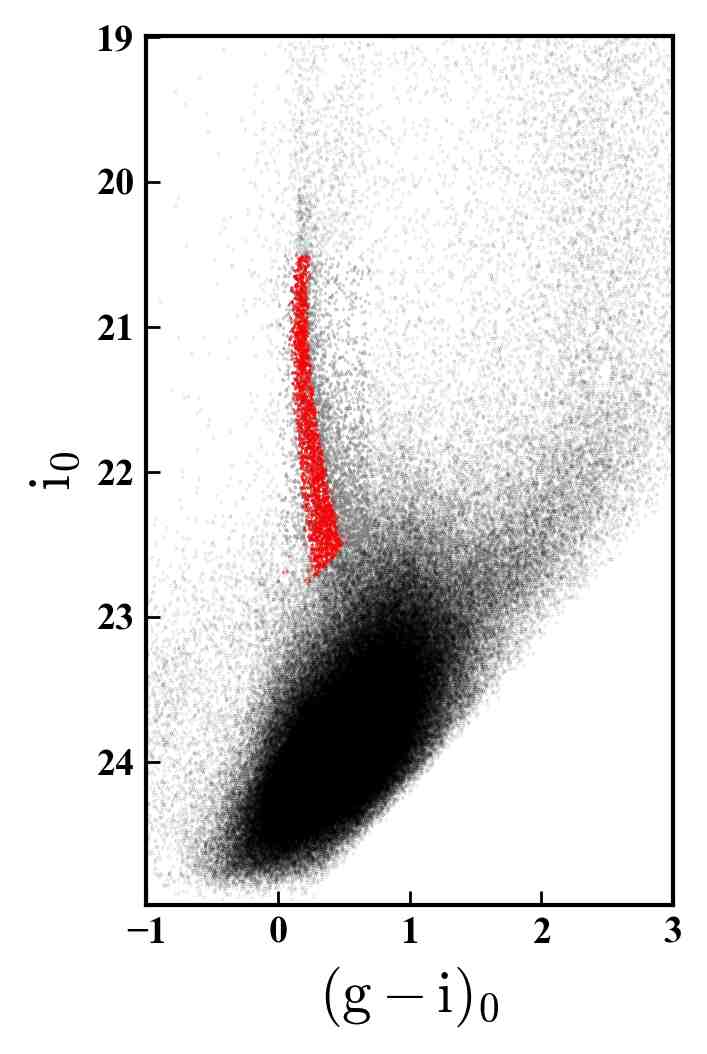}
        \hspace{1mm}
        \includegraphics[align=c,width=6.5cm]{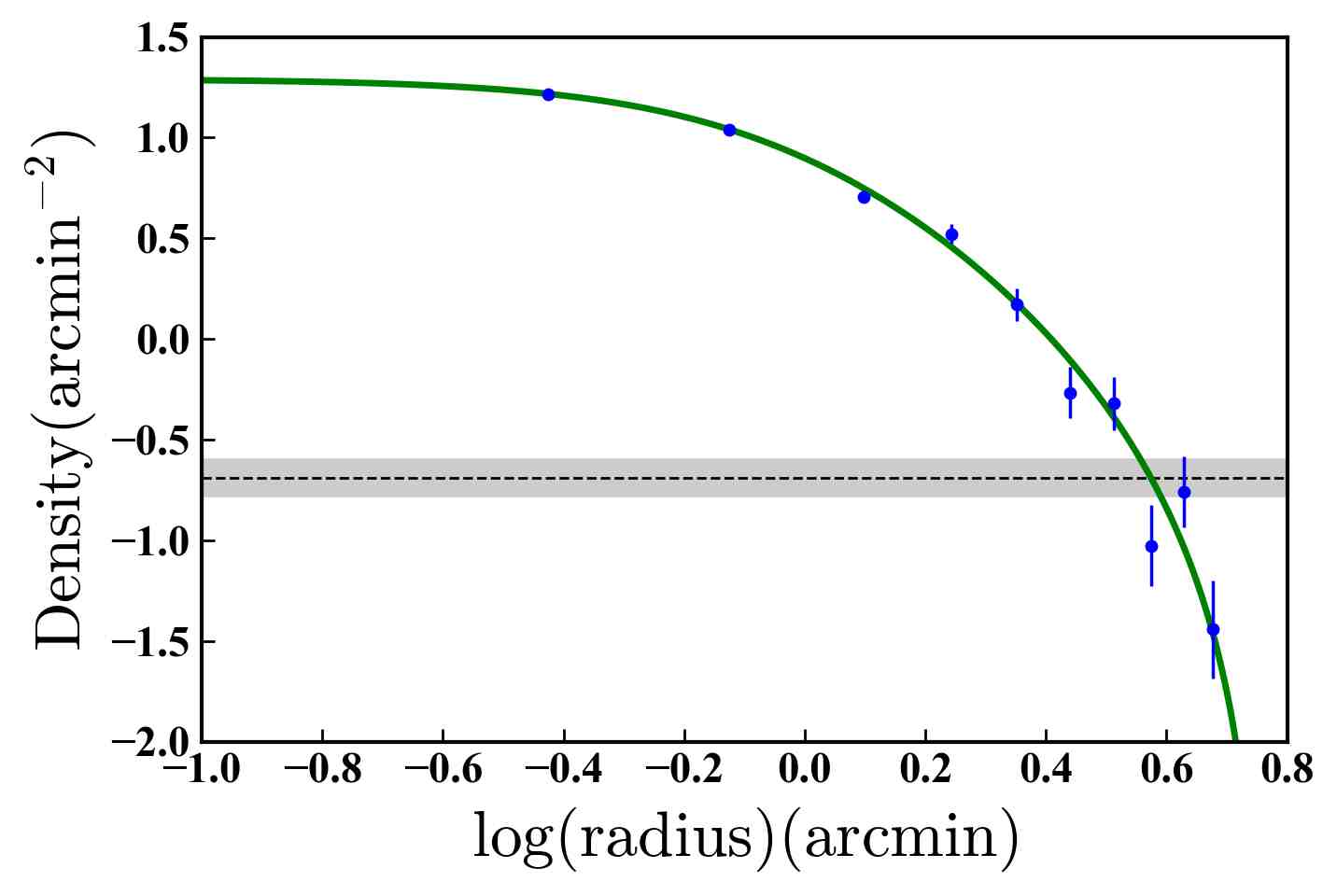}
    \end{minipage}\\
    \vspace{2mm}
    \begin{minipage}{0.99\textwidth}
        \centering
        \includegraphics[align=c,height=7.1cm]{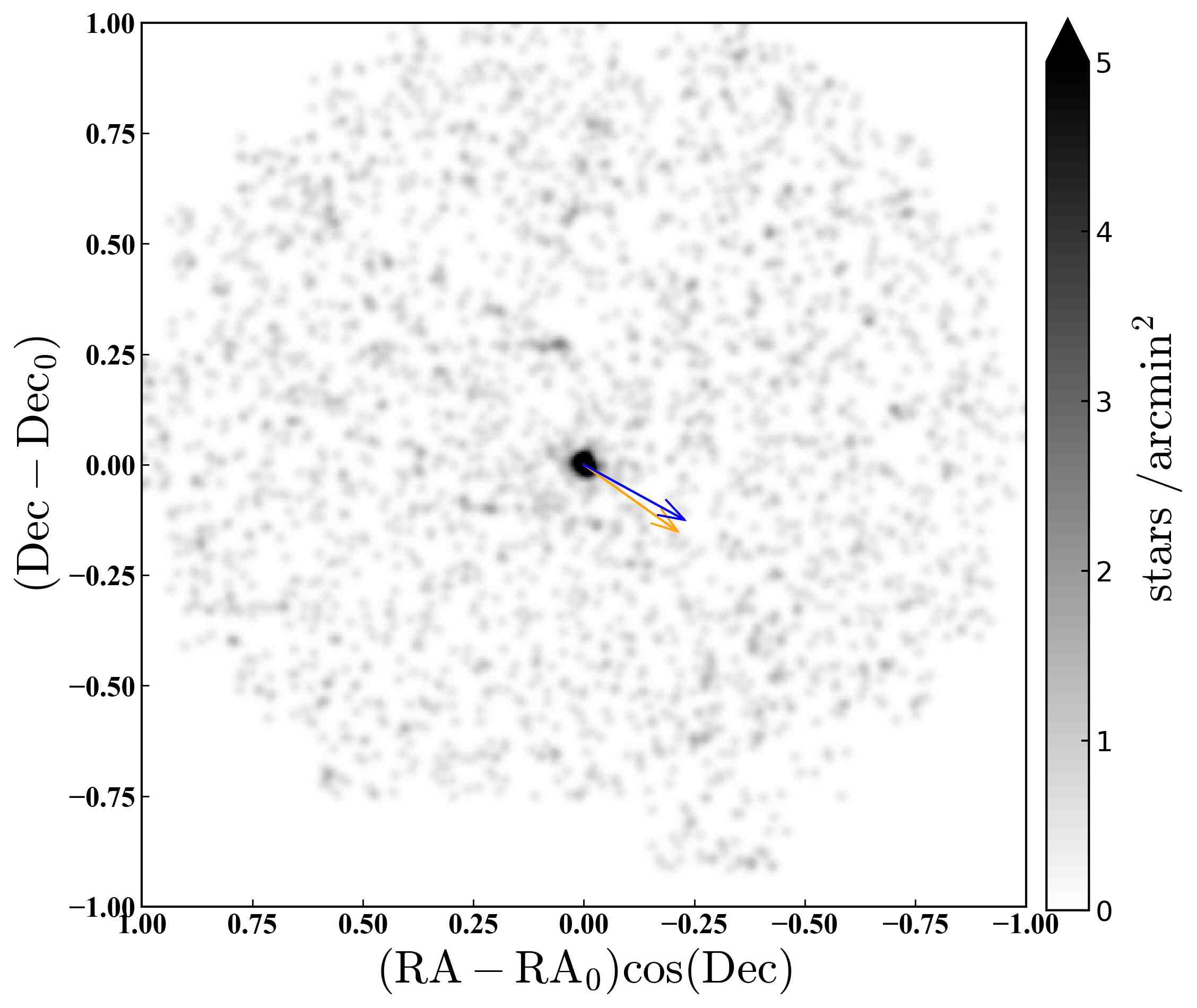}
        \hspace{2mm}
        \includegraphics[align=c,height=7.1cm]{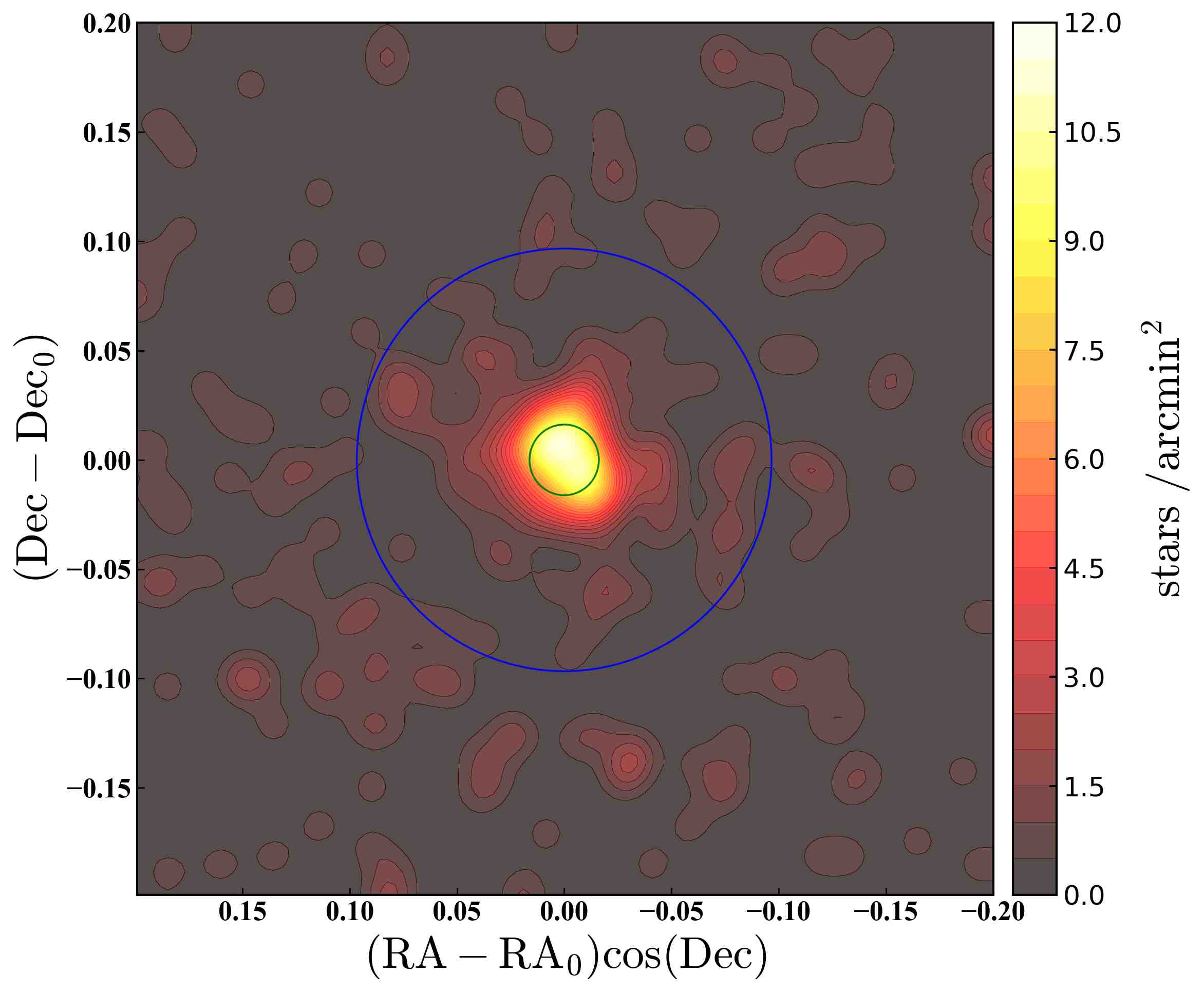}
    \end{minipage}
\caption{Same as Figure \ref{f:NGC1904}, but for Whiting 1.}
\label{f:whiting1}
\end{figure*}    

Whiting 1 is a remote outer halo cluster with $R_{\rm gc} \approx 35.2$\ kpc. It was first revealed as a halo object by \citet{carraro:05}, and later as an almost certain member of the Sagittarius stream by \citet{carraro:07} and \citet{carballo:17}.  \citet{carraro:07} also demonstrated that Whiting 1 is amongst the youngest globular clusters in the Milky Way system, with an age $\approx 6.5$\ Gyr. 

Our results for this cluster are shown in Figure \ref{f:whiting1}. Our photometry traces around $2$ magnitudes of the upper main sequence and MSTO.  As with NGC 7492, the known presence of surrounding Sagittarius populations \citep[see also][]{sollima:18} complicates the identification of cluster members, and we are careful to precisely define the selection region on the CMD \citep[although the main sequences of both the cluster and Sagittarius populations inevitably overlap to some extent -- see e.g.,][]{munoz:18a}. Neither our radial density profile, nor our two-dimensional density map, shows any evidence for extended structure around Whiting 1. The radial profile is well described by a King-type model with $r_c = 0.97 \pm 0.08\arcmin$ and $r_t = 5.8 \pm 0.3\arcmin$, which is slightly larger than the Jacobi radius $r_J \approx 4.3\arcmin$ predicted by \citet{balbinot:18}. Outside our nominal limiting radius, the density map shows uniformly distributed background fluctuations, most likely due to residual contamination from the Sagittarius stream (which the flatness of our full-field map shows must be wider than $\sim 2\degr$ on the sky at this location).

Deep CFHT/MegaCam photometry for Whiting 1 was presented by \citet{munoz:18b}, who observed no evidence for extra-tidal structure and measured a limiting radius $r_t = 6.2 \pm 0.6\arcmin$, consistent with the results presented here.

\subsection{AM 1}
\label{ss:am1}

\begin{figure*}
    \begin{minipage}{0.99\textwidth}
        \centering
        \includegraphics[align=c,height=6.6cm]{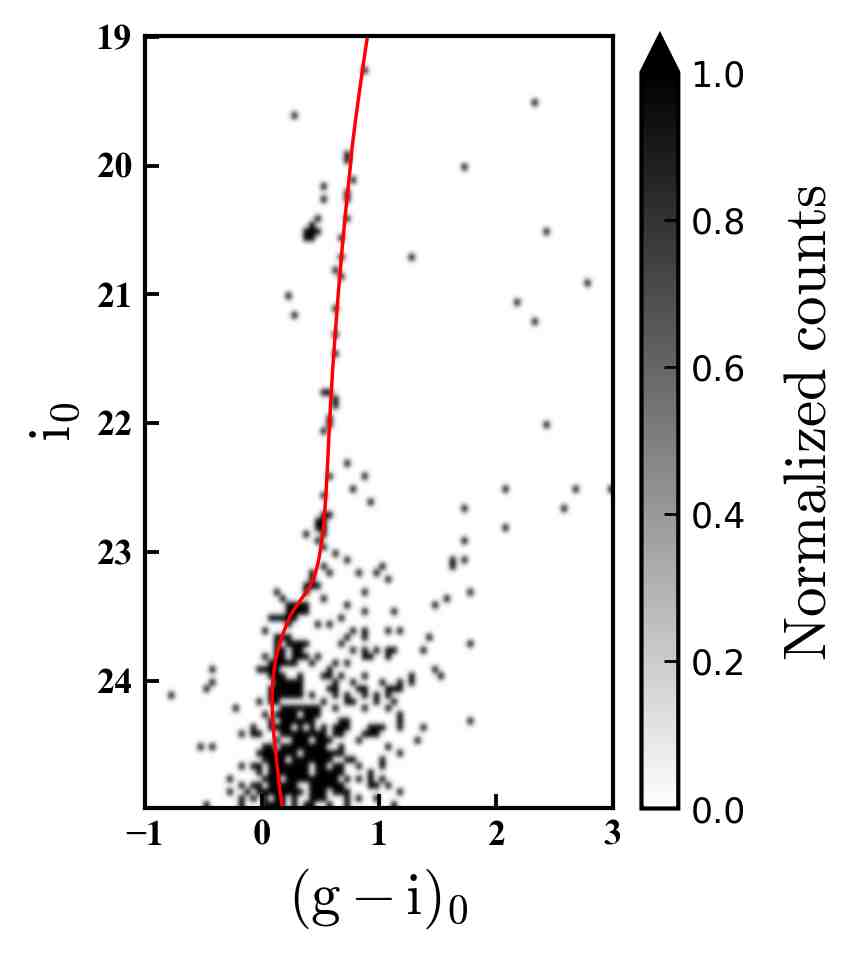}
        \hspace{-1mm}
        \includegraphics[align=c,height=6.6cm]{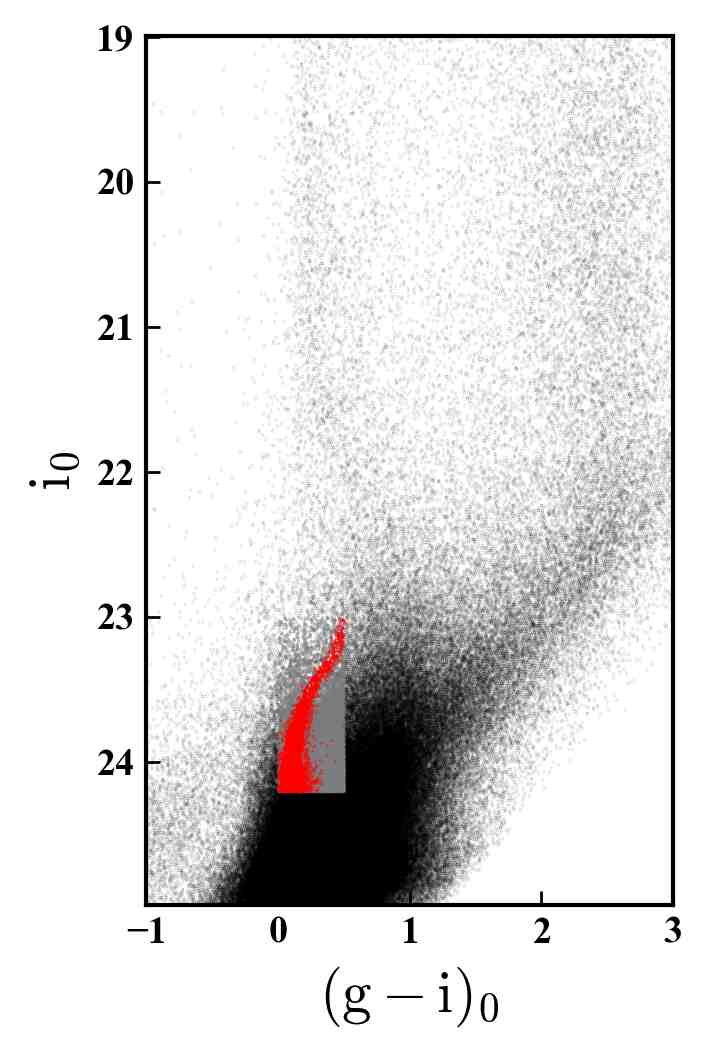}
        \hspace{1mm}
        \includegraphics[align=c,width=6.5cm]{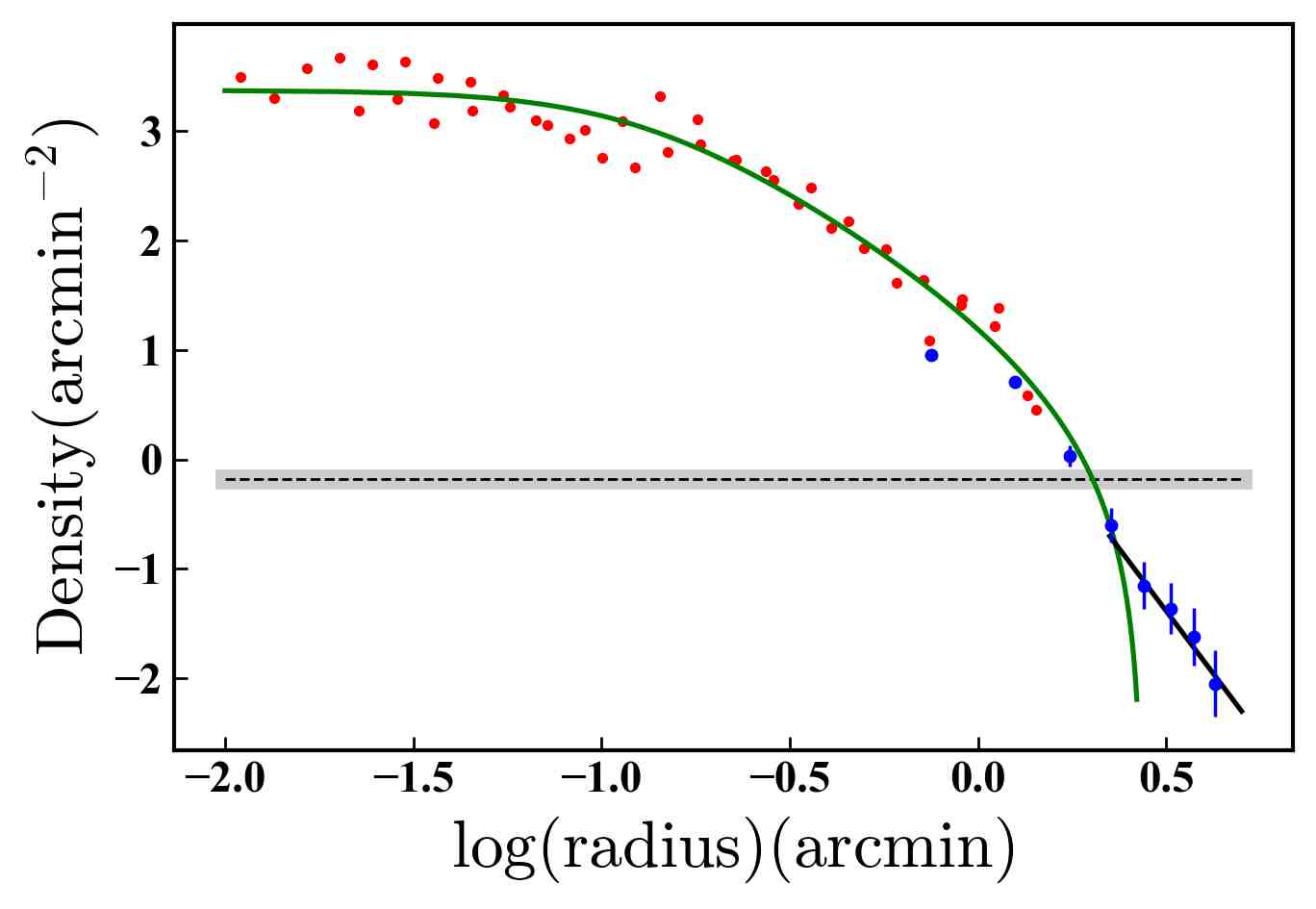}
    \end{minipage}\\
    \vspace{2mm}
    \begin{minipage}{0.99\textwidth}
        \centering
        \includegraphics[align=c,height=7.1cm]{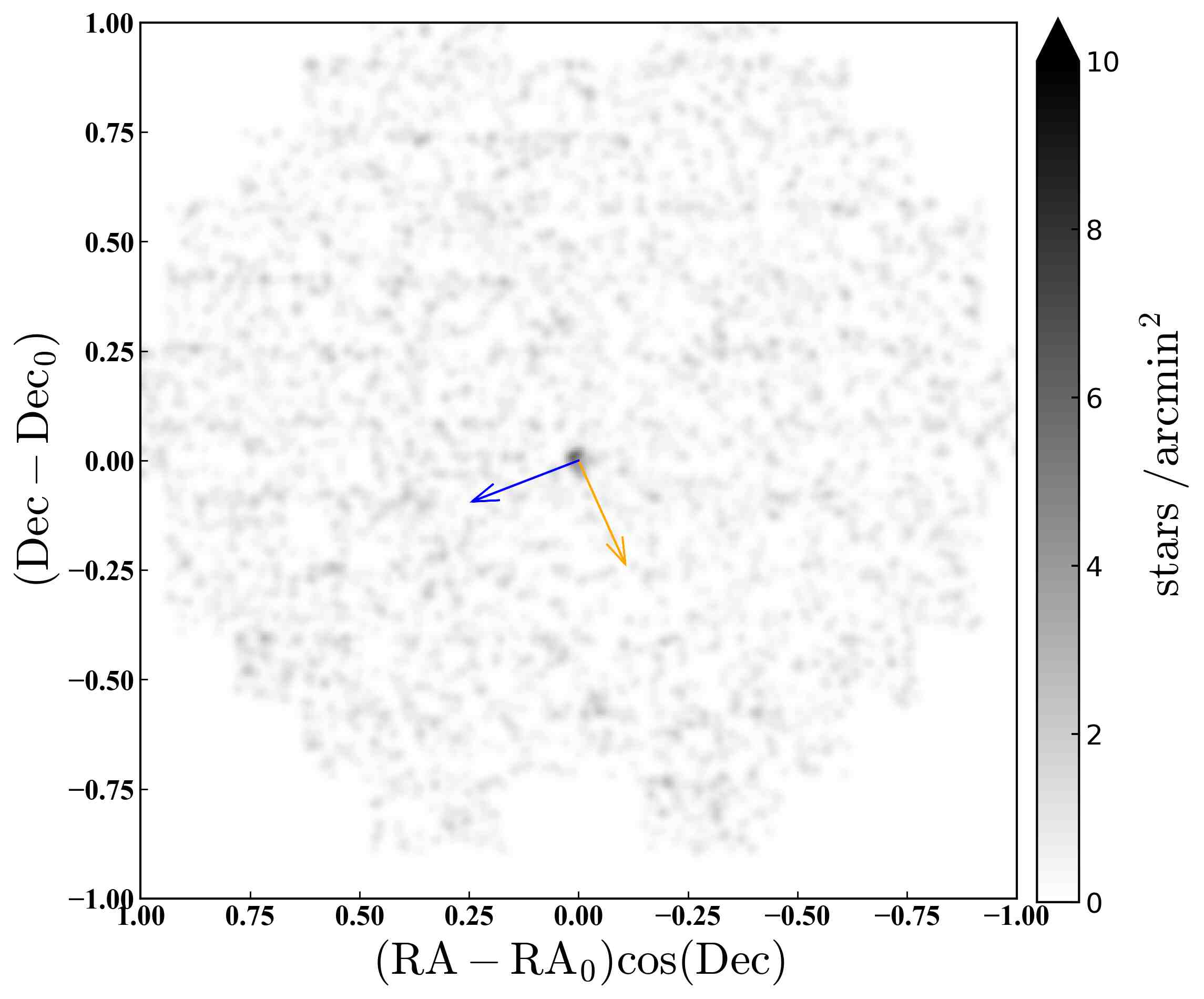}
        \hspace{2mm}
        \includegraphics[align=c,height=7.1cm]{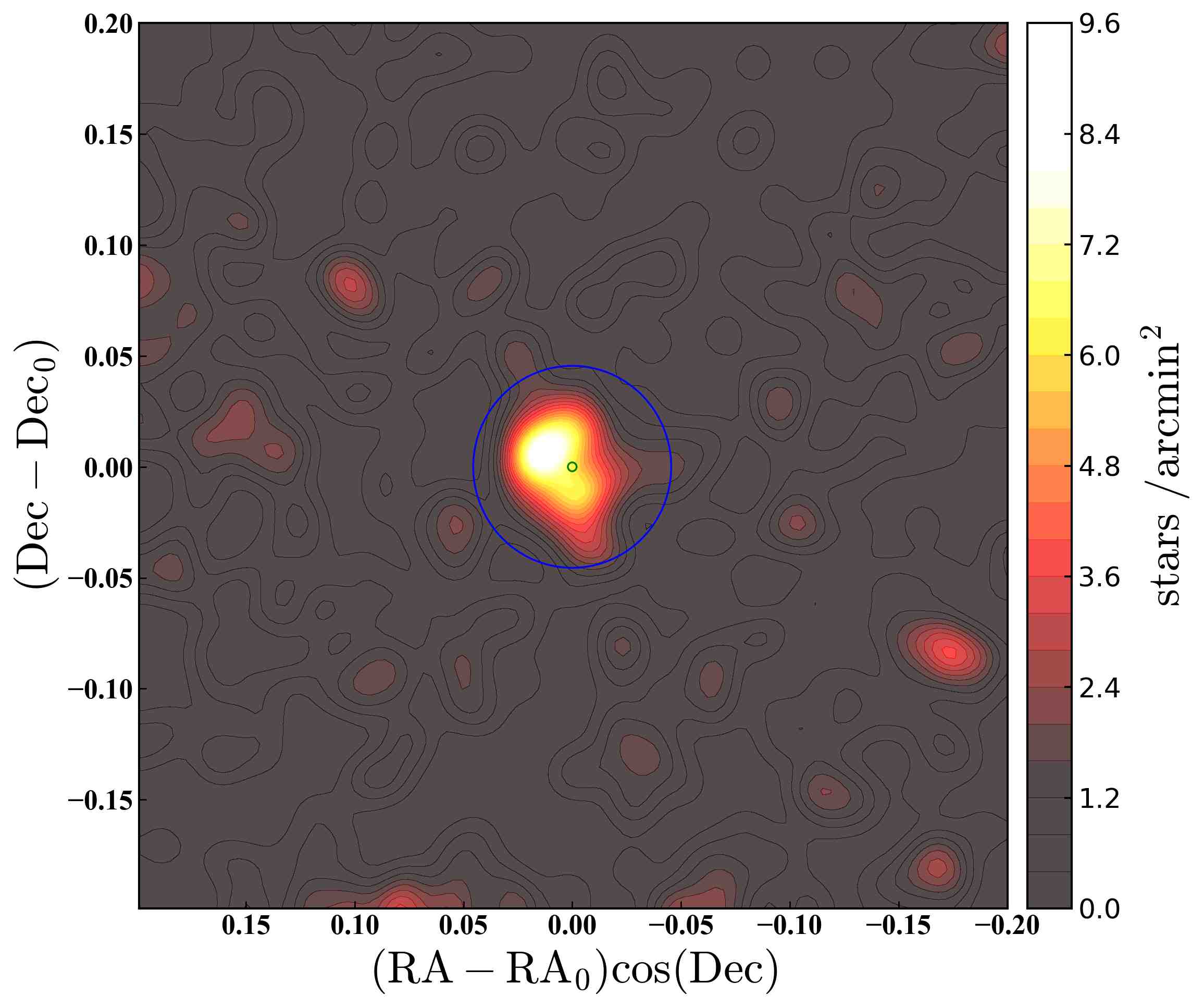}
    \end{minipage}
\caption{Same as Figure \ref{f:NGC1904}, but for AM 1.}
\label{f:am1}
\end{figure*}    

AM 1 is one of the most remote globular clusters in the Milky Way (second only to Crater), with a Galactocentric radius $120.3$\ kpc. Very little is known about the Galactic halo at such distances. Various properties of AM 1, including its size, metallicity, and location, make it a good candidate for having been accreted from a now-destroyed dwarf satellite \citep[e.g.,][]{mackey:04}. Our results for AM 1 are presented in Figure \ref{f:am1}. Because it is so far away, our photometry for this cluster barely reaches past the MSTO, despite the good observing conditions and long integration times. Contamination at these faint levels is mainly due to unresolved background galaxies (especially at the relatively blue colour of the MSTO), so in selecting an appropriate member region on the CMD we were careful to avoid the worst of these sources. Nonetheless, for this target there is overall only comparatively low signal from cluster members relative to the background.

We are able to trace the cluster to a radius $\approx 3\arcmin$. The radial density profile appears somewhat more extended than would be expected from a King model. We attempt to fit a King model in the same way as previously described (e.g., for NGC 1904) and measure $r_c = 0.12\pm 0.01\arcmin$ and $r_t = 2.7\pm 0.1\arcmin$. Outside this radius the data can be described by a power law with a relatively steep fall-off, $\gamma = -4.6$. However, the two-dimensional density map does not show convincing evidence for populations in excess of the background fluctuations outside our King-model limiting radius. \citet{balbinot:18} calculate a Jacobi radius of $r_J \approx 5.9\arcmin$, which is significantly larger and reinforces the idea that AM 1 does not have any extra-tidal structure, at least to the limit of our photometry.

Our density map does suggest, at low significance, that AM 1 may be somewhat asymmetric inside its limiting radius, with possible extension in the NE-SW direction. Interestingly, the position angle of the Galactic centre with respect to AM 1 is $\phi_{\rm gc} = 204\degr$. However, our observations are affected by crowding at the centre of the cluster, and its large distance means that AM 1 is a small target such that the smoothing in our map could produce at least some of this apparent asymmetry. It would be worth revisiting this cluster again in future using wide-field space-based imaging to (i) mitigate the effects of crowding, and (ii) increase the signal from cluster members relative to the background though deeper photometry and better star/galaxy separation.

The structure of AM 1 has previously been measured by \citet{miocchi:13}, who were able to trace the profile to just over $3\arcmin$. They find a marginal preference for the more extended \citet{wilson:75} model than a \citet{king:66} model when fitting their data, consistent with our observation of a mild departure from a King-type shape in the cluster outskirts. AM 1 also appears in the study of \citet{sollima:18}, who measured a limiting radius of $\approx 3.6\arcmin$, as well as in the large sample of objects observed by \citet{munoz:18a,munoz:18b}, who also observe that the radial profile is more extended than can be described by a King-type model. Interestingly, they also observed a mildly elliptical shape for this cluster ($e = 0.16 \pm 0.06$) oriented in the same NE-SW direction observed here, but no evidence for extra-tidal structure (again consistent with our conclusions).

\subsection{Pyxis}
\label{ss:pyxis}

\begin{figure*}
    \begin{minipage}{0.99\textwidth}
        \centering
        \includegraphics[align=c,height=6.6cm]{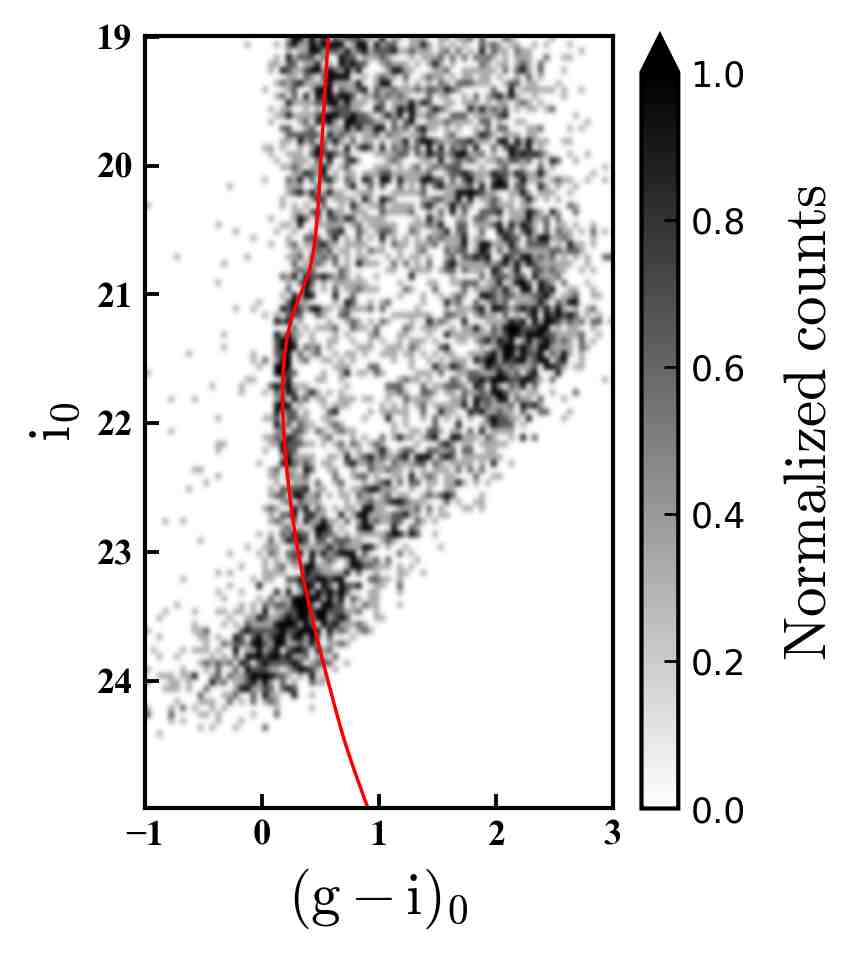}
        \hspace{-1mm}
        \includegraphics[align=c,height=6.6cm]{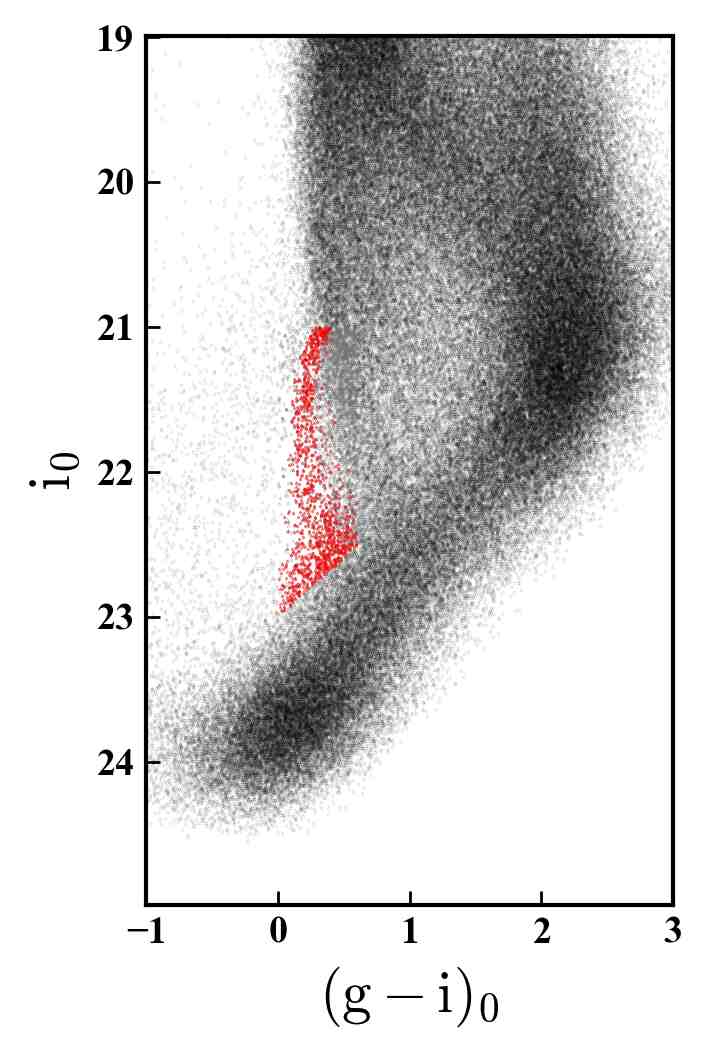}
        \hspace{1mm}
        \includegraphics[align=c,width=6.5cm]{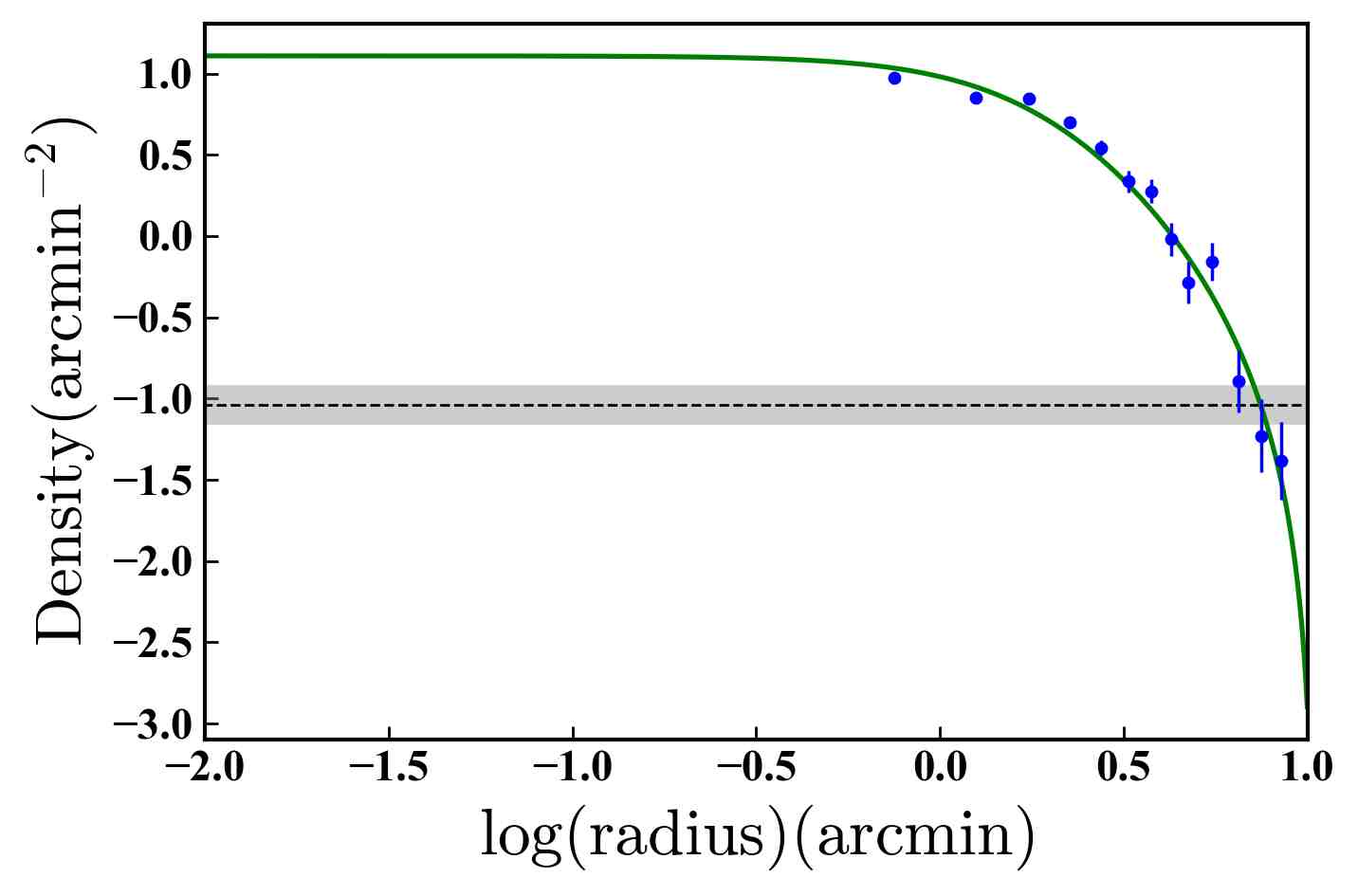}
    \end{minipage}\\
    \vspace{2mm}
    \begin{minipage}{0.99\textwidth}
        \centering
        \includegraphics[align=c,height=7.05cm]{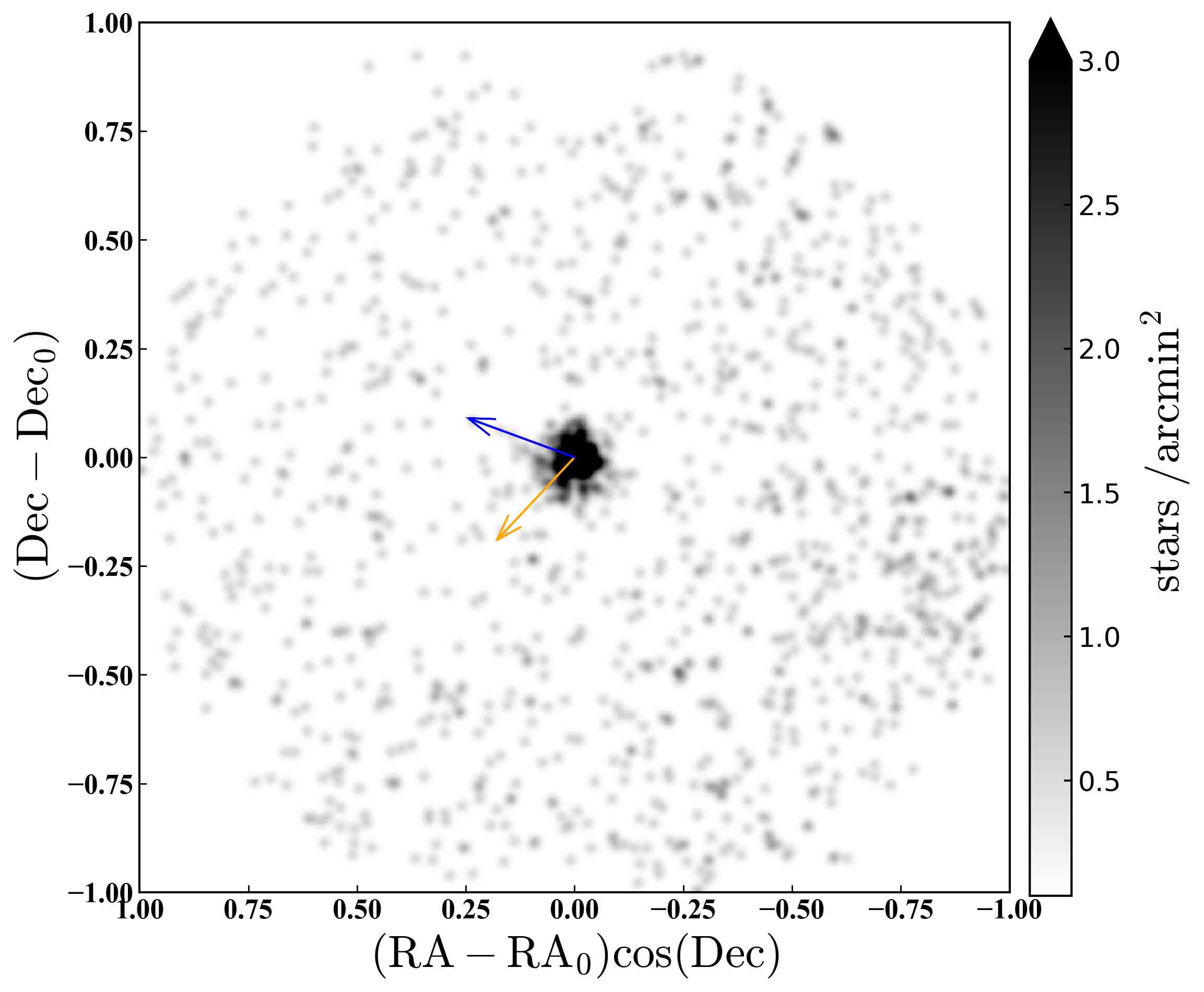}
        \hspace{2mm}
        \includegraphics[align=c,height=7.05cm]{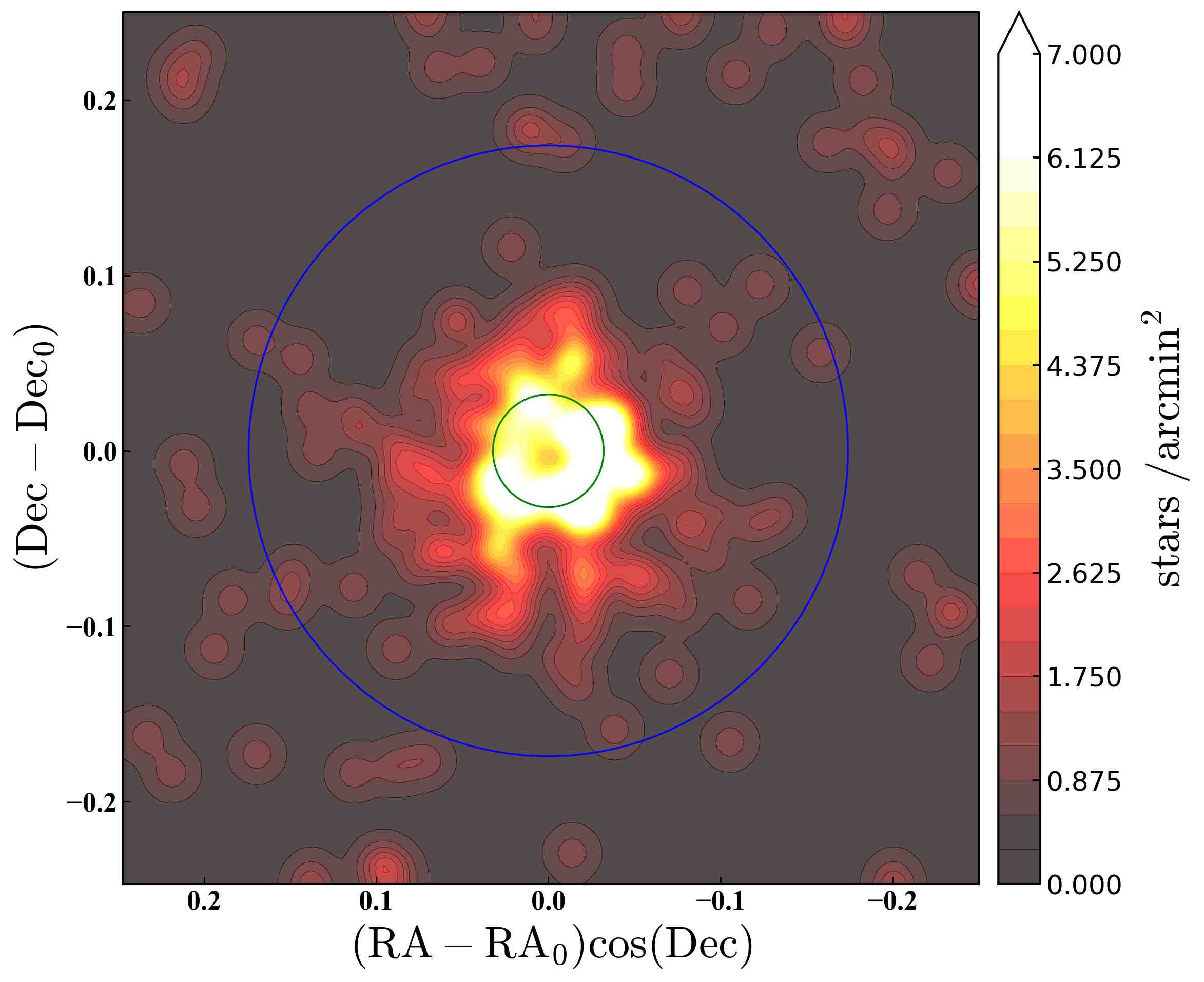}
    \end{minipage}
\caption{Same as Figure \ref{f:NGC1904}, but for Pyxis.}
\label{f:pyxis}
\end{figure*}

Pyxis is the second most remote object in our sample, with a Galactocentric distance of $38.6$\ kpc. Originally recognised as a candidate low-luminosity globular cluster by \citet{weinberger:95}, and subsequently confirmed as such by \citet{dacosta:95} and \citet{irwin:95}, Pyxis is a challenging target due to its low Galactic latitude and consequent heavy foreground contamination and relatively high reddening. Proper motion measurements by \citet{fritz:17} showed that Pyxis has a rather eccentric orbit and was likely accreted into the Milky Way from a now-destroyed progenitor system.

Our results for Pyxis are shown in Figure \ref{f:pyxis}. Despite the observational difficulties, our photometry clearly extends at least $\sim 1.5$\ mag below the MSTO. However, we restrict the member selection region to the MSTO and upper $\approx 1$\ mag of the main sequence in order to avoid the worst of the contamination across the DECam field of view. Our radial density profile traces the cluster to approximately $10\arcmin$ and is well fit by a King model with $r_c = 1.93\pm0.38\arcmin$ and $r_t = 10.5\pm0.5\arcmin$, although the overall contrast is significantly lower than for the more luminous and less contaminated clusters in our sample\footnote{For example, the dynamic range of the radial density profiles for clusters such as NGC 1904, 2298, 6864, and 6981 is $\sim 5-7$ orders of magnitude, compared to $\sim 2.5$ for Pyxis.}. This matches the predicted Jacobi radius of $r_J \approx 12.4\arcmin$ from \citet{balbinot:18} quite closely. 

Although Pyxis does not obviously possess any structure outside its limiting radius, our density map suggests that the distribution of members in its outskirts may be mildly asymmetric, with structures plausibly visible to larger radii in the south and east than the north and west. Interestingly, the direction of the Galactic centre with respect to Pyxis is almost exactly south-east ($\phi_{\rm gc} \approx 137\degr$). Independent Magellan/Megacam photometry for Pyxis has been presented by \citet{munoz:18b}, who measured a limiting radius of $8.2 \pm 0.5\arcmin$ in good agreement with our determination; however, these authors do not observe any sign of an asymmetric outer shape. Nonetheless, this target would be worth revisiting in future with deeper photometry and/or kinematic information. 

\subsection{IC 1257}
\label{ss:IC1257}

\begin{figure*}
    \begin{minipage}{0.99\textwidth}
        \centering
        \includegraphics[align=c,width=5.9cm,height=6.6cm]{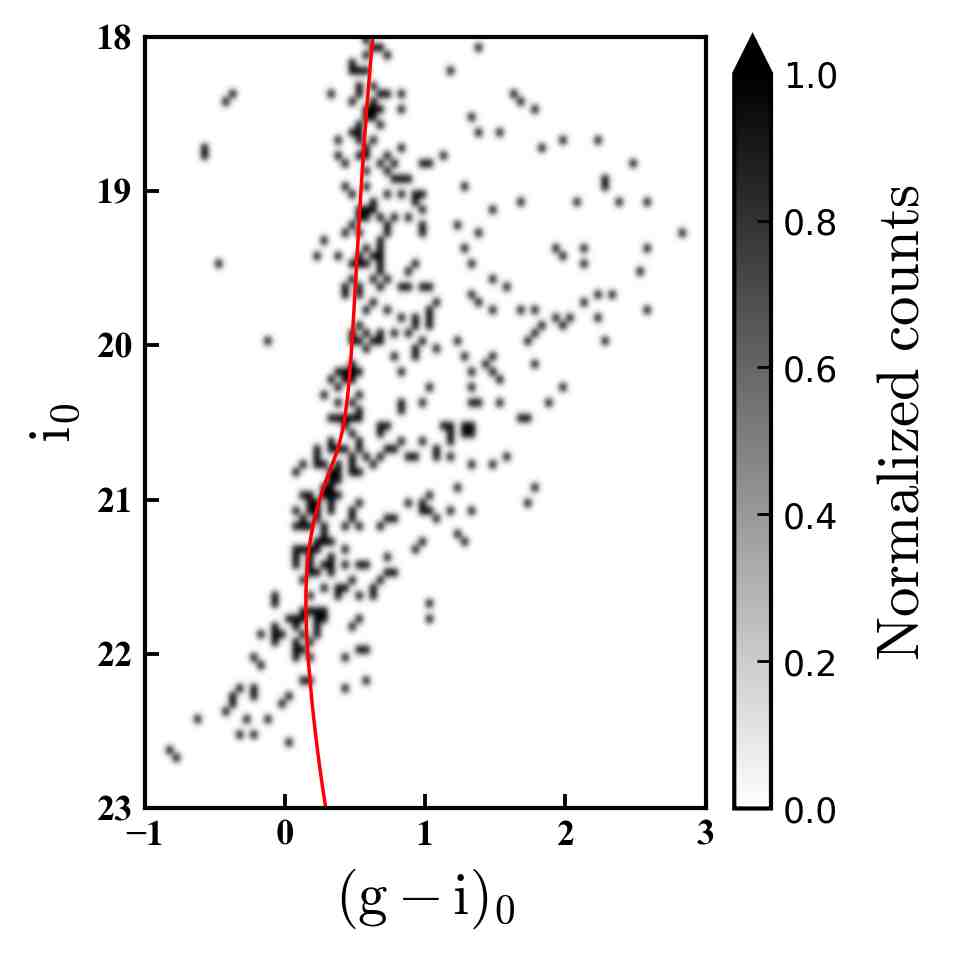}
        \hspace{-1mm}
        \includegraphics[align=c,width=4.4cm,height=6.6cm]{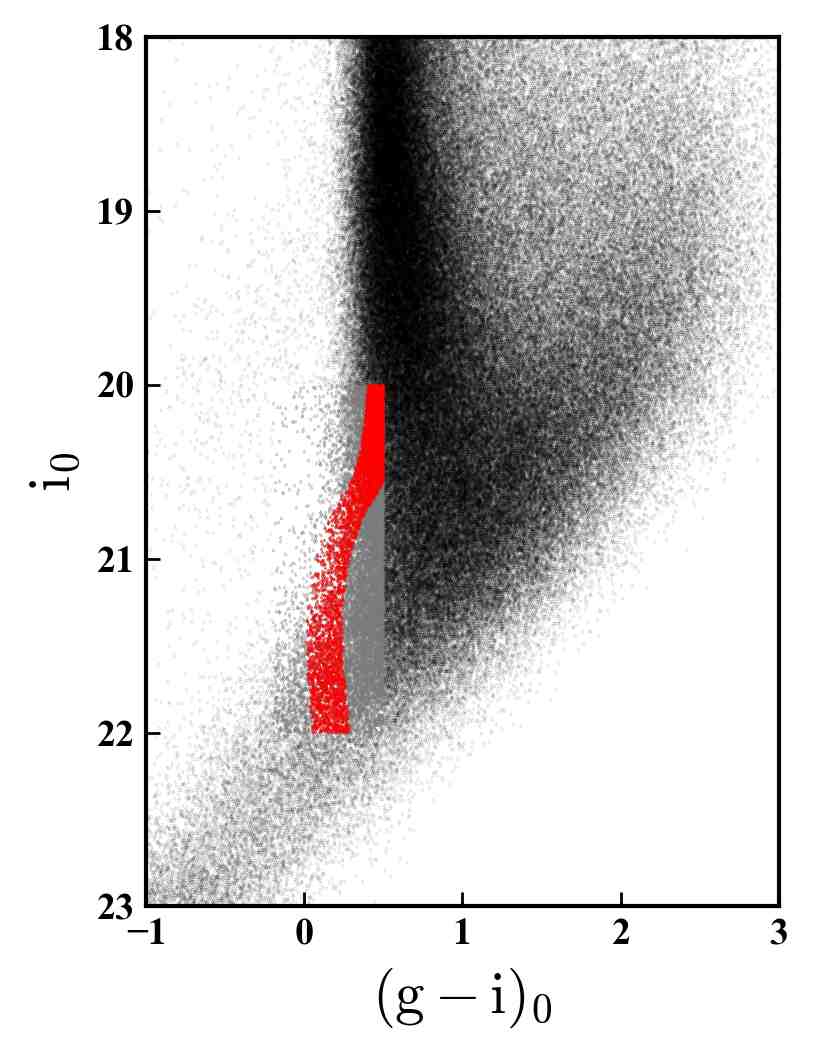}
        \hspace{1mm}
        \includegraphics[align=c,width=6.5cm]{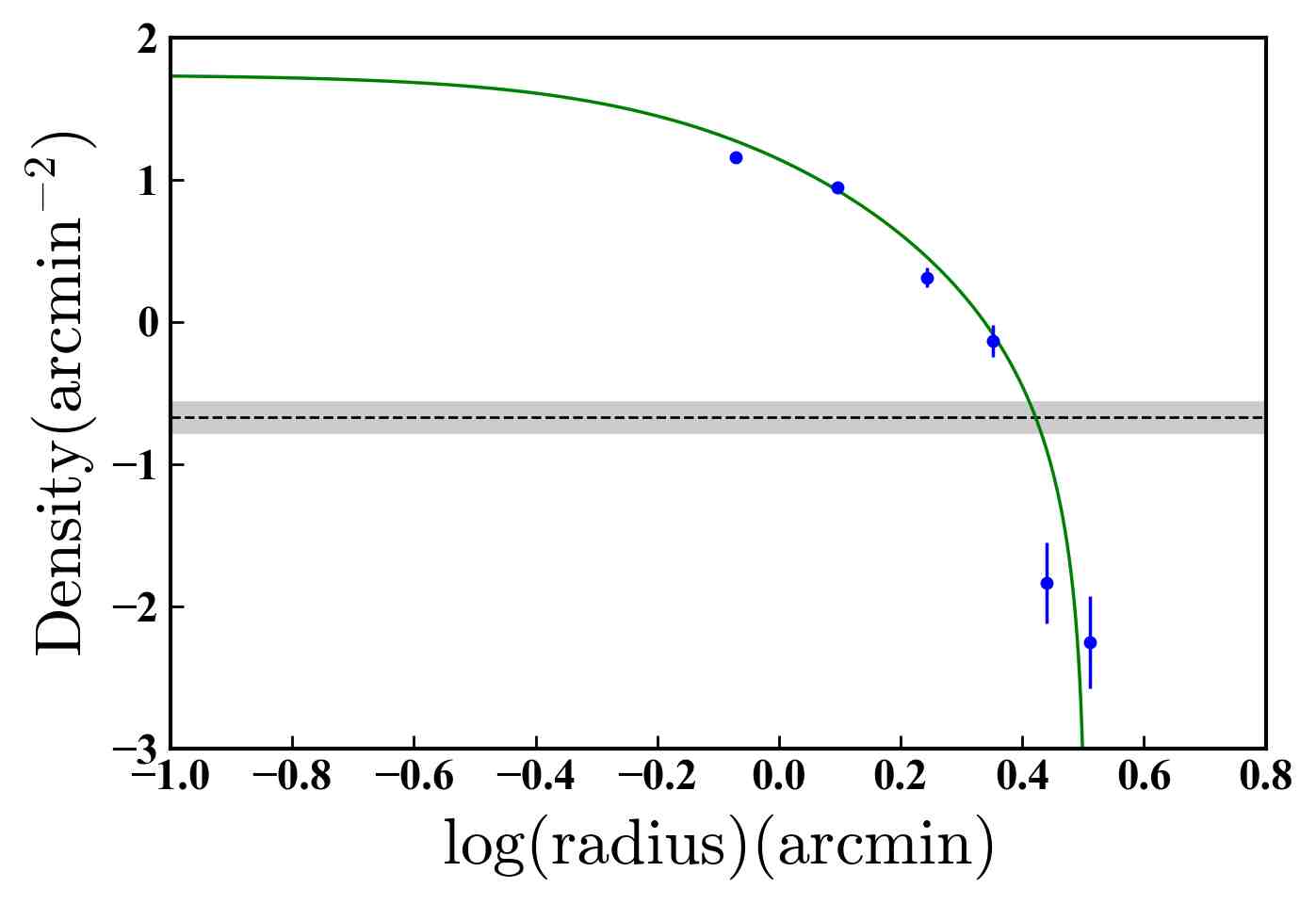}
    \end{minipage}\\
    \vspace{2mm}
    \begin{minipage}{0.99\textwidth}
        \centering
        \includegraphics[align=c,height=7.05cm]{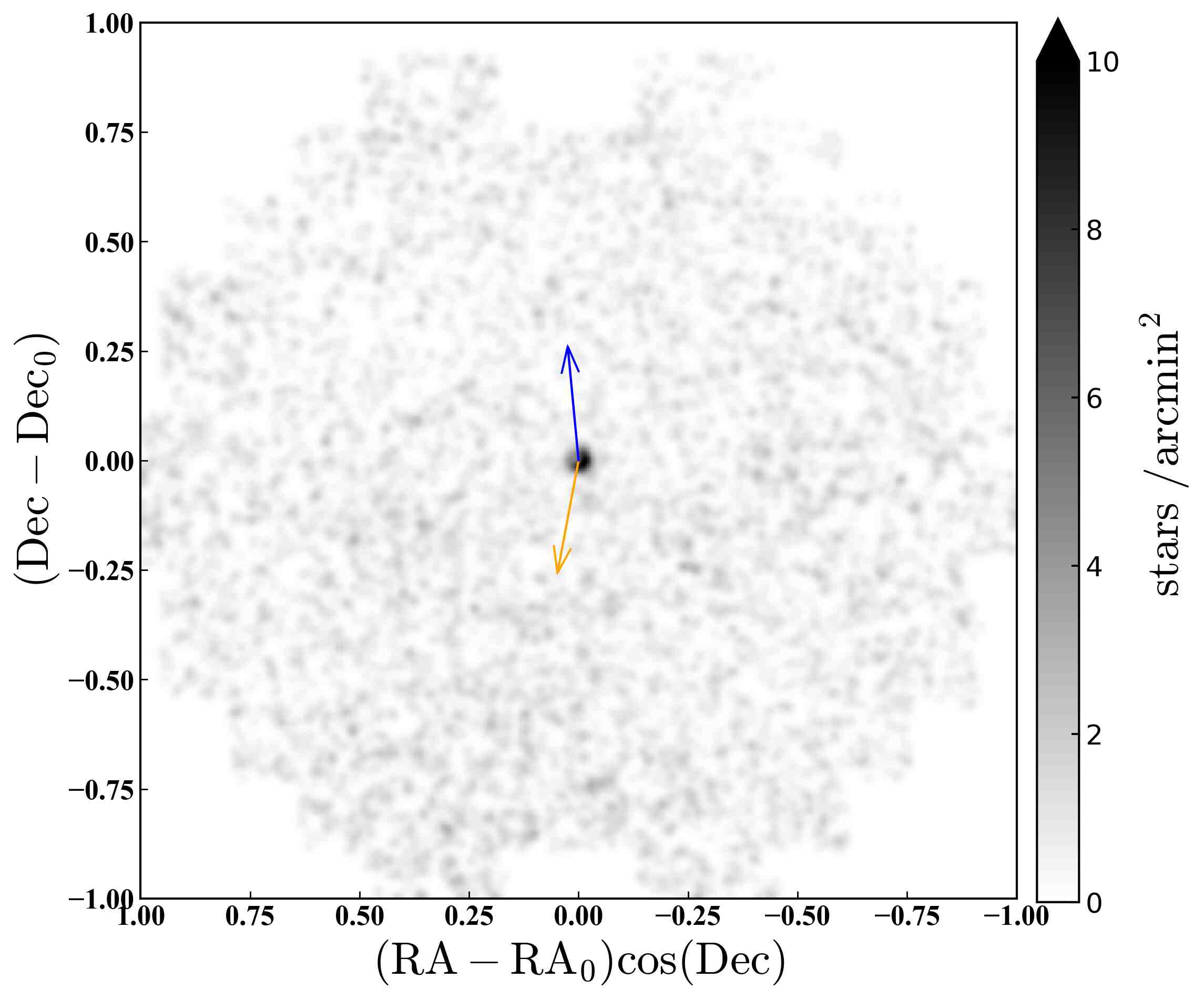}
        \hspace{2mm}
        \includegraphics[align=c,height=7.05cm]{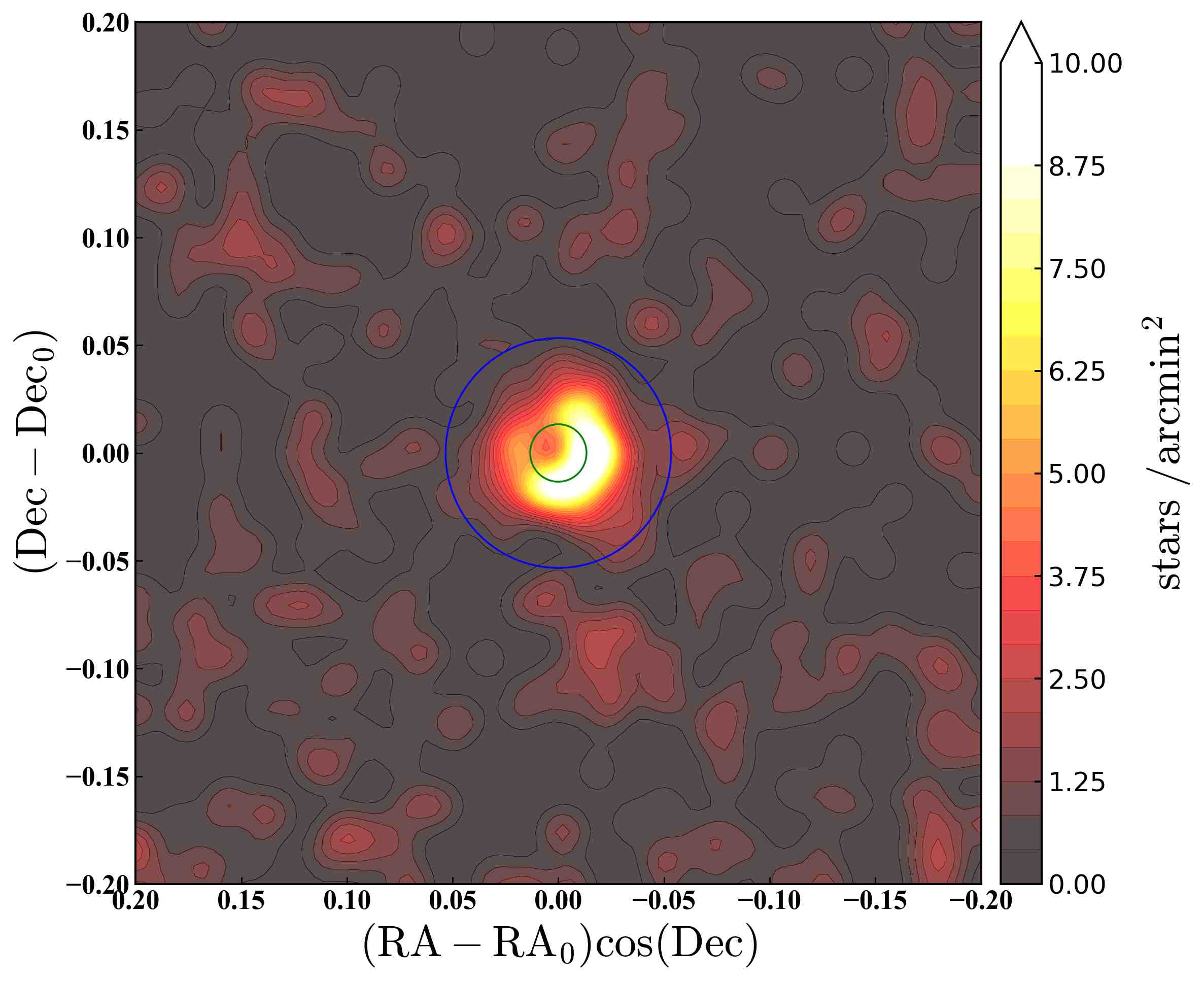}
    \end{minipage}
\caption{Same as Figure \ref{f:NGC1904}, but for IC 1257.}
\label{f:IC1257}
\end{figure*}    

IC 1257 is a poorly studied cluster due to its location behind the Galactic plane.  This object sits at a Galactocentric radius of $19.3$\ kpc, and was first recognised as a halo globular cluster by \citet{harris:97}.  Heavy foreground contamination and line-of-sight extinction render IC 1257 a very challenging target; however, the sparsity of existing data made it a worthwhile addition to our sample. 

Our results for IC 1257 are presented in Figure \ref{f:IC1257}. Our photometry barely reaches the MSTO; however, as some of the bluest stars in the cluster, members in this region are relatively well separated from the majority of foreground stars on the CMD. Despite this, our density map reveals moderately high residual contamination. We are able to trace the cluster to approximately $3\arcmin$ from its centre, and find the shape of its radial density profile to be well described by a King model with $r_c = 0.80 \pm 0.09\arcmin$ and $r_t = 3.2\pm0.3\arcmin$. To the best of our knowledge these are the only structural measurements of IC 1257 in optical passbands \citep[e.g., no data are listed by][]{harris:10}; however, \citet{bonatto:08} present infrared measurements from 2MASS data, finding a smaller core radius $r_c = 0.2\pm0.1\arcmin$ and larger limiting radius $r_t = 7.1\pm2.0\arcmin$. This suggests that crowding and the high background level may affect one or both profiles.

\section{Discussion}
\label{s:discussion}
	
\subsection{The present work in context}
We have used deep wide-field DECam imaging to explore the peripheries of nine globular clusters in the outer halo of the Milky Way. Apart from Whiting 1, which is known to be embedded in the Sagittarius stream \citep[e.g.,][]{carraro:07,carballo:17}, and NGC 7492, which is projected against different wraps of the same stream \citep[e.g.,][]{carballo:18b}, we see no evidence for co-located stellar populations around any of our clusters that might indicate the presence of coherent tidal debris from a destroyed host dwarf. While it is possible that our method for eliminating the effects of residual non-member contamination in our density maps (Section \ref{ss:2dmap}) could lead to the removal of cluster-like stellar populations near the edge of a given field-of-view, inspection of the raw CMDs for the outskirts of each field does not show any convincing evidence for the presence of such populations.

Our results are consistent with previous efforts \citep[e.g.,][]{carballo:14,sollima:18}, as well as recent {\it Gaia} results on the accretion history of the Milky Way. These latter suggest that although multiple merger events can be identified using phase space information for field stars \citep[e.g.,][]{helmi:18,belokurov:18,myeong:19,koppelman:19,naidu:20}, apart from Sagittarius they all occurred many Gyr ago and their debris is now spatially well mixed in the Galactic halo. We now know\footnote{Although we were not aware at the time of observation.} that the majority of the clusters in our sample can be identified with these ancient events \citep[e.g.,][]{massari:19}: NGC 1904, NGC 2298, NGC 6864, NGC 7492, and IC 1257 were likely accreted with {\it Gaia}-Enceladus-Sausage progenitor, while NGC 6981 is plausibly associated with the event that created the \citet{helmi:99} streams. Only AM 1 and Pyxis, at large Galactocentric radii, are not associated with these events; their orbital properties (high energy, high eccentricity, polar or retrograde motion) suggest that they were likely accreted from different low-mass progenitors that did not contribute significant field populations to regions nearer the Sun \citep[otherwise they would be detectable with {\it Gaia}, see][]{massari:19}. Additional study of the regions surrounding these two difficult targets would therefore be worthwhile.

We also found that none of the clusters in our sample exhibit obvious tidal tails, although two -- NGC 1904 and NGC 6981 -- have extended outer structures with power-law radial density profiles similar to other clusters that have been identified as possessing stellar envelopes \citep[e.g., NGC 1261, NGC 1851, and NGC 7089 -- see][]{olszewski:09,kuzma:16,kuzma:18}. We explore the nature of such envelopes in the following two sections. 

In general our observations of the outer structures of the nine clusters in our sample are consistent with previous analyses, with a couple of notable exceptions. First, we are unable to reproduce the detection of tidal tails around NGC 7492 by several previous works \citep{leon:00,lee:04,navarrete:17}. A similar issue was encountered by \citet{munoz:18b}, who suggested that this disparity in outcomes might be due to the presence of the Sagittarius stream populations that surround NGC 7492 in projection and potentially complicate member selection. One possible difference between the studies which do, and do not, detect tidal tails around this cluster is photometric depth: the data of \citet{leon:00} do not appear to reach the main-sequence turn-off, while the Pan-STARRS PS1 photometry used by \citet{navarrete:17} extends at most $\sim 1$\ mag below the turn-off.  In contrast, our photometry reaches more than $\sim 2.5$\ mag below the turn-off, while that of \citet{munoz:18b} extends about $4$\ mag below the turn-off. The exception is the study of \citet{lee:04}, whose photometry is comparable to ours; however, taken at face value, this pattern suggests that deeper photometry is plausibly better able to separate cluster members from field stars -- likely because the luminosity function for cluster members steeply increases the further down the main sequence one is able to trace.

The second discrepancy worth noting is that for NGC 2298.  In this case, all recent deep wide-field ground-based studies, including the present work, agree that there is little evidence for distortion or extra-tidal features in the outskirts of this cluster (see the discussion in Section \ref{ss:NGC2298}). In contrast, however, two recent {\it Gaia}-based studies -- those of \citet{sollima:20} and \citet{ibata:21} -- have revealed extended tidal tails belonging to NGC 2298 and spanning at least $\sim 12$\degr on the sky. Note that, as discussed in the following two sections (see also Appendix \ref{a:class}), a similar discrepancy arises for a significant number of additional clusters -- for example, NGC 1261, NGC 1851 and NGC 7089 (all three of which were known to possess extended envelopes but not long tails), NGC 2808, NGC 6397, NGC 7099, and others. This suggests that the power of {\it Gaia} astrometry to cleanly filter cluster members from contaminants across very large regions of sky, especially when combined with algorithms tuned for finding long stream-like structures \citep[such as the {\sc streamfinder} algorithm of][]{ibata:19b,ibata:21}, may strongly exceed the ability of purely photometric ground-based studies to detect very low surface-brightness structures. 

It is likely also true that \citep[as noted by][]{bonaca:21} many clusters with long tidal tails actually exhibit surprisingly little detectable structure in close proximity to the clusters themselves. Good examples are NGC 4590, which both \citet{ibata:21} and \citet{bonaca:21} associate with the Fj\"{o}rm stream of \citet{ibata:19b} but where the {\it Gaia}-based study of the cluster outskirts by \citet{sollima:20} detected no structure, and NGC 5272, which \citet{bonaca:21} associate confidently with the Sv\"{o}l stream of \citet{ibata:19b} but where \citet{sollima:20} again sees no structure. This may reflect changes in the mass-loss rate as a function of time, but careful modelling is required to test this assertion. Clearly more factors are at work than a simple correlation between mass-loss and orbital phase -- calculations in the next section show that NGC 4590 and NGC 3201 (with comparably long tails to NGC 4590 -- the Gj\"{o}ll stream -- but which are easily detectable in the vicinity of the cluster) share almost identical eccentricity, apocentre radius, inclination, and phase, with the sole difference being that NGC 3201 is on a retrograde orbit and NGC 4590 on a prograde orbit.

It is worth noting that traditional ground-based studies of cluster outskirts are not yet obsolete.  These have been notably successful in in identifying systems with large tidal tails (such as Pal 5 and NGC 5466) prior to the {\it Gaia} era, and they are still likely superior for studies of more distant objects where {\it Gaia}'s faint limit for astrometry precludes the detection of many members. A good example is provided by the outer halo clusters Eridanus and Pal 15, which were revealed with DECam imaging to possess tidal tails \citep{myeong:17} but which do not appear in e.g., the blind {\it Gaia} surveys of \citet{ibata:19b,ibata:21}. As discussed above, however, a key caveat that now needs to be recognised with such work is that even if there is no extra-tidal structure detectable in the vicinity of a cluster, it may still possess long tidal tails.

\subsection{Globular cluster orbits and extra-tidal structures}
\label{ss:orbits}
Given the above discussion, we decided to investigate the extent to which extra-tidal structures around globular clusters, in particular the presence or absence of tidal tails, can be linked to the properties of their orbits. This is not a new idea -- variations on this theme have previously been explored by numerous authors \citep[e.g.,][and many others]{grillmair:95,leon:00,jordi:10,carballo:12,balbinot:18}. Most recently, \citet{piatti:20b} compiled an exhaustive list of clusters for which the presence, or absence, of extra-tidal structure could be classified, and correlated their findings against various kinematic and internal properties. However, they found that none was a straightforward predictor for the development of tidal tails or other extra-tidal features.

We are motivated to revisit this problem for two key reasons. First, the recent release of improved {\it Gaia} astrometry in EDR3 has led to the discovery of a large number of new stellar streams in the Milky Way halo \citep[e.g.,][]{ibata:21}, many of which are apparently due to the escape of stars from globular clusters \citep[e.g.,][]{bonaca:21}. {\it Gaia} EDR3 has, moreover, led to substantially improved proper motion information {\it and} distance measurements for Galactic globular clusters themselves \citep{vasiliev:21,baumgardt:21}, both of which propagate into producing higher precision orbital data. Second, as discussed above, it is becoming increasingly clear that some detection methodologies for extra-tidal structures are apparently more robust than others. This suggests that it may be fruitful to pursue a new classification approach aimed at obtaining, as far as possible, a homogeneous data set of the highest quality observations, and employing a systematic set of hierarchical criteria for weighting results in cases of disagreement.

Addressing this problem requires (i) a set of clusters where the presence or absence of extra-tidal structure has been classified, and (ii) information on the orbital properties of all Milky Way globular clusters. Our classification scheme is described in full in Appendix \ref{a:class}; in short, we conducted a literature review of results on globular cluster extra-tidal structures and accept objects that have been analysed in one or more of the following ways: 
\begin{enumerate}
\item{Using algorithms tuned for detecting long stream-like structures in the Galactic halo using {\it Gaia} data \citep[e.g.,][]{ibata:19b,ibata:21}, and/or algorithms for associating known streams with globular clusters using orbit integration based on full 6D phase-space information \citep[e.g.,][]{bonaca:21};\vspace{1mm}}
\item{Using {\it Gaia} data to study the immediate outskirts of clusters, beyond the Jacobi radius \citep[e.g.,][]{carballo:19,sollima:20};\vspace{1mm}}
\item{Using wide-field ground-based data to explore the immediate outskirts of clusters, where the azimuthal coverage is complete and the photometry extends {\it at least} $2.5$\ magnitudes below the main-sequence turn-off \citep[e.g.,][and the present work]{munoz:18a,munoz:18b};\vspace{1mm}}
\item{Using wide-field ground-based data to explore the immediate outskirts of clusters, where the azimuthal coverage is complete, the photometry extends between $1.5-2.5$\ magnitudes below the main-sequence turn-off, and at least two independent studies have reached consistent conclusions.}
\end{enumerate}
The possible classifications are `T' (a cluster with tidal tails), `E' (a cluster with an envelope), and `N' (a cluster with no extra-tidal structure). In cases of disagreement between studies of a given cluster that fall in different categories, the above hierarchy is used to resolve the classification -- i.e., category (i) supersedes the other three, category (ii) supersedes the third and fourth categories, and so on. Clusters that have not been studied in one of the four ways listed above remain unclassified.

Table \ref{t:class} lists the results of this exercise. We are able to classify $46$ clusters altogether; a detailed accounting of our rationale behind the adopted classification for each is presented in Appendix \ref{a:class}. Of these $46$ objects, $27$ are classified `T', four are classified `E', and the remaining $15$ are `N'. 

\begin{table}
    	\centering
    	\caption{List of classified clusters -- their central coordinates (RA$_0$,\,Dec$_0$) from \citet{harris:10}, distance $D$ from the Sun \citep{baumgardt:21}, and the type of extra-tidal structure. The possible classifications are: `T' (a cluster with tidal tails), `E' (a cluster with an envelope), and `N' (a cluster with no detection of extra-tidal structure). Question marks indicate tentative classifications, while Whiting 1 and Pal 12 are marked with asterisks to indicate their special status embedded in the Sagittarius stream. Full details of our classification scheme are provided in Appendix \ref{a:class}.}
    	\label{t:class}        
    	\begin{tabular}{lcccc}
    		\hline
    		Name &  RA$_0$ & Dec$_0$ & $D$  & Class\\
    		 & (deg) & (deg) & (kpc) & \\
    		\hline
   		    NGC 288 & $13.188$ & $-26.583$ & $8.99\pm0.09$ & T \\
    		NGC 362 & $15.809$ & $-70.849$ & $8.83\pm0.10$ & T \\
    		Whiting 1 & $30.737$ & $-3.253$ & $30.59\pm1.17$ & N$^*$ \\
    		NGC 1261 & $48.068$ & $-55.216$ & $16.40\pm0.19$ & T\\
    		Pal 1 & $53.333$ & $+79.581$ & $11.18\pm0.32$ & T\\
    		AM 1 & $58.760$ & $-49.615$ & $118.91\pm3.40$ & N? \\ 
     		Eridanus & $66.185$ & $-21.187$ & $84.68\pm2.89$ & T\\
     		NGC 1851 & $78.528$ & $-40.047$ & $11.95\pm0.13$ & T\\
    		NGC 1904 (M79) & $81.044$ & $-24.524$ & $13.08\pm0.18$ & E?\\
    		NGC 2298 & $102.248$ & $-36.005$ & $9.83\pm0.17$ & T\\
    		NGC 2419 & $114.535$ & $+38.882$ & $88.47\pm2.40$ & N\\
    		Pyxis & $136.991$ & $-37.221$ & $36.53\pm0.66$ & N\\ 
    		NGC 2808 & $138.013$ & $-64.863$ & $10.06\pm0.11$ & T\\
    		E 3 & $140.238$ & $-77.282$ & $7.88\pm0.25$ & T\\
    		NGC 3201 & $154.403$ & $-46.412$ & $4.74\pm0.04$ & T \\
    		NGC 4590 (M68) & $189.867$ & $-26.744$ & $10.40\pm0.10$ & T \\
    		NGC 5024 (M53) & $198.230$ & $+18.168$ & $18.50\pm0.18$ & T?\\
    		NGC 5053 & $199.113$ & $+17.700$ & $17.54\pm0.23$ & E\\
    		NGC 5139 ($\omega$ Cen) & $201.697$ & $-47.480$ & $5.43\pm0.05$ & T\\
    		NGC 5272 (M3) & $205.548$ & $+28.377$ & $10.18\pm0.08$ & T? \\
    		AM 4 & $209.090$ & $-27.167$ & $29.01\pm0.94$ & N\\
    		NGC 5466 & $211.364$ & $+28.534$ & $16.12\pm0.16$ & T\\
    		NGC 5694 & $219.901$ & $-26.539$ & $34.84\pm0.74$ & E\\
    		NGC 5824 & $225.994$ & $-33.068$ & $31.71\pm0.60$ & T? \\ 
    		Pal 5 & $229.019$ & $-0.121$ & $21.94\pm0.51$ & T \\
    		NGC 5897 & $229.352$ & $-21.010$ & $12.55\pm0.24$ & N \\
    		NGC 5904 (M5) & $229.638$ & $+2.081$ & $7.48\pm0.06$ & T \\
    		Pal 14 & $242.752$ & $+14.958$ & $73.58\pm1.63$ & T \\
    		NGC 6101 & $246.450$ & $-72.202$ & $14.45\pm0.19$ & T\\
    		NGC 6205 (M13) & $250.422$ & $+36.460$ & $7.42\pm0.08$ & N \\
    		NGC 6229 & $251.745$ & $+47.528$ & $30.11\pm0.47$ & N\\
    		Pal 15 & $254.963$ & $-0.539$ & $44.10\pm1.14$ & T\\
    		NGC 6341 (M92) & $259.281$ & $43.136$ & $8.50\pm0.07$ & T \\
    		NGC 6362 & $263.979$ & $-67.048$ & $7.65\pm0.07$ & T\\
    		NGC 6397 & $265.175$ & $-53.674$ & $2.48\pm0.02$ & T\\
    		NGC 6752 & $287.717$ & $-59.985$ & $4.12\pm0.04$ & N \\
    		NGC 6809 (M55) & $294.999$ & $-30.965$ & $5.35\pm0.05$ & N\\
    		NGC 6864 (M75) & $301.520$ & $-21.921$ & $20.52\pm0.45$ & N \\
    		NGC 6981 (M72) & $313.365$ & $-12.537$ & $16.66\pm0.18$ & E \\
    		NGC 7006 & $315.365$ & $-12.537$ & $39.32\pm0.56$ & N\\
    		NGC 7078 (M15) & $322.493$ & $+12.167$ & $10.71\pm0.10$ & N \\
    		NGC 7089 (M2) & $323.363$ & $-0.823$ & $11.69\pm0.11$ & T \\
    		NGC 7099 (M30) & $325.092$ & $-23.180$ & $8.49\pm0.09$ & T\\
    		Pal 12 & $326.662$ & $-21.253$ & $18.49\pm0.30$ & N$^*$ \\
    		Pal 13 & $346.685$ & $+12.772$ & $23.48\pm0.40$ & T\\
    		NGC 7492 & $347.111$ & $-15.611$ & $24.39\pm0.57$ & N?\\
    		\hline
    	\end{tabular}
    \end{table}

To determine the orbital properties of the complete Milky Way globular cluster population, we used the 6D phase space information from \citet{vasiliev:21}. This includes line-of-sight velocities from \citet{baumgardt:19}, distances from \citet{baumgardt:21}, and proper motions computed using {\it Gaia} EDR3. To calculate the cluster orbits we use the {\sc galpy} galactic dynamics package \citep{bovy:15} to integrate the motion of each cluster forward in time for $10$\ Gyr in the Milky Way potential from \citet{mcmillan:17}, which has a total mass of $1.3\times 10^{12}\,M_\odot$. We also checked that the results outlined below do not change if we use the default {\tt MWPotential2014} potential from \citet{bovy:15}. 

\begin{figure*}
\centering
\includegraphics[width=0.75\textwidth]{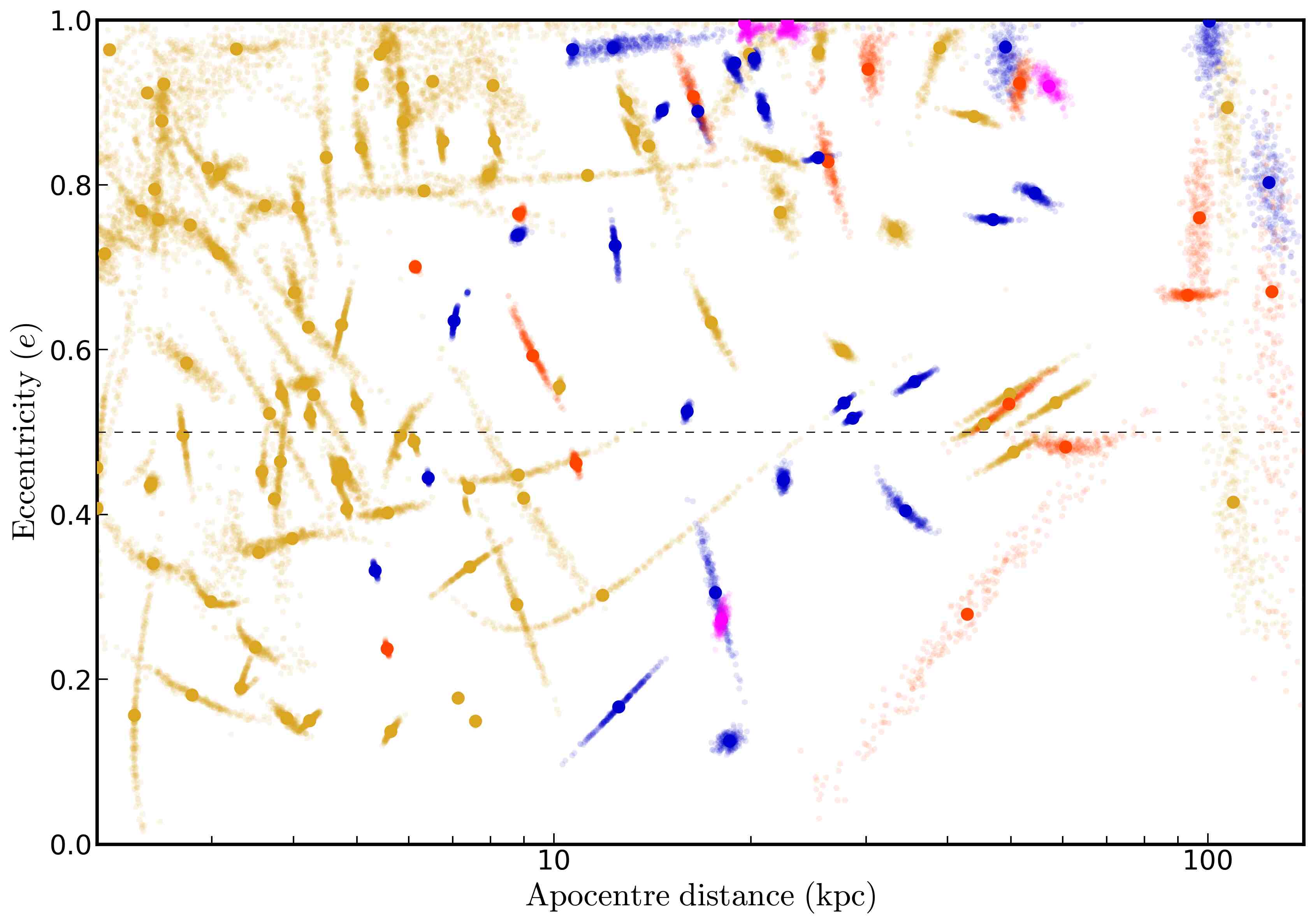}\\
\includegraphics[width=0.75\textwidth]{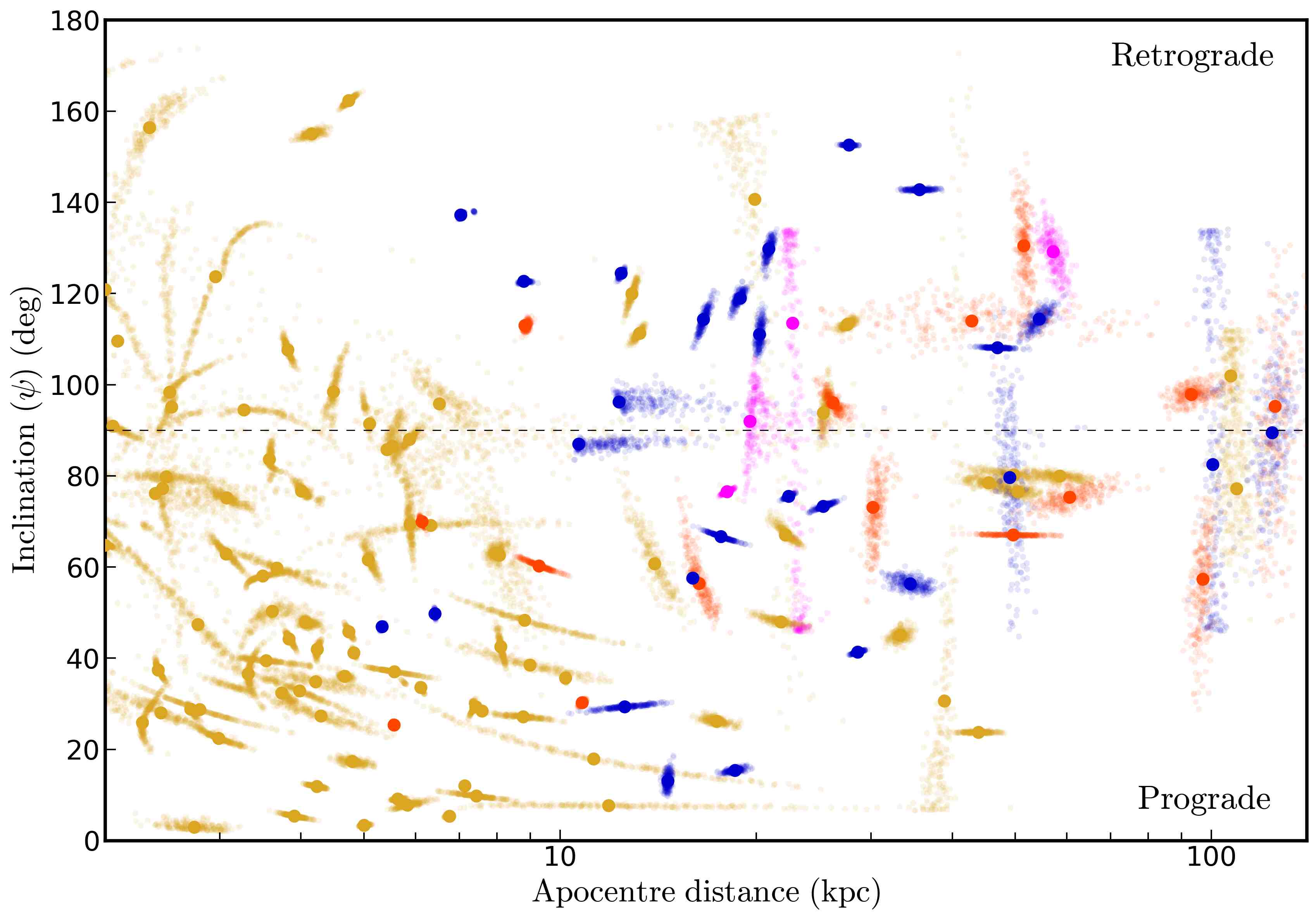}
\caption{Orbital eccentricity (upper panel) and inclination (lower panel) as a function of apocentre radius for Milky Way globular clusters. Objects classified `T' in our scheme (i.e., those with tidal tails) are marked with blue, those classified `E' (i.e., clusters with envelopes) are marked with magenta, and those classified `N' (no extra-tidal structure) are marked with orange. Objects with insufficient data to be classified under our scheme are marked with gold. Each cluster is surrounded by a cloud of points indicating the distribution of orbital parameters according to the observational uncertainties in the six-dimensional phase space. In the upper panel, an eccentricity $e=0$ indicates a circular orbit, while an $e=1$ indicates a radial orbit. In the lower panel, inclination is defined such that $\psi=0\degr$ is a prograde in-plane orbit, $\psi=90\degr$ is a polar orbit, and $\psi=180\degr$ is a retrograde in-plane orbit.}
\label{f:eccinc}
\end{figure*}

Having integrated the orbits, we estimate the eccentricity, inclination, actions, and various other parameters for each cluster. To determine the uncertainties on these quantities, we follow \citet{vasiliev:19} and \citet{vasiliev:21} in drawing random samples from the errors in the six input observables and computing new orbits. In doing this we assume that the errors in the input positions are negligible, and that the remaining four input parameters are independent apart from the two components of proper motion, for which \citet{vasiliev:21} provide the covariance coefficient.

In Figure \ref{f:eccinc} we plot orbital eccentricity and inclination as a function of apocentre radius. Clusters with tidal tails are plotted in blue, those with envelopes in magenta, and those with no detected extra-tidal structure in orange. Objects not classified under our scheme are plotted in gold. Each cluster is surrounded by a cloud of $250$ points indicating the distribution of orbital parameters calculated according to the observational uncertainties in the six-dimensional phase space as described above. An eccentricity of zero indicates a circular orbit, while an eccentricity of one indicates a radial orbit. Inclination is defined such that $\psi=0\degr$ is a prograde orbit in the plane of the Milky Way disk, $\psi=90\degr$ is a polar orbit, and $\psi=180\degr$ is a retrograde in-plane orbit.

Figure \ref{f:eccinc} reveals several things. First, we see that clusters with tails or an envelope all have an apocentre radius $\ga 5$\ kpc. This has previously been noticed by \citet{piatti:21b}, who suggested it might be due to the kinematically chaotic nature of the orbits of globular clusters in the inner parts of the Milky Way. However, Figure \ref{f:eccinc} indicates that clusters with apocentre radii inside $5$\ kpc do not have the same level of high-quality data as clusters with larger apocentre radii. This is likely a result of the difficulty in observing these objects due to their location in the Galactic bulge or inner thick disk, and means we cannot rule out that the apparent lack of inner clusters with tails or an envelope is purely a selection bias.

Figure \ref{f:eccinc} also shows that clusters with tidal tails or envelopes are commonly (although not exclusively) in moderately eccentric or very eccentric orbits -- of the $31$ objects classified `T' or `E', $23$ have $e>0.5$, and $16$ of these have $e>0.75$. Clusters with tails or envelopes are also typically (although again not exclusively) in orbits that are quite highly inclined relative to the Galactic disk -- only four of the $31$ objects classified `T' or `E' have $\psi<40\degr$ or $\psi>140\degr$. Moreover, just over half of `T' or `E' clusters have $\psi>90\degr$, indicating retrograde orbits.

\begin{figure}
\centering
\includegraphics[width=\columnwidth]{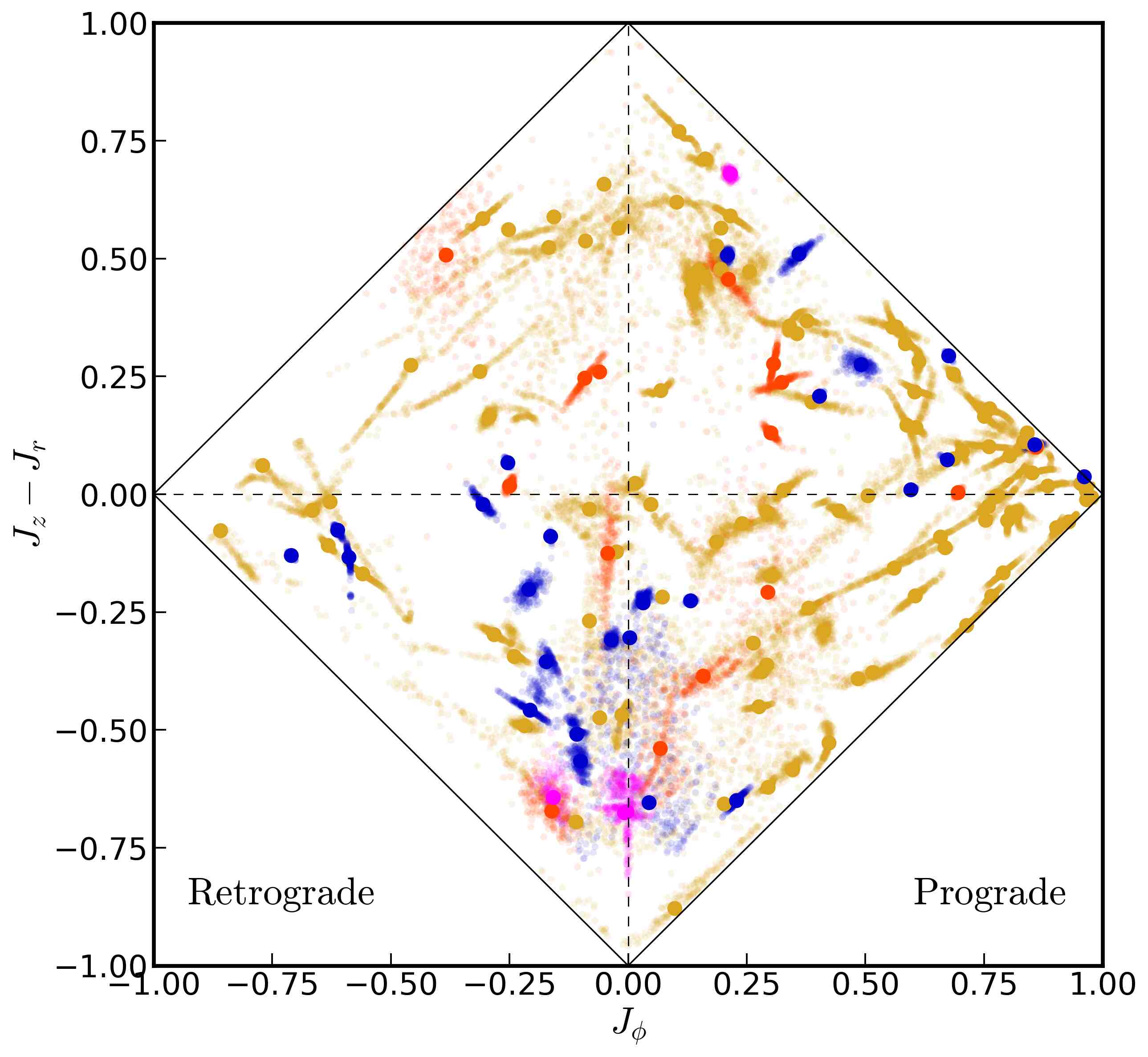}
\caption{This plot shows the action-space map of Milky Way globular clusters with orbits integrated in the \citet{mcmillan:17} potential. The axes are normalised to the total action $J_{\rm tot} = |J_\phi| + J_z + J_r$. As in the previous figure, objects classified `T' in our scheme (i.e., those with tidal tails) are marked with blue, those classified `E' (i.e., clusters with envelopes) are marked with magenta, and those classified `N' (no extra-tidal structure) are marked with orange. Objects with insufficient data to be classified under our scheme are marked with gold. Clusters with negative $J_\phi$ are in retrograde orbits.}
\label{f:actions}
\end{figure}

Figure \ref{f:actions} shows the action-space map for globular clusters.  In this plot, the $x$-axis shows the azimuthal component of the action $J_\phi$, while the $y$-axis shows the difference between the vertical and the radial actions $J_z-J_r$. Both axes are normalised by the total action $J_{\rm tot} = |J_\phi| + J_z + J_r$ so that they span the range $-1$ to $+1$. Clusters with negative $J_\phi$ are on retrograde orbits, and vice versa. Objects lying close to the boundaries of the plot at negative values on the $y$-axis have nearly in-plane orbits ($J_z \approx 0$), while those lying close to the boundaries of the plot at positive values on the $y$-axis have nearly circular orbits ($J_r \approx 0$).

From this plot, we can see that clusters with tails or envelopes fall into three distinct groups: clusters on prograde orbits with positive $J_z-J_r$ (i.e., relatively low eccentricity), clusters with $J_\phi$ close to zero and negative $J_z-J_r$ (i.e., relatively high eccentricity), and a small set of three clusters with very retrograde orbits and $J_z-J_r$ close to zero. The first group includes clusters such as Pal 1, Pal 5, and E 3, and most likely represents objects belonging to the Milky Way thick disk or {\it in situ} halo (although we note that the accreted structure identified as Wukong by \citet{naidu:20} also lies in this region of phase space). The second group of clusters has properties most similar to the so-called {\it Gaia} Enceladus-Sausage (GSE) structure \citep[e.g.,][]{helmi:18,belokurov:18} (although again, \citet{naidu:20} identify other partially-overlapping structures in this region of phase space), and includes clusters like NGC 1851, 1904, 2298, 2808, and 7089. The likelihood that many of the clusters with evident extra-tidal structures and highly radial orbits were accreted during the GSE event has previously been noted by e.g., \citet{piatti:19} and \citet{piatti:22}. Finally, the third group of objects sits in the region of phase space occupied by the so-called Sequoia structure \citep[e.g.,][]{myeong:19} and two other coincident stellar groups with different mean metallicities \citep{naidu:20}. This small group includes $\omega$ Cen, NGC 6101, and NGC 3201.

It is not clear whether the fact that many apparently accreted clusters in the Milky Way possess extra-tidal structures is directly linked to their status as accreted objects (for example, because they evolved in different galaxies than the Milky Way for at least part of their lives, or because they experienced a tidally-destructive merger event), or whether it is due to the characteristic orbits they ended up on after being accreted (many of which are retrograde and/or highly eccentric). Detailed modelling of the behaviour of clusters during and after accretion events could help shed light on this question.

\begin{figure*}
\centering
\includegraphics[width=0.75\textwidth]{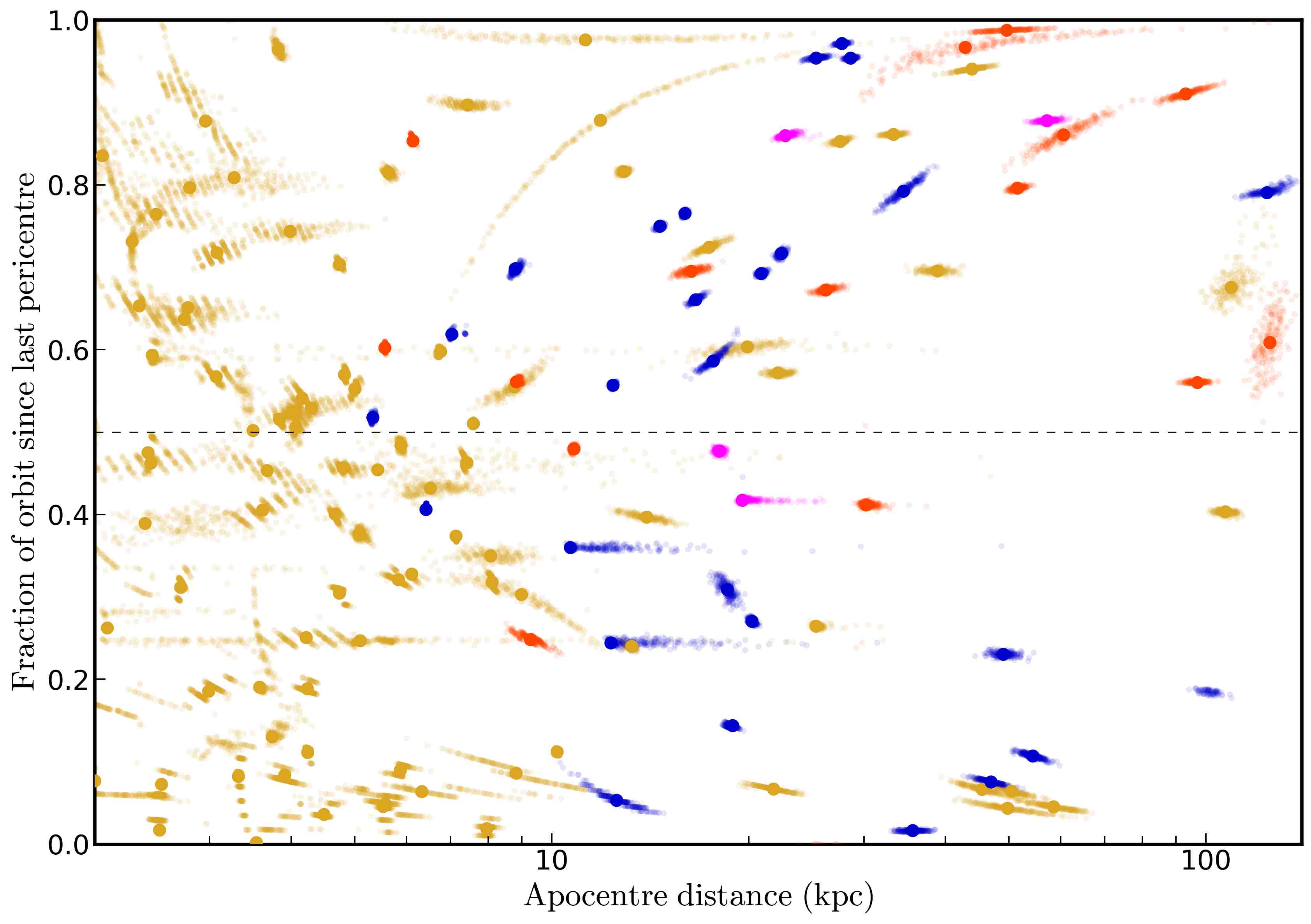}\\
\includegraphics[width=0.75\textwidth]{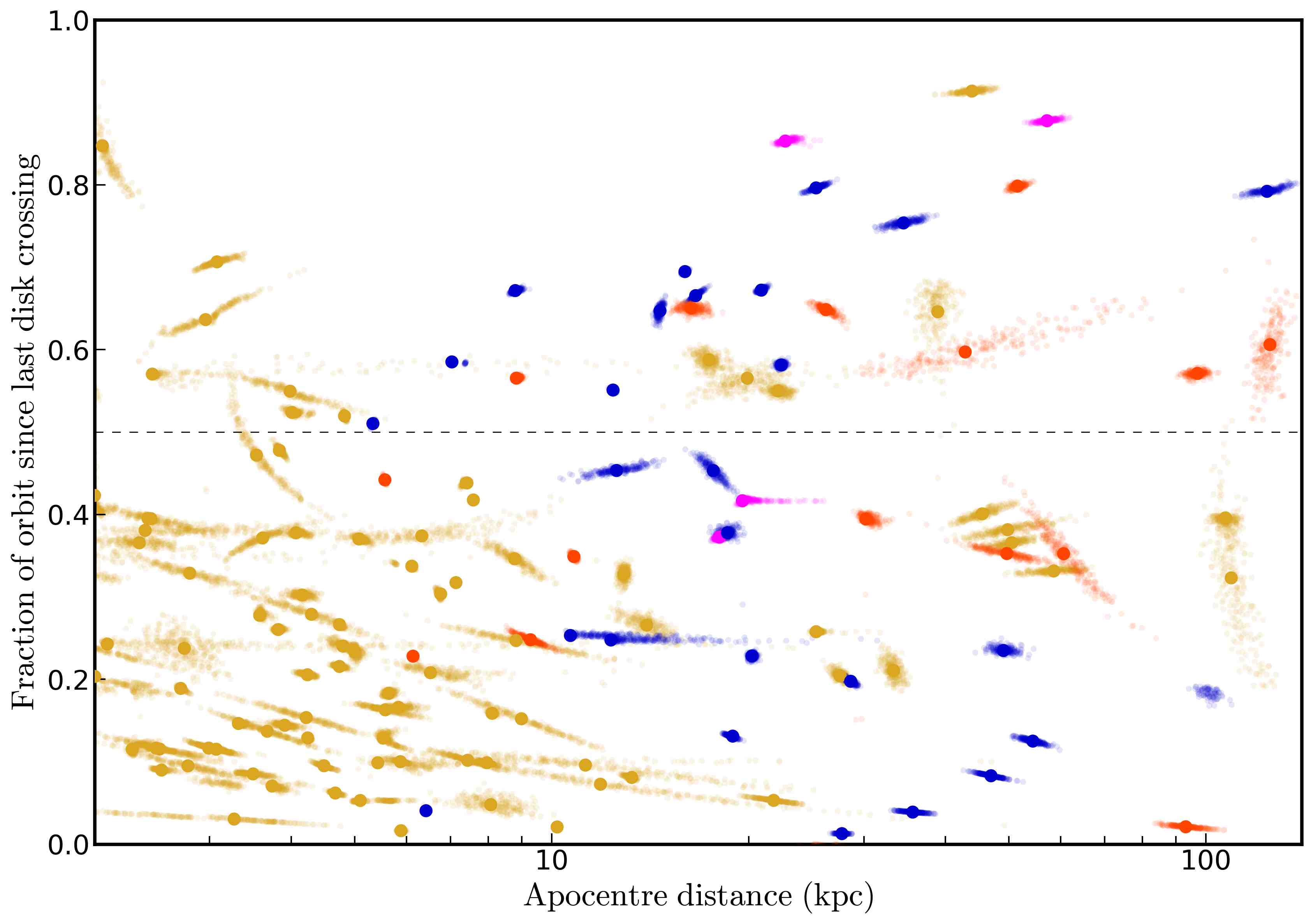}
\caption{Time since most the recent pericentre passage (upper) and since the last disk plane crossing (lower panel) as a function of apocentre radius for Milky Way globular clusters. These times are expressed as a fraction of the total orbital period for each cluster. As before, objects classified `T' in our scheme are marked with blue, those classified `E' are marked with magenta, and those classified `N' are marked with orange. Objects with insufficient data to be classified under our scheme are marked with gold. The apparent quantization of points in the upper panel is due to the finite time step during the orbit integration.}
\label{f:events}
\end{figure*}

One thing we can check with the present sample is whether there is any signature that orbital phase affects the production or observability of globular cluster extra-tidal structure. To do this we used {\sc galpy} to integrate the cluster orbits backwards in time by $4$\ Gyr, and determined how long ago each cluster experienced its most recent pericentre passage and crossing of the Galactic disk plane. We show the results of these calculations in Figure \ref{f:events}, as a fraction of the total orbital period for each object. 

There are no strong trends evident in these plots, either for clusters with identified extra-tidal structures or for those with firm non-detections. Around half of the clusters with tails or envelopes are relatively close to apocentre (i.e., sitting between $\sim0.3-0.5$ of an orbit since pericentre), but there are also plenty of clusters with tails or envelopes that are near pericentre. There is a mild indication that clusters with non-detections are preferentially past apocentre, with $12$ of $15$ such objects ($80\%$) more than half an orbit since the most recent pericentre passage. The spread in time since the last crossing of the disk plane appears relatively even, with roughly half of clusters with detected extra-tidal structure sitting either side of midline on the plot (i.e., more than half an orbit has elapsed since the most recent disk crossing, for around half the clusters). However, it is possible that a recent disk passage favours creation and/or detection of tidal tails, as $12$ of $27$ clusters with tails ($44.5\%$) have traversed less than a quarter of an orbit since their last crossing.

So far our analysis has considered all Milky Way globular clusters. However, this approach likely introduces some observational biases, for the following reasons. First, as previously noted, there are no clusters in our classified sample with apocentre radii smaller than $\sim 5$\ kpc. Any cluster on a tighter orbit than this barely moves outside the Galactic bulge, and observations are therefore complicated by both heavy field contamination, and significant and variable line-of-sight extinction. To minimise this type of bias we now only consider clusters that presently sit at a Galactocentric radius larger than $5$\ kpc. 

Second, the majority of clusters in our sample classified as possessing tidal tails have detections stemming from {\it Gaia} observations. While {\it Gaia} is clearly very good at detecting this type of structure for nearby clusters, the fact that it cannot measure reliable astrometry for stars fainter than $G\approx 20$ means that its efficiency likely declines for more distant systems. Good examples are Eridanus, Pal 13, and Pal 15, which have clear tails seen in DECam data \citep{myeong:17,shipp:20} that are not detected in e.g., the blind surveys of \citet{ibata:19b,ibata:21}. The most distant cluster for which \citet{ibata:21} observe robust tails with {\it Gaia} EDR3 is Pal 5, at a distance of $\sim 22$\ kpc. We can therefore speculate that {\it Gaia} would identify any system with long tails out to that distance, and consequently further restrict our sample to include only clusters with present-day line-of-sight distance $< 22$\ kpc.

\begin{figure*}
\centering
\includegraphics[width=\columnwidth]{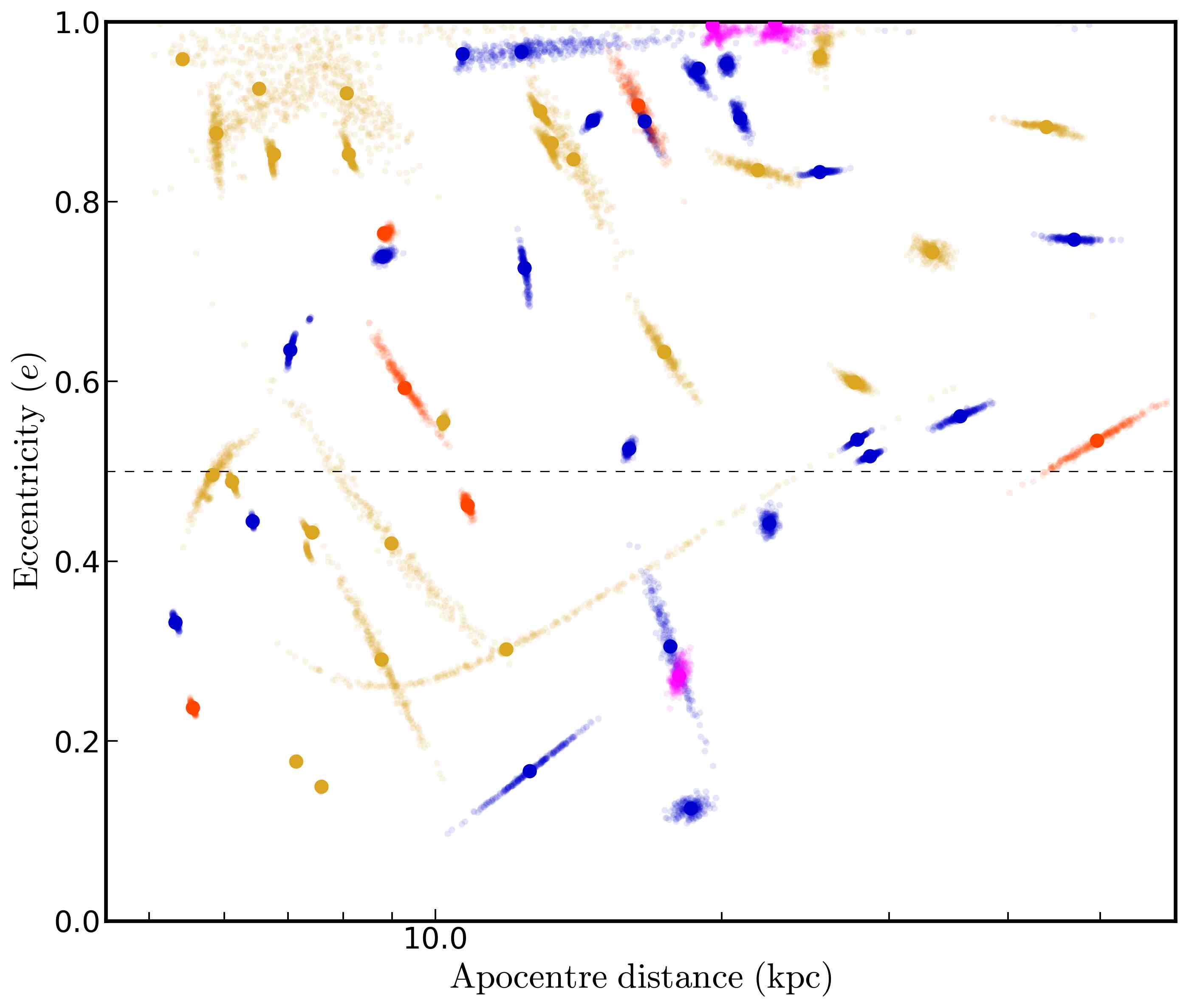}\hspace{1mm}
\includegraphics[width=\columnwidth]{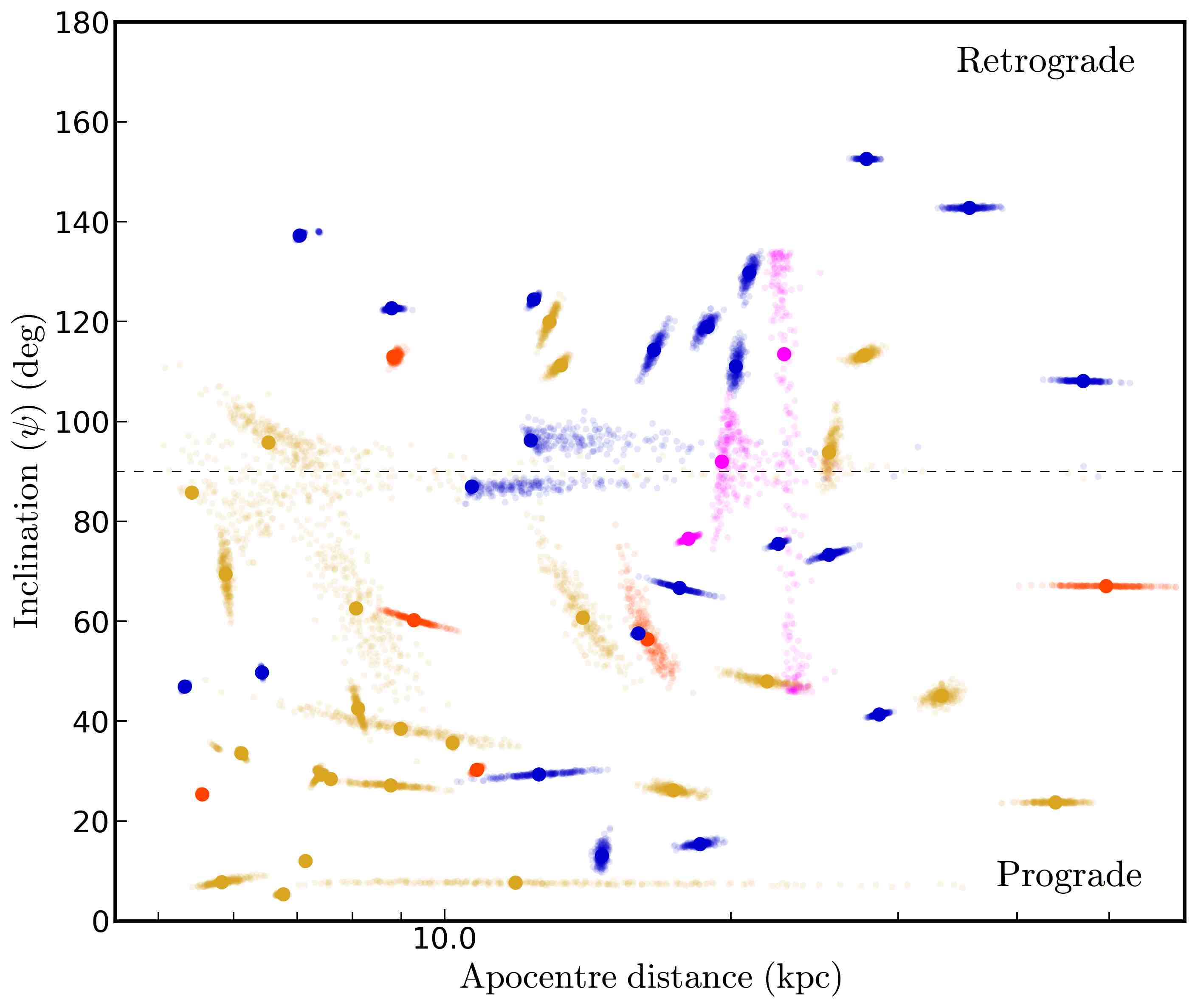}\\
\vspace{1mm}
\includegraphics[width=\columnwidth]{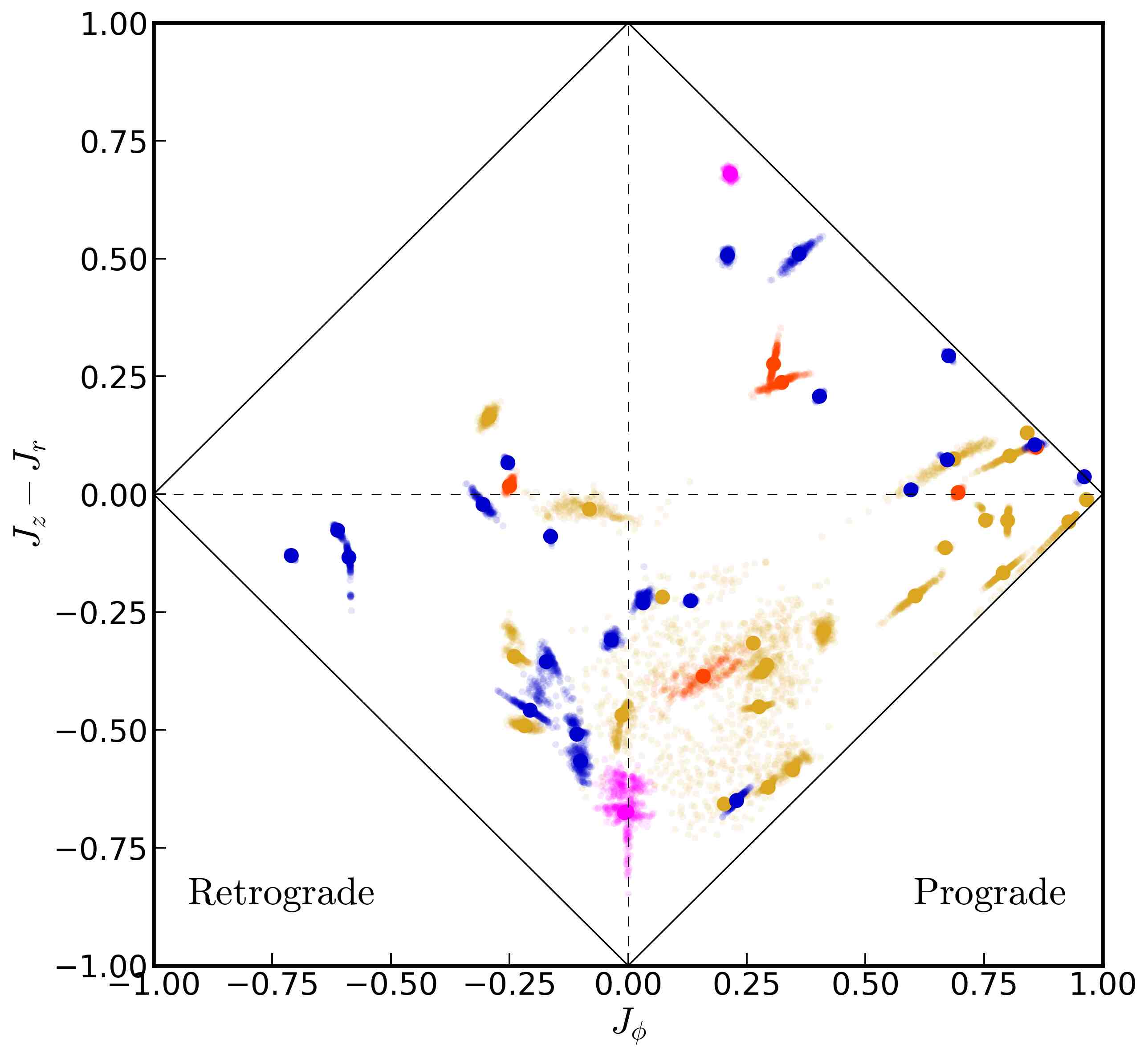}\\
\vspace{1mm}
\includegraphics[width=\columnwidth]{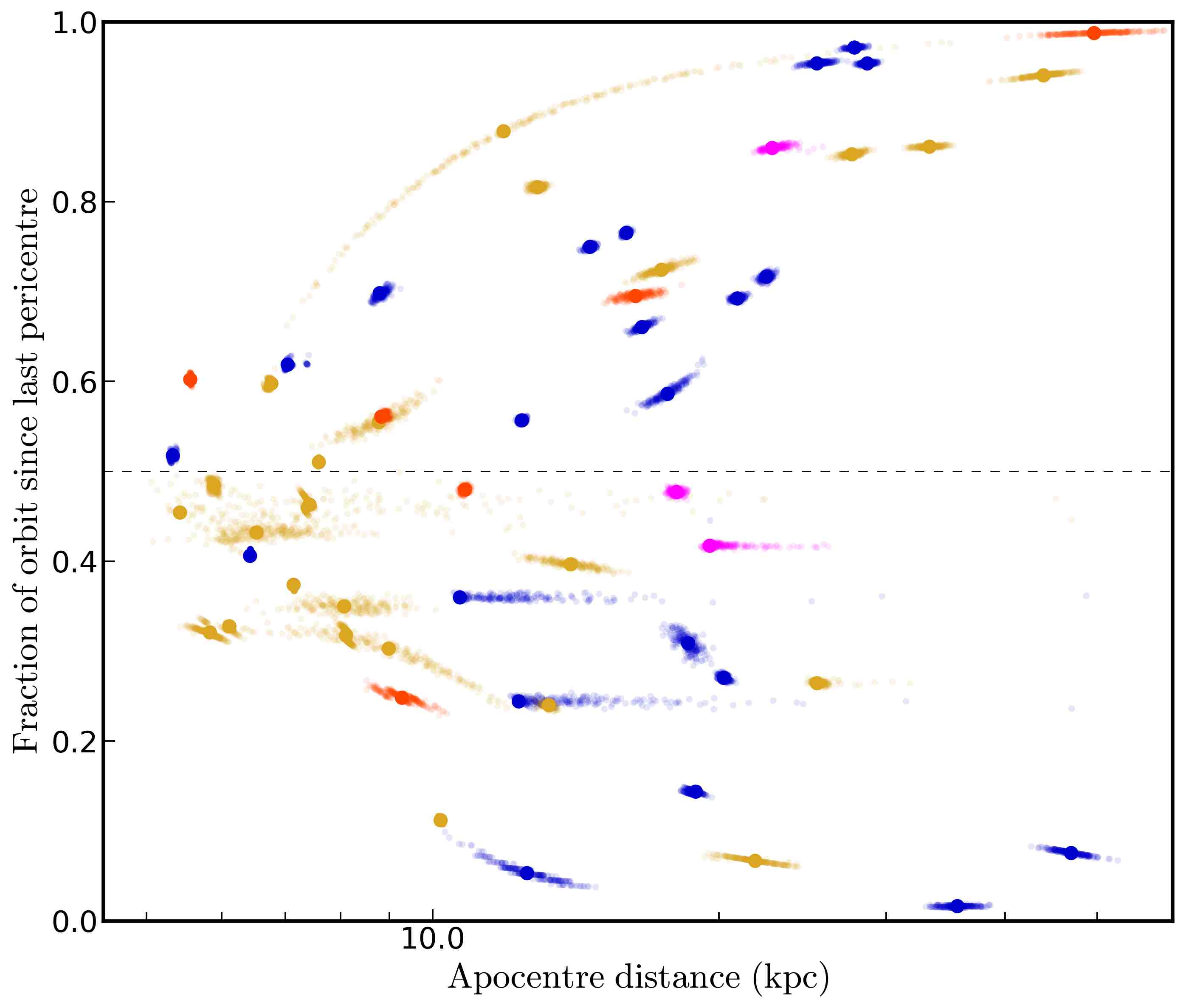}\hspace{1mm}
\includegraphics[width=\columnwidth]{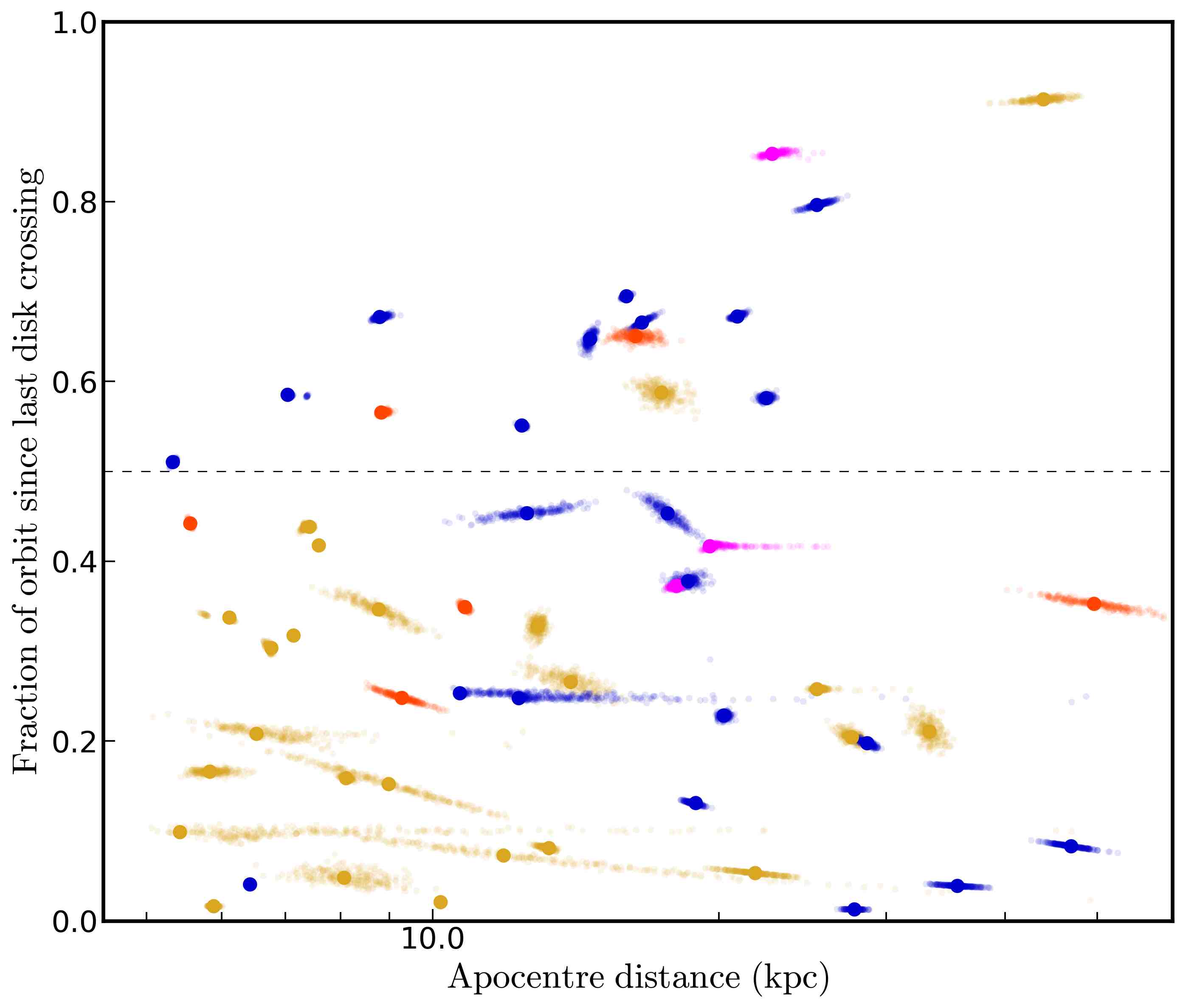}
\caption{These panels show the same plots as Figures \ref{f:eccinc}, \ref{f:actions}, and \ref{f:events}, but now for the restricted cluster sample described in the text.}
\label{f:repeat}
\end{figure*}

With these two cuts in place we are comparing our classified sample with a global set of objects that share similar orbital properties, and where selection and detection biases are reduced or minimised. The restricted cluster set contains a total of $55$ clusters, of which $31$ are in our classified list. These comprise $22$ with tails, $3$ with envelopes, and $6$ with non-detections. 

Figure \ref{f:repeat} shows the same plots as in Figures \ref{f:eccinc}-\ref{f:events} but now only for the restricted sample. We can see from these plots that some of the trends we previously noted when considering the full sample are enhanced. In particular, this selection accentuates the fact that tail clusters tend to be on polar or retrograde orbits. Of the $55$ clusters in the restricted set, $24$ ($43.5\%$) have polar or retrograde orbits (i.e., with $\psi \ga 70\degr$). However, if we consider the clusters with extra-tidal structures, this rises to $68\%$ ($17$ of $25$ objects). The fraction of clusters with tidal tails that have traversed less than a quarter of an orbit since the most recent disk crossing remains quite high ($9$ of $22$ objects, $41\%$), but is comparable to the overall population ($25$ of $55$ objects, $45.5\%$). However, the fraction of clusters with extra-tidal structure that have traversed more than half an orbit is high -- $44\%$ ($11$ of $25$ objects) as compared with $27\%$ ($15$ of $55$ objects) for the overall population.

In summary, we observe that clusters with tidal tails or extended envelopes are typically on moderately to very eccentric orbits that are highly inclined to the Galactic plane, and often retrograde. There is also a mild preference for clusters exhibiting extra-tidal structure to have traversed at least half an orbit since their last crossing of the Galactic disk plane. However, these are not strict rules. It is very easy to find exceptions -- for example, we clearly observe a set of clusters on relatively low-eccentricity, low-inclination prograde orbits, that also possess tidal tails. Similarly, there are plenty of clusters with non-detections that share similar orbital properties with clusters that do possess extra-tidal structure. The fact that we do not find an orbital parameter that is a strong predictor for the formation of extra-tidal structure in globular clusters matches the conclusion reached by \citet{piatti:20b}, even though recent analyses (especially using {\it Gaia} data) have allowed us to update many of their classifications. Since it appears that many clusters with tidal tails or stellar envelopes are associated with accretion events in the Milky Way halo (especially the GSE event), we speculate that the variation in the outer structures of objects with notionally similar orbital properties could be due to the inhomogeneous dynamical histories of clusters in the Milky Way system.

\subsection{Globular cluster streams with no remaining progenitor}
\citet{li:21} have recently presented orbital parameters for six stellar streams in the outer Milky Way halo (all at Galactocentric radius larger than $10$\ kpc) with properties indicating that they stem from completely disrupted globular clusters. Although we are beholden to small number statistics, it is interesting to note that these streams typically have less eccentric orbits than many of the clusters in our sample with tidal tails (only one of the six has $e>0.6$) and mostly exhibit prograde motion (four of the six). However, they all possess highly inclined orbits ($60\degr \leq \psi \leq 110\degr$), which is a trait common to a substantial proportion of our tidal tail clusters. 

\citet{li:21} also note a possible correlation between metallicity and eccentricity for their streams, such that more metal-poor streams possess lower eccentricity; however, we do not find a similar correlation in our sample of intact clusters with tails -- for all $[$Fe$/$H$]<-1$ we observe the full range of eccentricities $0.2\la e \la 1.0$ (this is true even if we restrict ourselves to the subset of clusters with Galactocentric radius larger than $10$\ kpc). In particular, our sample has a number of clusters with metallicity similar to the metal-poor \citet{li:21} streams, but possessing highly elliptical orbits (for example, NGC 2298, NGC 5466, and Pal 15). 

While this comparison is currently somewhat limited, plenty of additional streams with disrupted cluster progenitors are known \citep[cf.][]{bonaca:21} such that more detailed examination will be possible in the near future.  This has the potential to reveal key information about the tidal disruption of clusters, especially in the context of different dynamical histories.

\subsection{The nature of extended stellar envelopes}
As described earlier, ground-based wide-field photometric studies have revealed large diffuse stellar envelopes around several globular clusters. The most striking of these are NGC 1851, NGC 5824, and NGC 7089 \citep[all with radial extent $\ga 210$\ pc, see][]{olszewski:09,kuzma:16,kuzma:18}, while NGC 1261 possesses a smaller example \citep[with radial extent $\approx 100$\ pc, as in][]{kuzma:18}. The origin of these structures has proven difficult to identify. One suggestion is that they comprise the ghostly remains of now-destroyed host dwarfs -- precisely the type of external stellar populations we have searched for in the present paper. Another is that they are due to the build-up of stars in the process of escaping a cluster due to tidal erosion \citep[so-called "potential escapers" -- e.g.,][]{kupper:10,daniel:17,deboer:19}.  

While compiling the set of classified clusters described in the previous section, we noticed that the most prominent examples of systems with stellar envelopes (including the four clusters mentioned above) have now all been identified with {\it Gaia} as possessing long tidal tails \citep{ibata:21,bonaca:21}. This strongly suggests that a substantial stellar envelope or radially extended structure is indeed indicative of stars being tidally removed from a cluster. Furthermore, it suggests that the two clusters identified in the present work as possessing envelope-type features in their outskirts (NGC 1904 and NGC 6981), as well as the two additional clusters identified as `E' in our classification scheme (NGC 5053 and NGC 5694) are likely to possess tidal tails\footnote{Indeed, as already suggested for NGC 1904 by \citet{shipp:18}.} and would be fruitful targets for future examination using {\it Gaia} astrometry.

\section{Conclusions}
\label{s:conclusions}
We have presented the results of deep wide-field imaging, using the mosaic camera DECam, of the outskirts of the halo globular clusters NGC 1904, NGC 2298, NGC 6864, NGC 6981, NGC 7492, Whiting 1, AM 1, Pyxis, and IC 1257. Apart from Whiting 1 (which is known to be embedded in the stream from the Sagittarius dwarf) and NGC 7492 (which is projected against different wraps of the same stream), we see no evidence for adjacent stellar populations that would support the idea that any of these clusters is associated with coherent tidal debris from a destroyed host dwarf galaxy. Since many of the clusters in our sample are thought to be accreted objects, this is consistent with {\it Gaia} measurements of halo field stars that indicate such events occurred many Gyr ago and their debris is now spatially well mixed in the Galactic halo.

In addition, our deep DECam data reveals no evidence for tidal tails around any of the clusters in our sample. However, both NGC 1904 and NGC 6981 do appear to possess outer envelope-type structures. In general our results are consistent with previous comparable measurements in the literature, although we are unable to reproduce the detection of tidal tails around NGC 7492 by \citet{navarrete:17}. Our work represents the first time the outer structure of IC 1257 has been studied, although our photometry does not extend far past the main-sequence turn-off for this difficult target.

NGC 2298 is a particularly interesting cluster. Although deep, wide-field imaging studies of this system, including the present work, have in general failed to uncover any significant outer structure, recent {\it Gaia}-based studies by \citet{sollima:20} and \citet{ibata:21} have revealed tidal tails spanning $\sim 12\degr$ on the sky. Indeed, {\it Gaia} astrometry has proven extremely effective for detecting stream-like structures in the halo, including many associated with globular clusters. In particular, we note that most of the clusters known to possess prominent stellar envelopes or unusually extended structures (such as NGC 1851, NGC 5824, and NGC 7089, and to a lesser extent NGC 1261) have all been shown by {\it Gaia}-based work to possess long tidal tails. This is experimental confirmation that the presence of an extended envelope is indicative of stars being tidally removed from a cluster.

Motivated by the slew of recent {\it Gaia} discoveries, and by the existence of full six-dimensional phase space information for all Galactic globular clusters, we carried out a literature survey to identify a robust sample of systems known to possess tidal tails or stellar envelopes, and then investigated the extent to which the presence or absence of these structures can be linked to cluster orbital properties. We find that clusters with tidal tails or extended envelopes are typically on moderately to very eccentric orbits that are highly inclined to the Galactic plane, and often retrograde. There is also a mild preference for clusters exhibiting extra-tidal structure to have traversed at least half an orbit since their last crossing of the Galactic disk plane. 
However, these are neither necessary nor sufficient conditions for clusters to possess extra-tidal structure. We clearly observe a set of tidal-tail clusters on relatively low-eccentricity, low-inclination prograde orbits; similarly, there are plenty of clusters with non-detections that share similar orbital properties with clusters that do possess extra-tidal structure. We speculate that this lack of consistency may stem, at least in part, from the inhomogeneous dynamical histories of clusters in the Milky Way system -- many objects with extra-tidal structure appear to be associated with at least two of the main accretion events that formed the Milky Way halo ({\it Gaia}-Enceleadus-Sausage and Sequoia).

\section*{Acknowledgements}
DM gratefully acknowledges support from an Australian Research Council Future Fellowship (FT160100206). 

This project used data obtained with the Dark Energy Camera (DECam), which was constructed by the Dark Energy Survey (DES) collaboration. Funding for the DES Projects has been provided by the US Department of Energy, the US National Science Foundation, the Ministry of Science and Education of Spain, the Science and Technology Facilities Council of the United Kingdom, the Higher Education Funding Council for England, the National Center for Supercomputing Applications at the University of Illinois at Urbana-Champaign, the Kavli Institute for Cosmological Physics at the University of Chicago, Center for Cosmology and Astro-Particle Physics at the Ohio State University, the Mitchell Institute for Fundamental Physics and Astronomy at Texas A\&M University, Financiadora de Estudos e Projetos, Funda\c{c}\~{a}o Carlos Chagas Filho de Amparo \`{a} Pesquisa do Estado do Rio de Janeiro, Conselho Nacional de Desenvolvimento Científico e Tecnol\'{o}gico and the Minist\'{e}rio da Ci\^{e}ncia, Tecnologia e Inova\c{c}\~{a}o, the Deutsche Forschungsgemeinschaft and the Collaborating Institutions in the Dark Energy Survey.

The Collaborating Institutions are Argonne National Laboratory, the University of California at Santa Cruz, the University of Cambridge, Centro de Investigaciones En\'{e}rgeticas, Medioambientales y Tecnol\'{o}gicas–Madrid, the University of Chicago, University College London, the DES-Brazil Consortium, the University of Edinburgh, the Eidgen\"{o}ssische Technische Hochschule (ETH) Z\"{u}rich, Fermi National Accelerator Laboratory, the University of Illinois at Urbana-Champaign, the Institut de Ci\`{e}ncies de l’Espai (IEEC/CSIC), the Institut de F\'{i}sica d’Altes Energies, Lawrence Berkeley National Laboratory, the Ludwig-Maximilians Universit\"{a}t M\"{u}nchen and the associated Excellence Cluster Universe, the University of Michigan, NSF’s NOIRLab, the University of Nottingham, the Ohio State University, the OzDES Membership Consortium, the University of Pennsylvania, the University of Portsmouth, SLAC National Accelerator Laboratory, Stanford University, the University of Sussex, and Texas A\&M University.

Based on observations at Cerro Tololo Inter-American Observatory, NSF’s NOIRLab (NOIRLab Prop. ID 2013A-0617, 2013B-0617, and 2014A-0621, all PI: D. Mackey; and 2014A-0620, PI: A. Casey), which is managed by the Association of Universities for Research in Astronomy (AURA) under a cooperative agreement with the National Science Foundation.

The national facility capability for SkyMapper has been funded through ARC LIEF grant LE130100104 from the Australian Research Council, awarded to the University of Sydney, the Australian National University, Swinburne University of Technology, the University of Queensland, the University of Western Australia, the University of Melbourne, Curtin University of Technology, Monash University and the Australian Astronomical Observatory. SkyMapper is owned and operated by The Australian National University's Research School of Astronomy and Astrophysics. The survey data were processed and provided by the SkyMapper Team at ANU. The SkyMapper node of the All-Sky Virtual Observatory (ASVO) is hosted at the National Computational Infrastructure (NCI). Development and support of the SkyMapper node of the ASVO has been funded in part by Astronomy Australia Limited (AAL) and the Australian Government through the Commonwealth's Education Investment Fund (EIF) and National Collaborative Research Infrastructure Strategy (NCRIS), particularly the National eResearch Collaboration Tools and Resources (NeCTAR) and the Australian National Data Service Projects (ANDS).

\section*{Data Availability}
The data underlying this article will be shared on reasonable request to the corresponding author.



\bibliographystyle{mnras}
\bibliography{ms} 



\appendix
\section{Cluster classifications}
\label{a:class}
In this Appendix we provide full details of our classification scheme for the extra-tidal properties of Milky Way globular clusters. We first explain the key criteria underpinning our scheme (which differ somewhat from those in previous efforts), and then provide a note for each cluster that satisfies these criteria, with a detailed accounting of our rationale for the adopted classification. Finally, we briefly discuss the objects which have previous literature studies, but that we are unable to include in our present classified set given the available data.

\subsection{Classification scheme}
As discussed in Section \ref{s:discussion}, one issue that has plausibly affected previous compilations of the extra-tidal properties of Galactic globular clusters is inhomogeneity -- that is, combining the results of studies with different methodologies, different data quality, different areal coverage, etc. Even within the small sample of clusters considered in the present work, disagreements are seen between e.g., studies which use deep ground-based photometry extending several magnitudes below the main-sequence turn-off, and those based on substantially shallower data (see NGC 2298 and NGC 7492). 

In an effort to mitigate this problem, and given the recent explosion in both {\it Gaia}-based studies of globular cluster outskirts (where astrometry can be used to filter non-cluster contaminants with high efficiency), and deep ground-based wide-field imaging extending well past the main-sequence turn-off (which enhances the signal due to cluster members while suppressing noise from contaminants), we adopt here a hierarchical set of criteria for classifying clusters. The main basis underpinning our scheme is to consider only systems which have been studied using either $>$4D {\it Gaia} phase-space information plus photometry, or deep wide-field imaging where photometry of cluster members extends at least $\sim 1.5$\ mag (and preferably $\ga 2.5$\ mag) below the main-sequence turn-off:  
\begin{enumerate}
\item{We place highest weight on studies using algorithms tuned for detecting long stream-like structures in the Galactic halo from {\it Gaia} data \citep[e.g.,][]{ibata:19b,ibata:21}, and/or associating known streams with globular clusters using orbit integration based on full 6D phase space information \citep[e.g.,][]{bonaca:21}. Example clusters with detected tails or streams in this category include NGC 1851, NGC 3201, and NGC 5024.\vspace{1mm}}
\item{We next consider studies using {\it Gaia} data to explore the immediate outskirts of clusters, beyond the Jacobi radius \citep[e.g.,][]{carballo:19,sollima:20}. Clusters in this category include NGC 362, E 3, and NGC 7099 (tails), as well as NGC 5897 and NGC 6205 (no structure). An example object where a study satisfying this category is superseded by one in the top category is NGC 4590, which has no obvious structure in its outskirts but is strongly associated with the Fj\"{o}rm halo stream.\vspace{1mm}}
\item{Lower weight is placed on studies such as the present work \citep[see also e.g.,][]{munoz:18a,munoz:18b}, that use traditional techniques to analyse deep, wide-field ground-based imaging of the immediate outskirts of clusters. To be eligible for this category, photometry must extend {\it at least} $2.5$\ magnitudes below the main-sequence turn-off and the areal coverage around the cluster must be close to complete. Example clusters in this category include Pal 14 (tails), NGC 1904 (envelope), and NGC 6809 (no structure). A case where a cluster with a study in this category is superseded by one from a higher category is NGC 7099, where deep DECam photometry does not reveal tidal tails detected by {\it Gaia}.\vspace{1mm}}
\item{Our lowest-weight category is reserved for clusters with deep wide-field imaging of their outskirts, but where the photometry extends between $\sim 1.5-2.5$\ magnitudes below the main-sequence turn-off. Again the areal coverage must be essentially complete. In this case, two independent studies satisfying these criteria must have reached consistent conclusions. Example clusters in this category include AM 1 and Pyxis (no structure). A case where a study in this category is superseded by a category above is NGC 7492.\vspace{1mm}}
\item{Clusters with data not satisfying any of the above remain unclassified. A list of such objects is at the end of the Appendix.} 
\end{enumerate}

\subsection{Classified clusters}
In the following list, a classification of `T' means a cluster with tidal tails or a tidal stream, `E' means a cluster with an extended stellar envelope but no obvious tails, and `N' means a cluster with no detection of extra-tidal structure. We offer frequent reference to the most recent similar compilation of Milky Way cluster structures, by \citet[][hereafter PCB20]{piatti:20b}. These authors used a similar three-type classification scheme (albeit under a different set of criteria) for clusters with tails (`G1'), an envelope (`G2'), or no extra-tidal structure (`G3'). For simplicity we discuss these using our current notation (i.e., `G1'$\equiv$`T', `G2'$\equiv$`E', and `G3'$\equiv$`N').

\begin{itemize}
\item{NGC 288: PCB20 classify this cluster as `T' on the basis of deep DECam imaging from the Dark Energy Survey (DES) \citep{shipp:18}. Recent {\it Gaia} measurements from DR2 \citep{kaderali:19,sollima:20} and EDR3 \citep{ibata:21} support this result with a very robust detection of tidal tails; \citet{ibata:21} show that these features span more than $10\degr$ on the sky. We maintain the `T' classification.\vspace{2mm}}

\item{NGC 362: \citet{carballo:19} provide a clear {\it Gaia} DR2 detection of short tails around this cluster, spanning approximately $\pm 1.5\degr$ on the sky. On this basis PCB20 classify NGC 362 as `T', and we keep the same classification.\vspace{2mm}}

\item{Whiting 1: This small cluster is embedded in the large tidal stream from the Sagittarius dwarf galaxy. The deep photometry presented in the present paper (see Section \ref{ss:whiting1}) revealed no evidence for extra-tidal structure belonging to the cluster, as did the comparable, but independent, data of \citet{munoz:18b}. We hence classify the cluster as `N', but note that it is somewhat of a special case because it has only very recently been liberated from Sagittarius.\vspace{2mm}}

\item{NGC 1261: \citet{kuzma:18} detected a power-law envelope around NGC 1261 using deep DECam photometry, although the radial density profile of this cluster is also arguably consistent with a very extended "{\sc limepy}" model contained within the expected Jacobi radius \citep[as in e.g.,][]{deboer:19}. On this basis PCB20 provide an `E' classification. However, DES photometry from \citet{shipp:18}, which is deeper than that of \citet{kuzma:18} and spans a wider area, reveals clear extra-tidal structure, and \citet{ibata:21} use {\it Gaia} EDR3 to demonstrate the presence of a long tail-like feature spanning up to $\pm 10\degr$ on the sky. Hence we adopt a `T' classification for this object.\vspace{2mm}}

\item{Pal 1: This is an extremely low-luminosity globular cluster. \citet{nieder:10} used SDSS photometry extending more than two magnitudes below the main-sequence turn-off to demonstrate the presence of tidal tails spanning $\pm 1\degr$ from Pal 1. They subsequently confirm the tidal distortion of the cluster using deep {\it Hubble Telescope} photometry. We maintain the `T' classification of PCB20.\vspace{2mm}}

\item{AM 1: This is among the most distant globular clusters in the Milky Way and is not in the PCB20 compilation. In the present paper (see Section \ref{ss:am1}) we find no evidence for extra-tidal structure; however, our photometry does not reach far past the main-sequence turn-off. Using comparable, but independent, photometry, \citet{munoz:18b} reach a similar conclusion. Consequently we tentatively adopt an `N' classification for AM 1, while noting that deeper imaging is required for robust confirmation.\vspace{2mm}}

\item{Eridanus: This is another extremely distant halo cluster, which PCB20 classify as `T' based on the DECam detection of tidal tails by \citet{myeong:17}. \citet{munoz:18b} present independent deep photometry for Eridanus, but are unable to confirm this detection. However, they acknowledge that their photometry is at least a magnitude shallower than that of \citet{myeong:17}, and an inspection of their two-dimensional density map (their Figure 7) reveals an elongation of the outermost contours with magnitude and direction consistent with the measurements of \citet{myeong:17}. On this basis we maintain the `T' classification.\vspace{2mm}}

\item{NGC 1851: This cluster has long been known to possess a large, extended power-law envelope reaching $\sim 2\degr$ from its centre \citep{olszewski:09,kuzma:18,carballo:14,carballo:18a}. The {\it Gaia} DR2 study of \citet{sollima:20}, spanning to a radius of $5\degr$, did not reveal any more extended structure than this; however, the more recent EDR3 work of \citet{ibata:21} suggests diffuse tail-like features spanning nearly $10\degr$ on the sky. Consequently, we classify this cluster as `T'.\vspace{2mm}}

\item{NGC 1904 (M79): The deep DECam photometry presented in this paper reveals a symmetric power-law envelope, although previous studies (as outlined in Section \ref{ss:NGC1904}) have shown that this cluster's radial density profile is also consistent with extended (non King-like) models and is contained within the expected Jacobi radius. However, deep DES photometry presented by \citet{shipp:18} suggests the presence of extra-tidal features around NGC 1904, extending to a radius of $\sim 1.5\degr$ from its centre and oriented approximately N-S. These could be tails; however, the {\it Gaia} DR2 map of \citet{sollima:20}, while appearing to show a similar structure, reveals no additional features out to $5\degr$. On this basis we tentatively maintain the `E' classification of PCB20.\vspace{2mm}}

\item{NGC 2298: As discussed in Section \ref{ss:NGC2298}, several early studies suggested the possible presence of mild extra-tidal or extended structure around this cluster \citep[e.g.,][]{leon:00,balbinot:11}. On this basis PCB20 classify NGC 2298 as `E', although we note that more recent investigations based on deeper ground-based photometry, including the present work, find little evidence to support this conclusion.  Both of recent the {\it Gaia}-based studies by \citet{sollima:20} and \citet{ibata:21}, however, clearly reveal the presence of substantial tidal tails oriented SE-NW; in particular, the latter authors show that these tails span $\sim 12\degr$ on the sky. We therefore classify this cluster `T'.\vspace{2mm}}

\item{NGC 2419: PCB20 classify this cluster as `E' based on the SDSS study of \citet{jordi:10}. However, \citet{ibata:13} provide much deeper CFHT/MegaCam data and a very precise investigation of the cluster's radial density profile that shows no evidence for an envelope \citep[see also][]{munoz:18a,munoz:18b}. We elect to classify this cluster as `N'.\vspace{2mm}}

\item{Pyxis: This cluster is not in PCB20. In the present paper (see Section \ref{ss:pyxis}) we find no evidence for extra-tidal structure using photometry that reaches $\sim 1.5$\ mag below the main-sequence turn-off. A similar conclusion was obtained by \citet{munoz:18b}, using slightly deeper photometry. Given the consistency between these two studies, we classify Pyxis `N'.\vspace{2mm}}

\item{NGC 2808: PCB20 classify this cluster as `E' based on \citet{carballo:18a} who show a power-law radial density decline outside the formal King tidal radius \citep[but entirely within the Jacobi radius, see also][]{deboer:19}, possibly with a mild SE-NW elongation. Using {\it Gaia} DR2, \citet{kundu:21} find evidence for extra-tidal stars in proximity to the cluster, although \citet{sollima:20}, also using {\it Gaia} DR2, sees no obvious nearby structure. However, \citet{ibata:21}, using {\it Gaia} EDR3, observe clear tidal tails belonging to NGC 2808, stretching SE-NW and spanning $\pm 10\degr$ on the sky. We therefore classify this cluster as `T'.\vspace{2mm}}

\item{E 3: This is a very low-luminosity cluster, not included in the compilation by PCB20. Recent work using {\it Gaia} DR2 by \citet{carballo:20} has, however, revealed short tidal tails (approximately $\pm 1\degr$ in length) and we therefore classify it as `T'.\vspace{2mm}}

\item{NGC 3201: PCB20 classify this object as `E' on the basis of extra-tidal stars from the RAVE catalogue detected by \citet{kunder:14}; more recently, using {\it Gaia} DR2, \citet{sollima:20} observed apparent tidal deformation in the outskirts of the cluster, while \citet{bianchini:19} found the signature of tidal tails extending to approximately twice the Jacobi radius \citep[see also][]{wan:21}. Also using {\it Gaia} DR2, \citet{palau:21} showed that NGC 3201 is in fact responsible for an extremely long ($\ga 100\degr$) tidal stream in the Milky Way halo, first identified as Gj\"{o}ll by \citet{ibata:19a}. This conclusion is supported by the subsequent studies of \citet{hansen:20}, \citet{ibata:21}, and \citet{bonaca:21}; we hence classify NGC 3201 as `T'.\vspace{2mm}}

\item{NGC 4590 (M68): PCB20 classify as this as `T' based on the Gaia DR2 detection of long tidal tails by \citet{palau:19}, who associated the cluster with the Fj\"{o}rm stream discovered by \citet{ibata:19a}. Both \citet{ibata:21} and \citet{bonaca:21} confirm this association, although interestingly \citet{sollima:20} finds no evidence for tail-like structure or tidal elongation within a radius of $5\degr$ from the centre of NGC 4590. We maintain the `T' classification for this cluster.\vspace{2mm}}

\item{NGC 5024 (M53): PCB20 class this object as `N' since neither \citet{jordi:10} nor \citet{carballo:12,carballo:14} detected any evidence for extra-tidal structure. However, by calculating orbital properties, \citet{bonaca:21} associate NGC 5024 with both the Sylgr \citep{ibata:19a} and Ravi \citep{shipp:18} halo streams. Since the analysis of \citet{bonaca:21} also identifies the robust NGC 3201-Gj\"{o}ll, NGC 4590-Fj\"{o}rm, and NGC 5139-Fimbulthul associations found by several other studies (see the relevant entries in this list), it seems likely that NGC 5024 constitutes a system with very long tails but no easily detectable structure around the cluster boundary (similar to NGC 4590). On this basis we tentatively assign a it `T' classification.\vspace{2mm}}

\item{NGC 5053: \citet{lauchner:06} claimed the detection of a tidal tail emanating from this cluster, from an analysis of SDSS photometry. \citet{jordi:10} observe mild irregularity in its outermost isophotes, also using SDSS data; however, they note that the presence of Sagittarius debris in the surrounding area complicates analysis of this object. The deeper photometry presented by \citet{carballo:12,carballo:14} does not appear to support the presence of significant extra-tidal material around NGC 5053; however, it spans at most one quadrant of the cluster outskirts. The very deep, wide-field photometry of \citet{chun:10} appears to show an envelope-type structure around the cluster, while the {\it Gaia} DR2 density profile presented by \citet{deboer:19} plausibly extends past the Jacobi radius. We therefore classify this cluster `E', matching the conclusion of PCB20.\vspace{2mm}}

\item{NGC 5139 ($\omega$ Cen): This is the most massive globular cluster in the Milky Way. PCB20 classify it as `T' based on the detection of long tidal tails using {\it Gaia} DR2 by \citet{ibata:19b,ibata:19a} -- the so-called Fimbulthul stream. This detection is reinforced by \citet{ibata:21}; the tails are also easily detectable in the region immediately surrounding the cluster, as shown by \citet{sollima:20} and \citet{kuzma:21}. We classify this cluster as `T'.\vspace{2mm}}

\item{NGC 5272 (M3): PCB20 classify this cluster as `N', based on the ground-based photometric study of \citet{carballo:14}. This classification is reinforced by \citet{sollima:20}, who find no significant tidal structure within a radius of $5\degr$ of the cluster. However, \citet{bonaca:21} associate NGC 5272 with the Sv\"{o}l stream identified by \citet{ibata:19a}, on the basis of a comparison between the stream and cluster orbits. Since their machinery also robustly detects other well-known cluster-stream associations (as discussed above), it seems plausible that like NGC 4590 and 5024, NGC 5272 may possess very long tails without any obvious structure around the immediate cluster boundary. We therefore tentatively assign it a `T' classification.\vspace{2mm}}

\item{AM 4: This is a very low luminosity halo cluster, which PCB20 classify as `N' based on the study by \citet{carballo:14}. More recently, \citet{munoz:18b} presented photometry extending $\sim 3.5$\ mag below the main-sequence turn-off. They see no evidence for tidal tails, although the outer isopleths in their density map are notably elliptical. We maintain the `N' classification for this cluster but note that additional study could prove fruitful.\vspace{2mm}}

\item{NGC 5466: This is a classic example of a globular cluster with long tidal tails \citep[e.g.,][]{belokurov:06,grillmair:06a,chun:10,jordi:10}; consequently, PCB20 classify it as `T'. \citet{ibata:21} are clearly able to see the tails of NGC 5466 with {\it Gaia} EDR3.  We maintain the `T' classification.\vspace{2mm}}

\item{NGC 5694: This cluster is in Milky Way's outer halo and has been studied by several authors, all with comparably deep, but independent, photometry extending $\ga 2$\ mag below the main-sequence turn-off \citep{correnti:11,carballo:12,carballo:14,munoz:18b}. All have found NGC 5694 to be remarkably extended, with a power-law density profile in its outskirts. However, the outer density contours appear round and undistorted \citep{correnti:11,munoz:18b}. While it is unlikely the cluster extends past the expected Jacobi radius \citep{deboer:19}, its properties are sufficiently similar to other well-known clusters with large photometrically-detected envelopes (e.g., NGC 1851, NGC 7089) that we maintain the `E' classification adopted by PCB20.\vspace{2mm}}

\item{NGC 5824: PCB20 classify this cluster as `N' based on the deep DECam imaging of \citet{kuzma:18}, who showed that although NGC 5824 is a remarkably extended cluster \citep[see also][]{munoz:18b}, its radial density profile can be well reproduced by a \citet{wilson:75} or \citet{gieles:15} "{\sc limepy}" model \citep[see also][]{deboer:19}. While there are no more recent studies of the cluster outskirts, \citet{bonaca:21} associate NGC 5824 with both the Triangulum and Turbio halo streams on the basis of their orbital properties \citep[see also the discussion in][]{li:21}. Given this, we tentatively classify NGC 5824 as `T'.\vspace{2mm}}

\item{Pal 5: This is the classic example of a globular cluster with exceptionally long tidal tails, first seen in SDSS by \citet{odenkirchen:01,odenkirchen:03}. The tails are easily found in Gaia EDR3 by \citet{ibata:21}, and we maintain the `T' classification of PCB20.\vspace{2mm}}

\item{NGC 5897: This cluster is not in the compilation of PCB20. Using {\it Gaia} DR2, \citet{sollima:20} sees no evidence for tails or extra-tidal structure, so we classify it as `N'.\vspace{2mm}}

\item{NGC 5904 (M5): PCB20 classify this as `T' based on the SDSS study of \citet{jordi:10} and the \citet{grillmair:19} Gaia DR2 detection of a long tail. Both \citet{ibata:21} and \citet{sollima:20} also clearly see this tail in their respective {\it Gaia} studies, and we maintain the `T' classification.\vspace{2mm}}

\item{Pal 14: PCB20 classify this distant cluster as `T' based on the detection of tails by \citet{sollima:11}. The very deep data of \citet{munoz:18b} are consistent with this detection, and we maintain the `T' classification.\vspace{2mm}}

\item{NGC 6101: This cluster is not in the list compiled by PCB20. However, \citet{ibata:21} recently revealed that it possesses tails spanning $\approx10\degr$ on the sky; we classify as `T'. \vspace{2mm}}

\item{NGC 6205 (M13): PCB20 classify this cluster as `N' based on \citet{jordi:10}; using {\it Gaia} DR2 \citet{sollima:20} also sees no evidence for tidal tails or extra-tidal extension and we maintain the `N' classification.\vspace{2mm}}

\item{NGC 6229: This cluster has been studied by \citet{munoz:18b}, who presented photometry extending $\sim 3.5$\ mag below the main-sequence turn-off, as well as \citet{carballo:12,carballo:14}, who used shallower photometry covering approximately one-third of the cluster outskirts. Neither study found evidence for extra-tidal structure or tidal distortion; \citet{deboer:19} shows that the profile is extended but well contained within the expected Jacobi radius. We maintain the `N' classification adopted by PCB20.\vspace{2mm}}

\item{Pal 15: PCB20 classify this cluster as `T' based on the DECam detection of tidal tails by \citet{myeong:17}. As with Eridanus, \citet{munoz:18b} present independent and somewhat shallower photometry but cannot reproduce this detection. However, the outermost contours in their two-dimensional density map (their Figure 15) exhibit mild elongation in a direction compatible with the tails found by \citet{myeong:17}. We default in favour of the deeper photometry and maintain the `T' classification.\vspace{2mm}}

\item{NGC 6341 (M92): PCB20 classify this object as `N' based on the SDSS-based study of \citet{jordi:10}. However, both \citet{ibata:21} and \citet{sollima:20} see evidence in {\it Gaia} data for a long tidal tail, and \citet{thomas:20} clearly detect the same feature (spanning $\sim17\degr$ on the sky) using deep CFHT photometry. We therefore classify NGC 6341 as `T'.\vspace{2mm}}

\item{NGC 6362: PCB20 classify this cluster as `E' based on the {\it Gaia} DR2 study of \citet{kundu:19}, who found evidence for extra-tidal stars. More recently, \citet{sollima:20} demonstrates a clear detection of short ($\sim1\degr$) tidal tails, also using {\it Gaia} DR2, so we classify as `T'.\vspace{2mm}}

\item{NGC 6397: This is one of the closest globular clusters, and is not in the compilation of PCB20. \citet{ibata:21} find evidence for a long tidal tail $\approx 25\degr$ in length, using {\it Gaia} EDR3 \citep[see also][]{kundu:21}. While \citet{boldrini:21} are unable to reproduce this detection, we elect to classify this cluster `T' based on the generally robust success of the \citet{ibata:19a,ibata:21} {\sc streamfinder} algorithm.\vspace{2mm}}

\item{NGC 6752: This is another cluster not in the PCB20 catalogue. \citet{sollima:20} finds no evidence for substantial tidal tails using {\it Gaia} DR2, but suggests a possible tidal elongation at low significance. With no further information presently available, we classify this cluster as `N'.\vspace{2mm}}

\item{NGC 6809 (M55): This cluster is not in PCB20. However, \citet{piatti:21b} recently presented DECam imaging reaching nearly $6$ magnitudes below the main-sequence turn-off, and found no evidence for extra-tidal structure. On this basis we classify as `N'.\vspace{2mm}}

\item{NGC 6864 (M75): PCB20 classify this object as `N', based on the work of \citet{carballo:12,carballo:14}. As discussed in Section \ref{ss:NGC6864}, neither the deeper DECam observations in the present paper, nor the majority of other more recent studies \citep[excepting][]{piatti:22}, suggest the presence of significant extra-tidal structure. On balance we maintain the `N' classification.\vspace{2mm}}

\item{NGC 6981 (M72): This cluster is not in the list compiled by PCB20. However, as discussed in Section \ref{ss:NGC6981}, both our deep DECam data, and the recent comparable study by \citet{piatti:21a}, support the idea that NGC 6981 likely possesses a low-luminosity envelope.  Hence we classify this cluster as `E'.\vspace{2mm}}

\item{NGC 7006: This outer halo cluster is in the sample of \citet{munoz:18b}, who presented photometry extending $\sim 4$\ mag below the main-sequence turn-off. Their two-dimensional density map shows no evidence for tidal distortion, and the profile is not as extended as those of NGC 5694 or NGC 5824. Previously, \citet{jordi:10} claimed the detection of an envelope-type structure around NGC 7006, leading PCB20 to adopt an `E' classification. However, based on the much deeper photometry of \citet{munoz:18b} we elect to classify as `N'.\vspace{2mm}} 

\item{NGC 7078 (M15): PCB20 classify this cluster as `E' based on the SDSS study by \citet{jordi:10}. However, the much deeper photometry presented by \citet{carballo:12,carballo:14} provides no significant detection of extra-tidal structure, nor does the {\it Gaia} DR2 study of \citet{sollima:20}. We therefore classify `N'.\vspace{2mm}}

\item{NGC 7089 (M2): PCB20 classify this object as as `N' based on the SDSS study of \citet{jordi:10}. However, the deep DECam work by \citet{kuzma:16} provided a robust detection of a large power-law envelope. While \citet{sollima:20} finds no evidence for tidal tails in {\it Gaia} DR2, \citet{ibata:21} present the discovery of extended ($\sim 22\degr$) tidal tails using {\it Gaia} EDR3. We therefore classify NGC 7089 as `T'.\vspace{2mm}}

\item{NGC 7099 (M30): This cluster is not in PCB20; however, \citet{sollima:20} claim the detection of high significance tails using {\it Gaia} DR2. While \citet{piatti:20c} is unable to reproduce these using deep DECam photometry, we decide in favour of the {\it Gaia} result and classify `T'.\vspace{2mm}}

\item{Pal 12: This is another small cluster that is embedded in the Sagittarius stream. \citet{musella:18} presented photometry extending at least $\sim 3$\ mag below the main-sequence turn-off, but did not find any evidence for extended or extra-tidal structure belonging to this cluster. We therefore follow PCB20 in adopting an `N' classification, although as with Whiting 1 we note this as a special case because Pal 12 has only recently left the Sagittarius system.\vspace{2mm}}

\item{Pal 13: This low-luminosity cluster is not included in PCB20. However, \citet{shipp:20} find evidence for tidal tails extending approximately $\pm 5\degr$ from its centre, using data from the DECaLS survey that reaches at least $\sim 2.5$ mag below the main-sequence turn-off. This is consistent with previous measurements from \citet{piatti:20a}, who observed cluster members beyond the Jacobi radius. While the very deep data of \citet{munoz:18b} did not reveal the tidal tails later detected by \citet{shipp:20}, this is likely due to the limited spatial coverage of their data compared with the contiguous DECaLS survey imaging. \citet{munoz:18b} do measure a relatively high ellipticity for Pal 13, with the position angle of its major axis oriented similarly to the tails found by \citet{shipp:20}. We classify this cluster `T'.\vspace{2mm}}

\item{NGC 7492: PCB20 classify this cluster as `T' based on the detection of tidal tails by \citet{navarrete:17} using Pan-STARRS PS1. However, as discussed in Section \ref{ss:NGC7492}, subsequent studies utilising deeper photometry -- such as the DECam data presented here, or the CFHT and Magellan data of \citet{munoz:18a,munoz:18b} -- have not been able to reproduce this detection, finding no evidence for extra-tidal structure around this cluster. With no means of reconciling these contradictory outcomes, we decide in favour of the deeper data and tentatively classify as `N'.\vspace{2mm}}
\end{itemize}

\subsection{Omitted clusters}
The following list shows the set of clusters with previous literature studies, but which we have decided not to include in our present classified set. We emphasise that our comments should not be read as criticism of the various studies that are mentioned; rather they are intended to offer an explanation of the reasons why existing measurements do not meet the (quite stringent) criteria in the classification scheme we have developed.

\begin{itemize}
\item{NGC 104 (47 Tuc): The region around this cluster was studied by \citet{piatti:17}. While the photometry presented in this study is very deep, background contamination from the Small Magellanic Cloud led the author to consider only the region around the main-sequence turn-off; as a consequence, this cluster does not pass the classification criteria outlined above.\vspace{2mm}} 

\item{Pal 2: This cluster is in the outer halo of the Milky Way, but sits at low Galactic latitude and is heavily reddened. It is in the sample of \citet{munoz:18a,munoz:18b}; however, their photometry does not quite reach the main-sequence turn-off.\vspace{2mm}}

\item{Pal 3: This is a very distant halo cluster. The deep wide-field photometry presented by \citet{munoz:18a,munoz:18b} extends barely a magnitude below the main-sequence turn-off. Shallower imaging was considered by \citet{sohn:03}, \citet{hilker:06}, and \citet{jordi:10}.\vspace{2mm}}

\item{Pal 4: This is another distant halo cluster; again, the wide-field photometry presented by \citet{munoz:18a,munoz:18b} does not reach far below the main-sequence turn-off. Shallower imaging is presented in \citet{sohn:03}, \citet{hilker:06}, and \citet{jordi:10}.\vspace{2mm}}

\item{Crater: This is the outermost known cluster in the Milky Way. While \citet{weisz:16} present {\it Hubble Telescope} data extending nearly $4$\ mag below the main-sequence turn-off, their imaging does not span sufficiently far from the cluster to allow detailed study of the region around and beyond its expected Jacobi radius.\vspace{2mm}}

\item{NGC 4147: Although \citet{jordi:10} studied this cluster with SDSS, their photometry does not extend far past the main-sequence turn-off. \citet{carballo:12,carballo:14} present much deeper data, but spanning only one quadrant of the cluster outskirts.\vspace{2mm}}

\item{Rup 106: This cluster is a difficult target, sitting at low Galactic latitude. The wide-field data presented by \citet{carballo:14} reaches only $\sim 1$\ mag below the main-sequence turn-off.\vspace{2mm}}

\item{NGC 5634: \citet{carballo:12,carballo:14} present photometry extending more than three magnitudes below the main-sequence turn-off for this cluster, but their imaging spans only one quadrant of the cluster otuskirts.\vspace{2mm}}

\item{NGC 6266 (M62): This is a metal-poor cluster situated towards the Galactic bulge. \citet{han:17} present an imaging survey of its outskirts; however, their photometry does not extend past the main-sequence turn-off and spans only one quadrant of the region surrounding the cluster. This object was also studied in the infrared by \citet{chun:15}, but their photometry again only reaches the main-sequence turn-off.\vspace{2mm}}

\item{NGC 6273 (M19): This is another metal-poor cluster in the Galactic bulge, studied by \citet{han:17}. As with NGC 6266, the presented photometry does not extend past the main-sequence turn-off, and spans only one quadrant of the cluster outskirts.\vspace{2mm}}

\item{IC 1257: Although we study this cluster in the present work, our photometry does not reach much below the main-sequence turn-off.\vspace{2mm}}

\item{NGC 6544: This is a metal-poor cluster situated towards the Milky Way bulge. \citet{cohen:14} used {\it Hubble Telescope} imaging of the cluster centre combined with wide-field infrared photometry of its outskirts and the surrounding area. While the central measurements extend well past the main-sequence turn-off, the wide-field data are much shallower and only just reach this level.\vspace{2mm}}

\item{NGC 6626 (M28): This is a metal-poor bulge cluster, imaged in the infrared by \citet{chun:15}. As with the other clusters in this study, the presented photometry does not reach past the main-sequence turn-off.\vspace{2mm}}

\item{NGC 6642: This is a third metal-poor bulge cluster in the sample of \citet{chun:15}. As with the other clusters in this study, the presented photometry does not reach past the main-sequence turn-off.\vspace{2mm}}

\item{NGC 6681 (M70): A third metal-poor cluster in the Galactic bulge studied by \citet{han:17}. Again, the photometry only reaches the main-sequence turn-off and covers only one quadrant of the cluster outskirts.\vspace{2mm}}

\item{NGC 6723: This is the fourth cluster in the sample of \citet{chun:15}; again the photometry does not reach past the main-sequence turn-off.\vspace{2mm}}

\item{NGC 6779 (M56): The area around this cluster was examined by \citet{piatti:19} using Pan-STARRS PS1 data; however, the considered photometry only reaches about $1$\ magnitude below the main-sequence turn-off.\vspace{2mm}} 

\end{itemize}

\bsp	
\label{lastpage}
\end{document}